\documentclass[pra,superscriptaddress,aps,nofootinbib]{revtex4}

\usepackage{amsmath,epsfig,amsfonts,graphicx,hyperref}
\usepackage{subfigure}
\usepackage{graphicx}
\usepackage[american]{babel}
\usepackage{amssymb}
\usepackage{amsfonts}
\usepackage{color}
\usepackage{comment}
\usepackage{marvosym}
\usepackage{ifsym}
\usepackage[normalem]{ulem}

\def\re{\text{Re}}
\def\im{\text{Im}}
\newcommand{\rmi}{\mathrm{i}}
\begin{document}

	\newcommand{\be}{\begin{equation}}
	\newcommand{\ee}{\end{equation}}
	\newtheorem{remark}{Remark}[section]
	\newtheorem{definition}{Definition}[section]
	\newtheorem{theorem}{Theorem}[section]
	\newtheorem{proposition}{Proposition}[section]
	\newtheorem{lemma}{Lemma}[section]

\def\re{\text{Re}}
	\def\im{\text{Im}}
	\def\labelitemi{$\blacktriangleright$}
	
\title{Exciting extreme events in the damped and AC-driven NLS equation through plane wave initial conditions}

	%
	\author{Sevastos Diamantidis}
	\affiliation{Department of Mathematics, University of the Aegean, Karlovassi, GR
		83200
		Samos, Greece}	
	\author{Theodoros P. Horikis}
	\affiliation{Department of Mathematics, University of Ioannina, Ioannina GR 45110,
		Greece}
	\author{Nikos I. Karachalios\footnote{Corresponding author. E-mail: karan@uth.gr}}
	\affiliation{Department of Mathematics, University of Thessaly, Lamia, GR
		35100, Greece}

\begin{abstract}
We investigate, by direct numerical simulations, the dynamics of the damped and forced nonlinear
Schr\"odinger (NLS) equation in the presence of a time periodic forcing and for certain
parametric regimes. It is thus revealed, that the wave-number of a plane-wave initial condition
dictates the number of emerged Peregrine type rogue waves at the early stages of modulation
instability. The formation of these events gives rise to the same number of transient
``triangular" spatio-temporal patterns, each of which is reminiscent of the one emerging in the
dynamics of the integrable NLS in its semiclassical limit, when supplemented with vanishing
initial conditions. We find that the $L^2$-norm of the spatial derivative and the $L^4$-norm
detect the appearance of rogue waves as local extrema in their evolution. The impact of the
various parameters and noisy perturbations of the initial condition in affecting the above
behavior is also discussed. The long time behaviour, in the parametric regimes where the extreme
wave events are observable, is explained in terms of the global attractor possessed by the system
and the asymptotic orbital stability of spatially uniform continuous wave solutions.
\end{abstract}
\maketitle

\textbf{\small{The damped and AC-driven nonlinear Schr\"odinger equation, a prototypical model for
complex dynamical behavior, when supplemented with the simplest case of plane wave initial
conditions, exhibits behavior which supports the universality of the Peregrine rogue wave and
semiclassical type dynamics. These dynamics may be used to explain extreme event formation in the
presence of non-integrable perturbations that describe damping and external forcing effects.}}

\section{Introduction}

Key element in the study of extreme events and rogue waves is the study of the mechanisms for their
generation and predicting the location and time of their occurrence \cite{pre0,pre1,pre2,pre3}.
This is particularly difficult as the mechanisms that generate these waves are complex and highly
nonlinear in nature. Furthermore, as modulation instability (MI) \cite{ZO} is believed to be the
main mechanism for their formation \cite{EPeli,Kharif1,Kharif2}, another complication is added, as
different initial states, under MI, will produce different events in both amplitude, and location.

It is argued and justified by numerous experimental observations, that in deep water and other
physical systems like optics, plasmas \cite{hydro,opt,plasma} etc., rogue wave dynamics are
governed by the nonlinear Schr\"odinger (NLS) equation and its variants. These rogue wave events
are modelled by the so-called Peregrine soliton \cite{H_Peregrine}, which is a rational solution of
the NLS equation, in its integrable form \cite{akh}.

However, when additional phenomena are added, such as viscosity, wind effects, opposing currents or
higher order effects \cite{Ex1,Ex2,Ex3,Ex4,Ex5,Ex6,Ex7,Ex8}, the integrable NLS system is
insufficient to describe the resulting dynamics. As such, additional, physically relevant terms,
are incorporated in the model in order to account for these different effects.  In this spirit, the
present paper investigates once more, the wealth of the dynamics arising from the variant of the
NLS equation with linear damping and a time-periodic forcing term:
	\begin{eqnarray}
	\label{eq1}
	\mathrm{i}{{u}_{t}}+\frac{\nu}{2}{{u}_{xx}}+\sigma|u|^2u =-\mathrm{i}\gamma
	u+\Gamma \exp(\mathrm{i}\Omega t).
	\end{eqnarray}
The parameter $\nu$ measures the second-order dispersion, $\sigma$ the strength of the
nonlinearity; both $\nu,\sigma>0$ are motivated by the focusing integrable limit. The parameter
$\gamma>0$ is the strength of the linear damping (loss). The parameter $\Gamma>0$ measures the
amplitude of time-periodic forcing and $\Omega\in\mathbb{R}$ is its time frequency; subscripts
denote partial derivatives. The model is of significance in numerous physical contexts,  from
nonlinear optics to plasma physics \cite{BoYu}, and is one of the prototypical partial differential
equations exhibiting complex spatiotemporal behavior \cite{nobe1,nobe2,NB86}. Generically, this is
described by the existence of a global attractor for the associated infinite dimensional system.
For the existence, finite dimensionality and regularity of the global attractor we refer to
\cite{Ghid88,Goubet1,Goubet2} (see also \cite[pg. 234-255]{RTem}), \cite{XW95,PL95,NK2002} for the
periodic initial-boundary value and Cauchy problems of the damped and forced NLS in 1D and higher
dimensional spatial domains. The complicated structure of the global attractor has been further
elucidated by the analysis of the bifurcations the system undergoes \cite{Li,Wig1,kai},
particularly in the presence of a rich variety of soliton solutions which have been approximated
analytically \cite{Bar1,barashenkov,Bar2,Bar3,Boris4,Bar5,Bar6,Bar7,Bar8}. In many cases such
soliton solutions are the attractors of the system, in their stability regime. 	

In the present paper, it is further revealed that the richness of the dynamics of \eqref{eq1} can
be further enhanced by the emergence of extreme wave events generated by the simplest continuous
wave (cw) initial condition
	\begin{eqnarray}
	\label{eq2}
	u(x,0)=u_0(x)=A\exp\left(-\frac{\mathrm{i}K\pi x}{L}\right),\;\;K\in\mathbb{N},
	\end{eqnarray}
	when the system is supplemented with periodic boundary conditions in the
	fundamental interval $\mathcal{Q}=[-L,L]$,
	\begin{eqnarray}
	\label{eq3}
	u(-L,t)=u(L,t).
	\end{eqnarray} More precisely, we show that the initial condition \eqref{eq2} creates dynamical
effects, which are --to our knowledge-- novel and can be relevant to mechanisms for the generation
of rogue waves. At this point, we have to stress that our investigations also suggest that these effects may be highly unstable under noisy perturbations of the specific initial condition \eqref{eq2}, as it will be explained below. These effects are the following:

In certain regimes for the parameters of the system, the initial condition \eqref{eq2}  dictates by
its wave number $K$ and for suitable initial amplitudes $A$, the emergence of $K$ Peregrine-type
rogue waveforms at the initial stages of its evolution. By a comparison against the profile of the
analytical Peregrine rogue wave (PRW) of the integrable limit ($\gamma=\Gamma=0$) it is judged that
the profiles of these PRW-type waveforms do not only have similar profiles, but also share for a
significant-time interval, growth and decay rates, remarkably close to that of the PRW of the
integrable limit. Their similarity is further judged by examining the spectrum of the emerged
structures which is identical to the analytical PRWs. Another new dynamical effect we identified,
is the following: the PRW-type waveforms which emerge as first events in the dynamics are on the
top of  $K$ triangular-type (``Christmas tree'') spatiotemporal patterns \cite{BS}. Despite the
fact that the initial condition does not decay at infinity, each one of these patterns  possesses a
structure similar to the single spatiotemporal pattern formed in the case of the integrable NLS, in
its semi-classical limit \cite{BM1,BM2, Rev1_b}. Hence, the results suggest the universality of
this dynamics  even in the particular example of the AC-damped and driven NLS \eqref{eq1}, when
supplemented with periodic boundary conditions \cite{Rev1_b,BioMan}.

We discuss the potential structural robustness of the above dynamical behavior, as the several
parameters of the system are varied, and we find that it persists up to certain threshold values
for the parameters.  We found that is far from the dynamics of the integrable limit as it is not
observed in this limit when the same initial condition is posed. Furthermore, we examined the
stability of the observed patterns when the initial condition \eqref{eq2} is perturbed by noise. We
found that the noise perturbation destruct the previous structures. The observed instability
motivated us for a repetition of the numerical experiment for the semiclassical limit NLS with  the
same parameters and vanishing initial condition given in \cite{basis, BM1}, perturbing now this
initial condition by noise. We found again an instability destructing the characteristic triangular
pattern of a ``rogue waves lattice'' which in the presence of noise is replaced by the evolution of
breather modes. As rogue waves are highly unstable solutions \cite{PNJ}, the findings suggest that
the presence of noise enhances the instability of the background sustaining the rogue waveforms. Therefore, the numerical findings for the noisy perturbed initial conditions suggest that the structured semiclassical type dynamics in both cases of systems -the damped and AC-driven NLS and the conservative NLS in its semiclassical limit-are very sensitive under noisy perturbations and that  spatiotemporal chaos dominates in their entire dynamical behaviour.

It is well known that the existence of localized modes as solitons and breathers, can be proved by
the construction of  variational problems which may identify the localized structures as extrema of
suitable energy functionals. This approach has been successfully applied  in continuous, as well as
in discrete models  \cite{JSh,TTao,Wein99,Pan1}.  In this regard, we observed the following
dynamical feature: the square of the $L^2$-norm of the spatial derivative
\begin{eqnarray}
\label{derN}
D(t)=\int_\mathcal{Q}|u_x(x,t)|^2dx,
\end{eqnarray}
and the
$L^4$-functional
\begin{eqnarray}
\label{L4N}
||u(t)||_{L^4}^4=\int_{\mathcal{Q}}|u(x,t)|^4dx,
\end{eqnarray}
detect the emergence of the rogue waves as local extrema in their evolution in a time
interval of enhanced instability, exactly at the time of their appearance.
We found that only these functionals exhibit this detection property of extreme
events as opposed to either of the standard conserved quantities of the
integrable limit, namely the momentum,
\begin{eqnarray}	
\label{mom2}
M(t)=\mathrm{Im}\int_{\mathcal{Q}}u(x,t)\overline{u_x}(x,t)dx,
\end{eqnarray}
the power (also called mass or wave
action),
\begin{eqnarray}
	\label{l2f}
	P(t)=||u(t)||^2_{L^2}=\int_{\mathcal{Q}}|u(x,t)|^2dx,
\end{eqnarray}
the Hamiltonian energy,
\begin{eqnarray}
\label{Hameneg} H(t)=\nu\int_{\mathcal{Q}}|u_x(x,t)|^2dx-\sigma\int_{\mathcal{Q}}|u(x,t)|^4dx.
\end{eqnarray}
The quantities \eqref{mom2}-\eqref{Hameneg} are three of the conserved quantities in an infinite
hierarchy in the integrable case. In fact it was the conservation of the Hamiltonian energy of the
integrable limit which motivated us to investigate the behaviour of its components $D(t)$ and
$||u(t)||_{L^4}^4$. For $D(t)$, the manifestation of local extrema can be explained by the steep
gradients of the Peregrine-type waveforms. For $||u(t)||_{L^4}^4$, the balance laws justify that it
is suitably vertically translated above $D(t)$ at least for a finite time interval, explaining the
detection of rogue waves at the same times. These observations can be potentially useful in guiding
future studies for the theoretical detection of extreme wave events by suitably formulated
variational problems involving the aforementioned functionals, rationalizing further the numerical
findings. Let us remark, that both functionals are physically relevant as defining the kinetic and
potential energy respectively, and their behavior was used extensively in studying the asymptotic
state of noisy induced MI \cite{Rad1}.  The functional $||u(t)||_{L^4}^4$ is also known as the
interaction Hamiltonian, being relevant in studying soliton interactions \cite{BCai}.

Finally, the long-time asymptotics of the solutions, as observed by the involved functionals, are
analytically justified in the parametric regimes we are interested in, by the global asymptotic
orbital stability of cw solutions. The analytical arguments combine the linear stability analysis
of spatially uniform, cw-solutions (see Theorem \ref{stab} which follows the lines of \cite{barashenkov}), with the well known results on the
existence of a global attractor possessed by the associated infinite dimensional dynamical system
and the notion of orbital stability (see Theorem \ref{fc}). We yet stress that while in general, the global attractor for the model has been proved to posses a complicated  structure, in certain
regimes of our study it can be characterized by an orbitally asymptotically stable cw-state.  These
arguments prove that the observed instabilities in the parametric regimes of rogue waves emergence
are only transient in the presence of this state. The results of relevant numerical experiments
were found to be in  excellent agreement with the analytical predictions.

The presentation of the paper is the following. In Section II, we report the numerical results on
the observation of rogue-wave dynamics. In Section III we discuss the relevant balance laws of
\eqref{eq1} for functionals,  recall some useful estimates for our analysis and justify the
functional type diagnostics for the detection of rogue-waves. In Section IV, we explain the
asymptotic behavior of the solutions observed in terms of the orbital stability of spatially
uniform cw solutions. The last Section IV, briefly discusses our conclusions and sketches a plan
for future investigations.

	\section{Numerical Results}
	\label{numerical}
	\subsection{Emergence of extreme events and spatiotemporal patterns.}
	\label{par1}

	We start the presentation of the numerical results by discussing the
	spatiotemporal dynamics of the system, varying the wave number $K\geq 2$ of the
	initial condition \eqref{eq2}, keeping its amplitude fixed, $A=1$; the case $K=1$ will be discussed at the end of
Section \ref{lta}.
	
\begin{figure}[!htb]
\centering
			\includegraphics[scale=0.3]{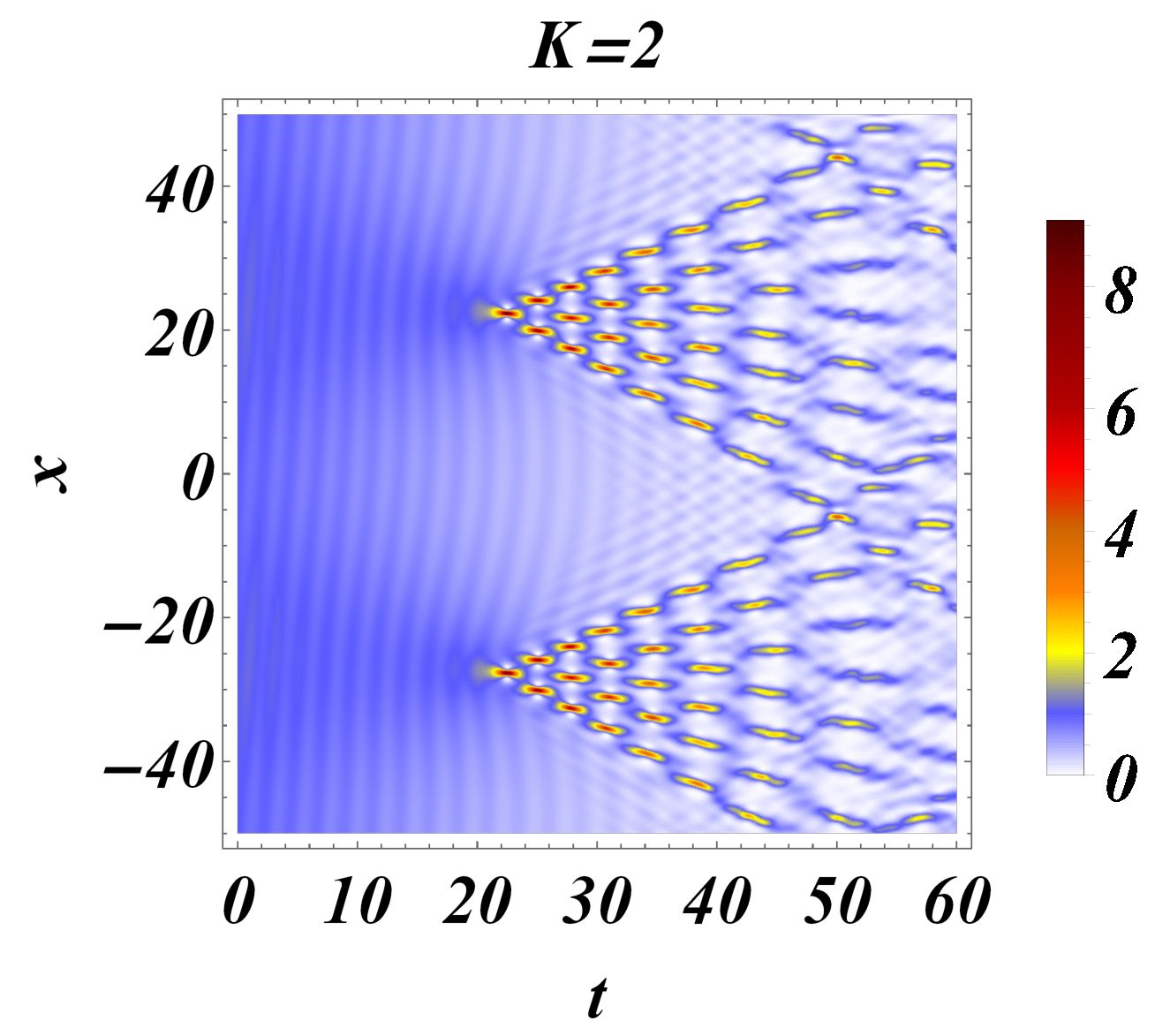}
			\includegraphics[scale=0.3]{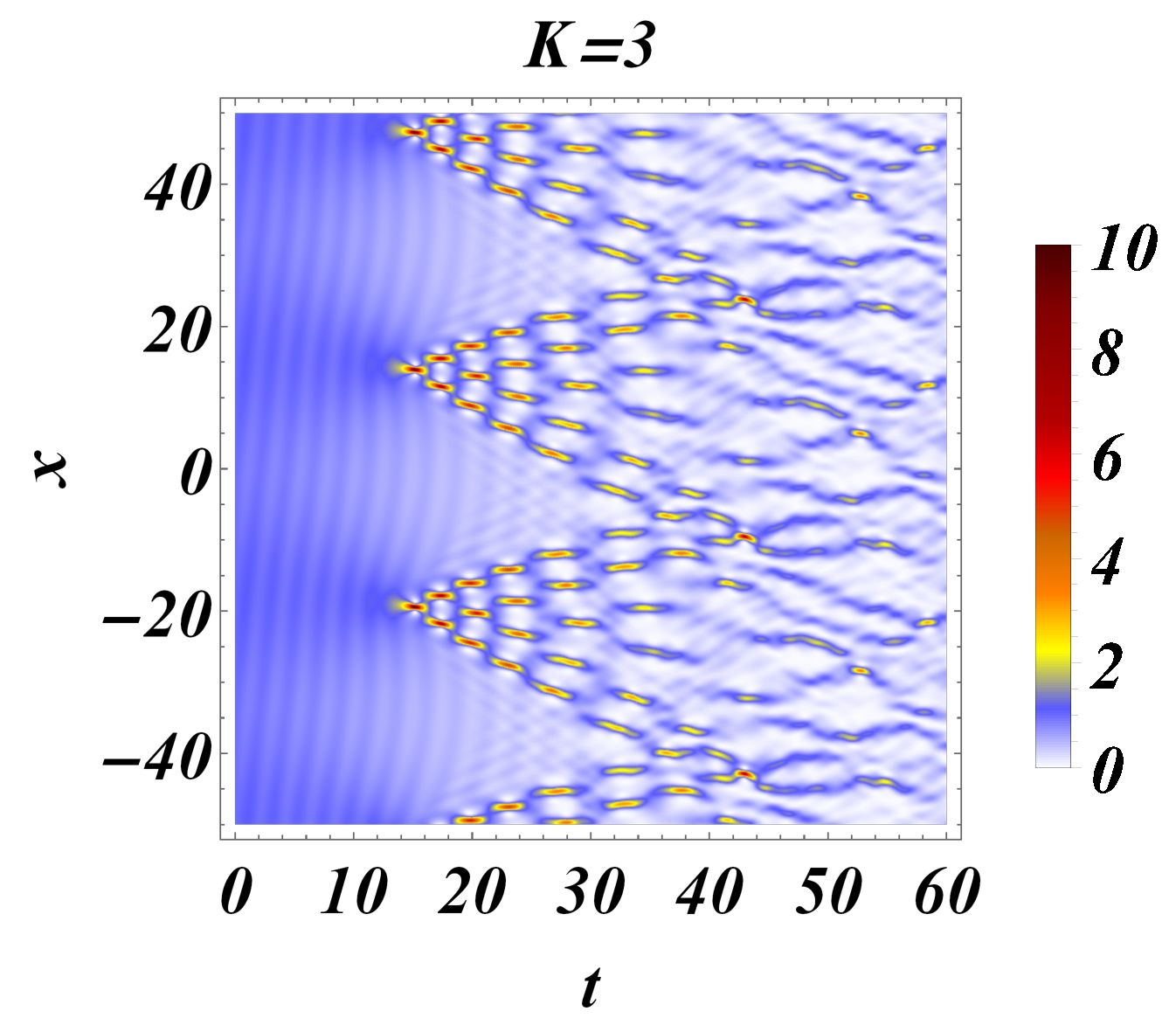}\\
			\includegraphics[scale=0.3]{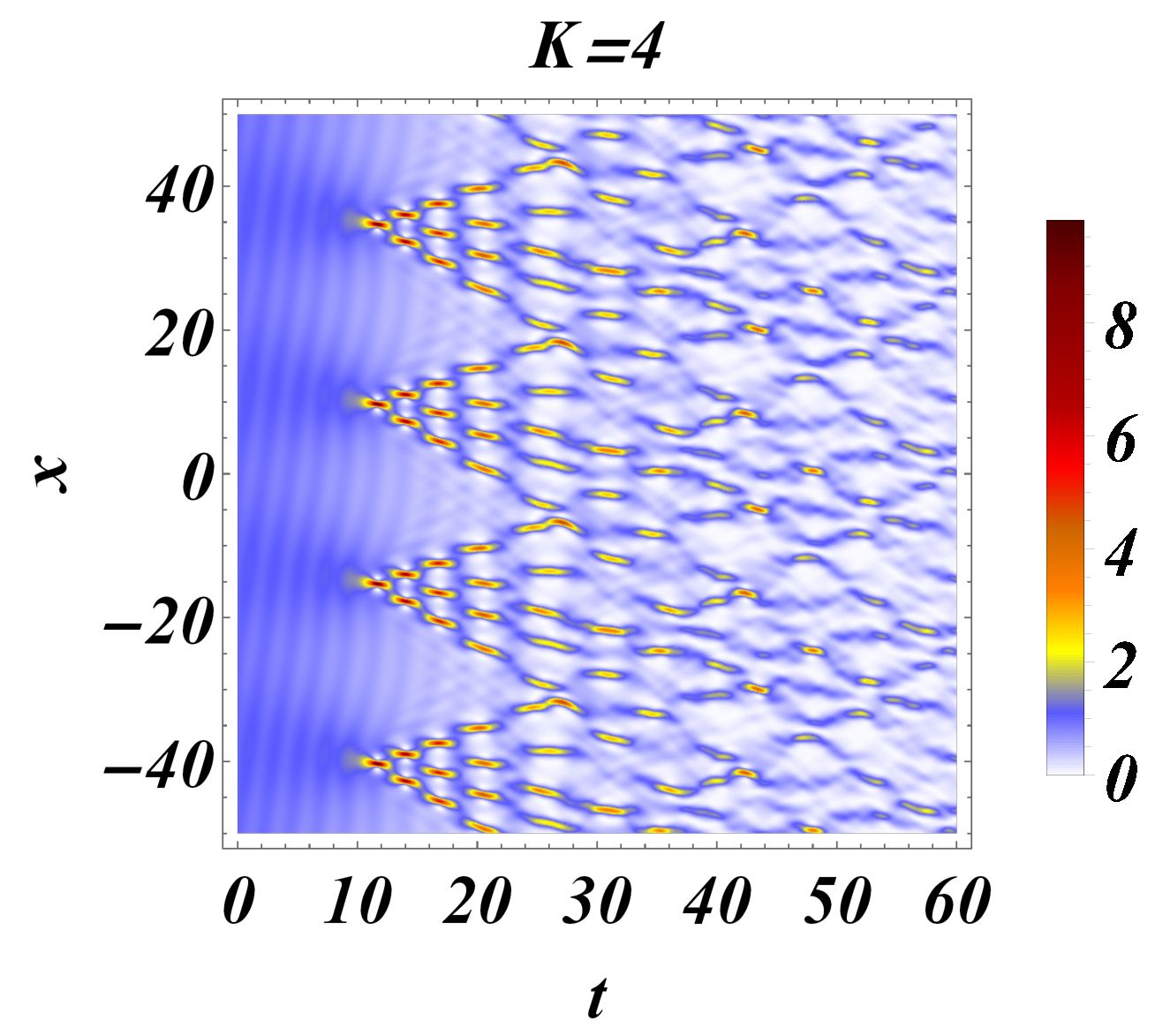}
			\includegraphics[scale=0.3]{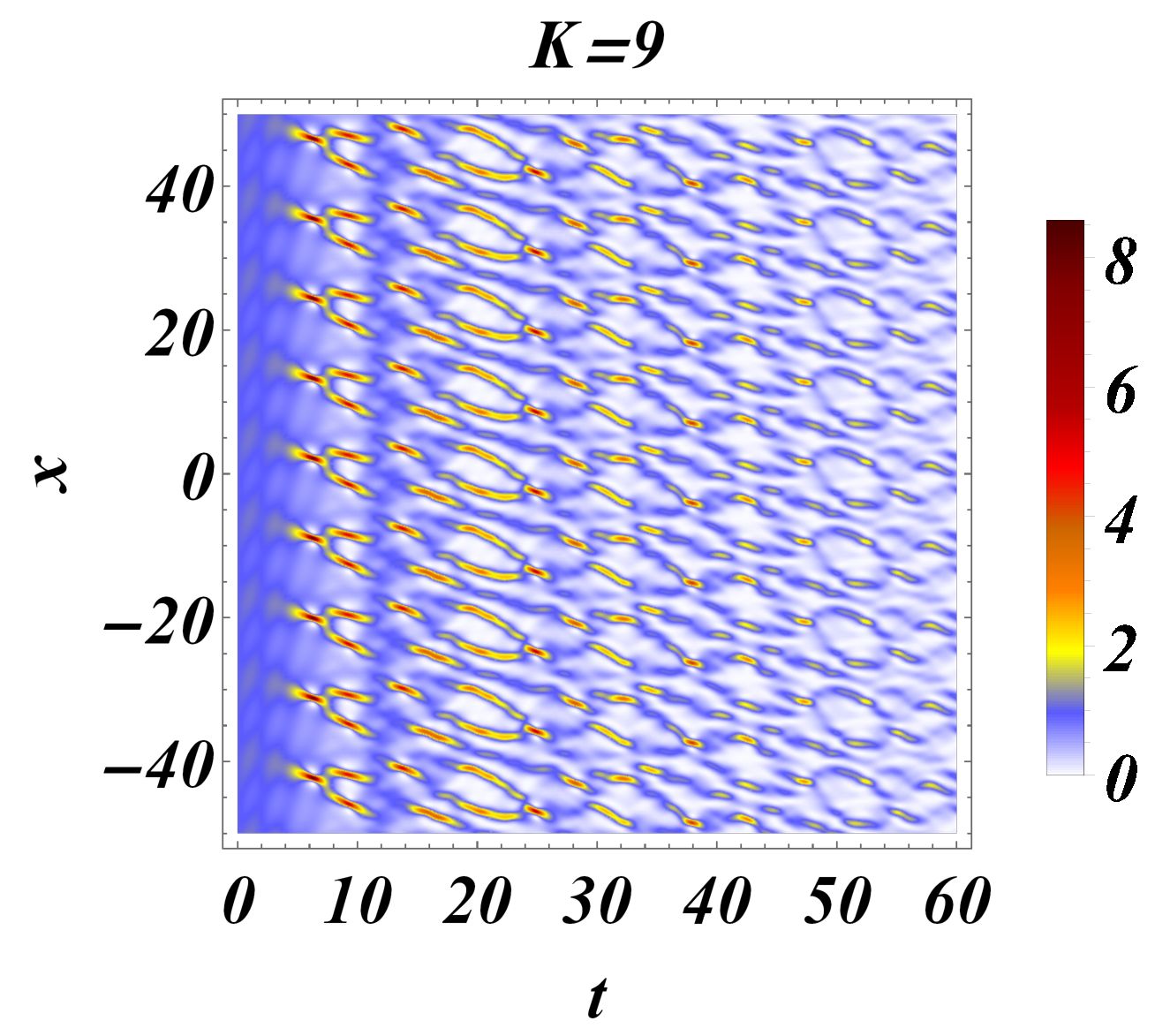}
		\caption{(Color Online) Dynamics of the cw-initial condition \eqref{eq2} for
			$A=1$, varying its wavenumber $K$. The panels show the  spatiotemporal evolution
			of the density $|u|^2$ for each $K$. Other parameters for
			the problem \eqref{eq1}-\eqref{eq3}:  $\nu=\sigma=1$, $\gamma=0.01$,
			$\Gamma=0.1$,  $\Omega=-2$  and $L=50$.}
		\label{figure1}
\end{figure}

	Fig. \ref{figure1} depicts the spatiotemporal evolution of the density $|u(x,t)|^2$ of the
	solution for various cases of the initial wave number $K\geq2$.  The parameters
	for Eq. \eqref{eq1} are $\nu=\sigma=1$, $\gamma=0.01$, $\Gamma=0.1$,
	$\Omega=-2$  and $L=50$.

\begin{figure}[!htb]
\centering
			\includegraphics[scale=0.4]{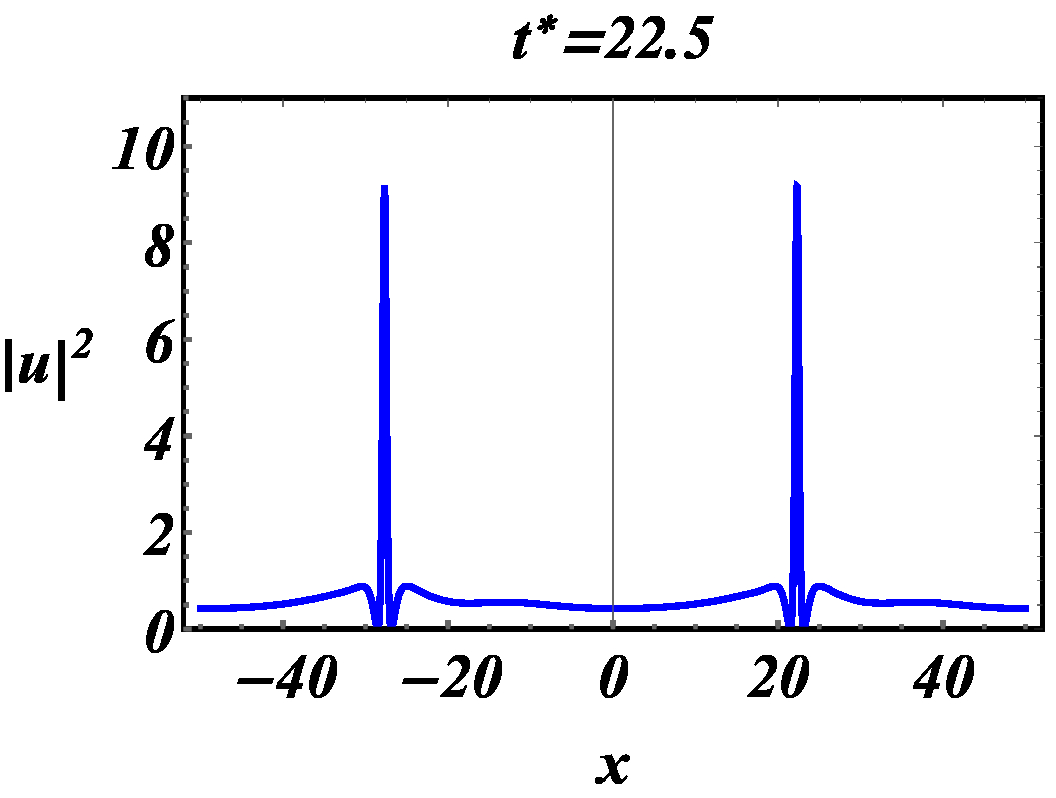}
			\includegraphics[scale=0.4]{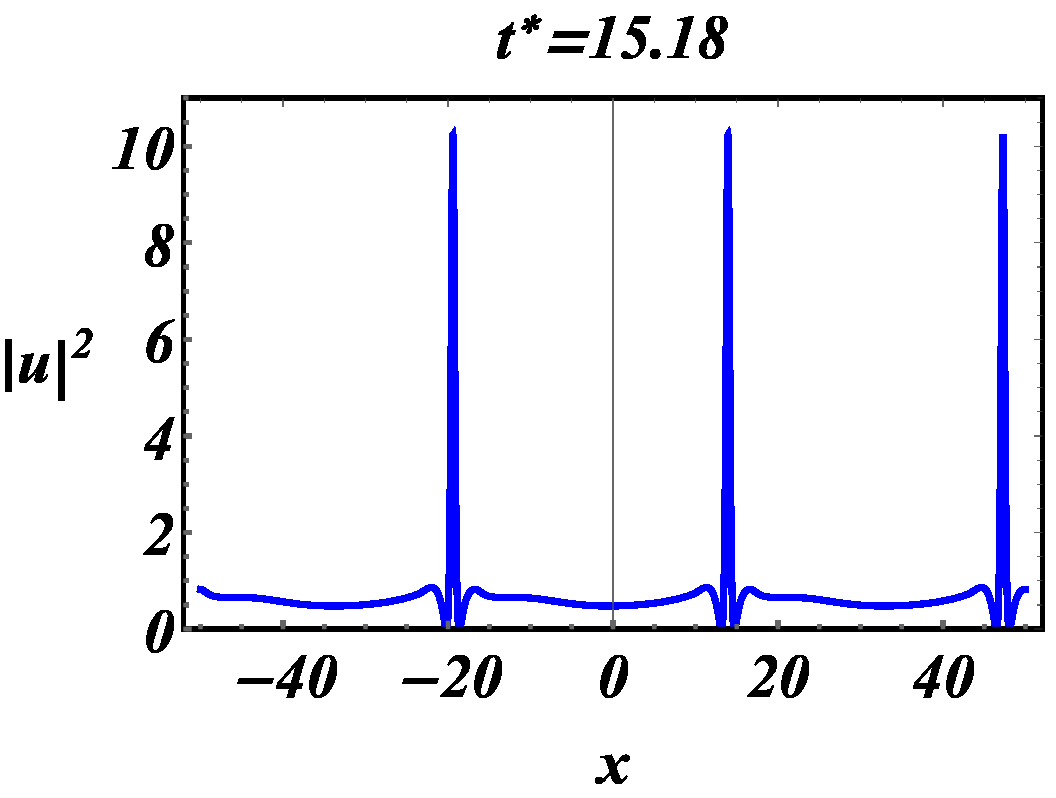}\vspace{0.2cm}\\
			\includegraphics[scale=0.4]{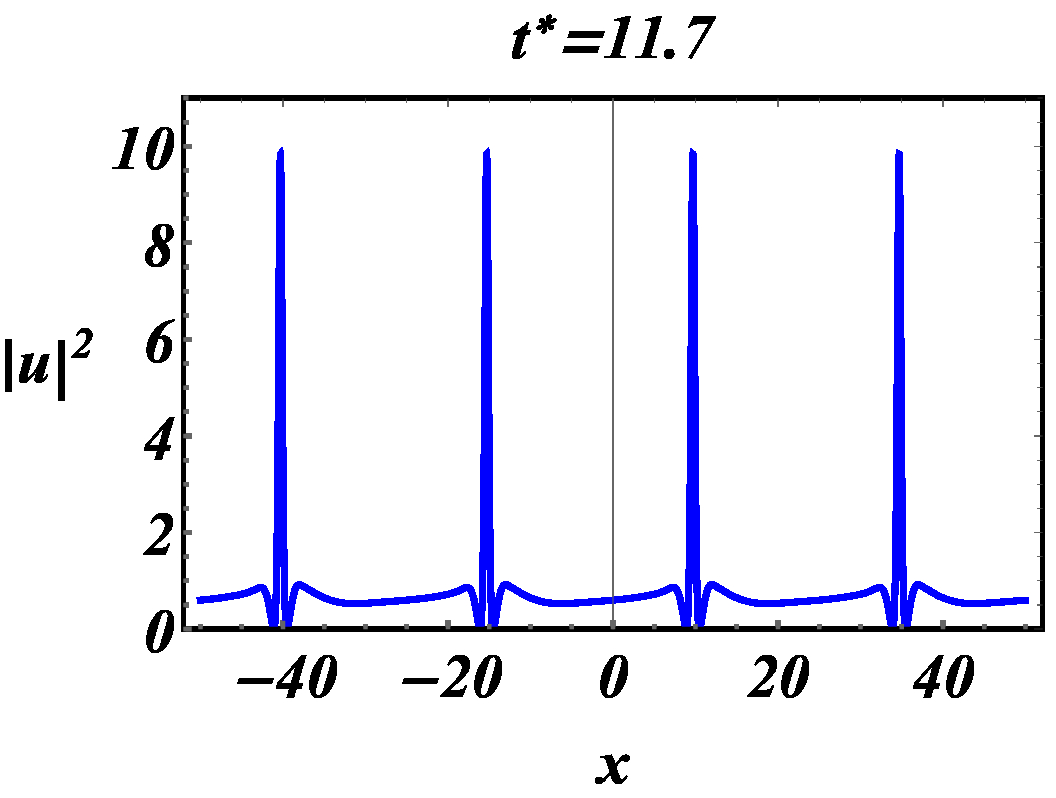}
			\includegraphics[scale=0.4]{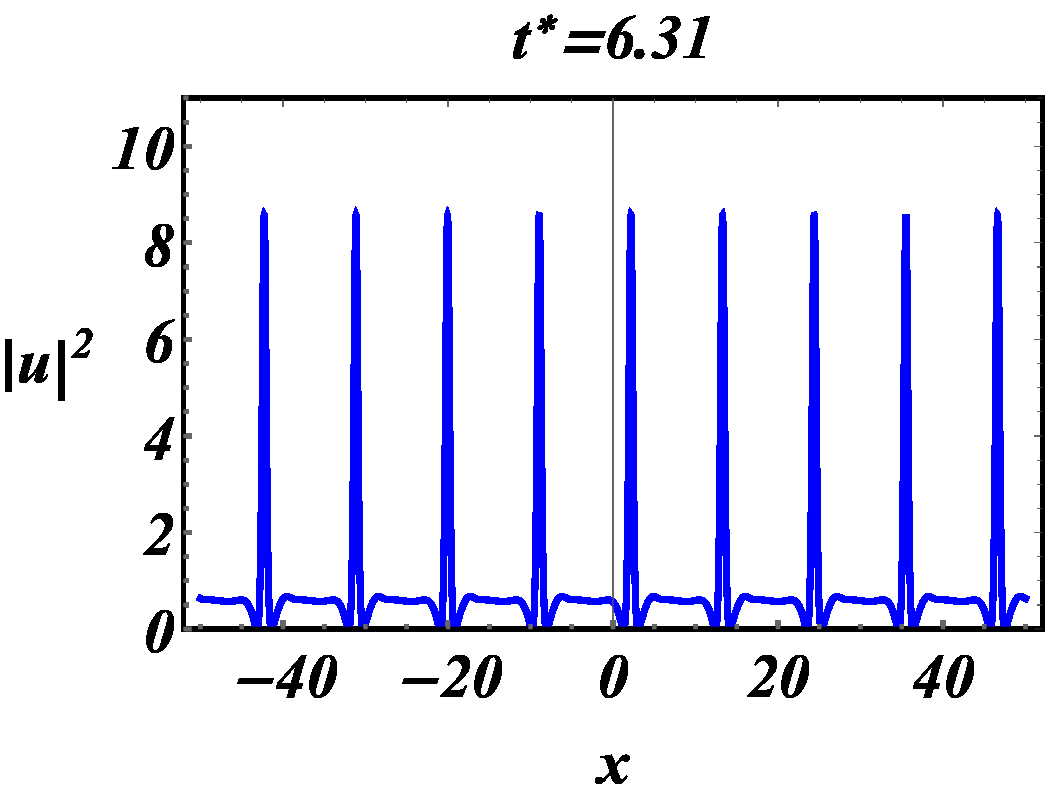}
		\caption{(Color Online) Snapshots of the density $|u|^2$ of the  solution, at
			the times of the emergence of the PRW-type waveforms (first events in the
			dynamics). These correspond to the first events shown in the contour plots and the related parameter values of
Fig. \ref{figure1}.}
		\label{figure2}
\end{figure}

Each one of the panels of Fig. \ref{figure1} is associated with one of the
snapshots portrayed in Fig. \ref{figure2}, showing the profiles of the first
localized structures emerged in the contour plots of Fig. \ref{figure2}.

Combining the results presented in both figures,  the following dynamical features
are revealed:
\begin{enumerate}
\item The first events in the spatiotemporal dynamics (first ``spots''
    observed in the contour plots of Fig. \ref{figure1}), correspond to
localized waveforms
		possessing extreme amplitude (at least eight-times higher than the initial one).
		The initial wave number $K$ gives rise to $K$-localized waveforms, which are
		formed on the top of a finite background, as shown in the snapshots of Fig.
		\ref{figure2}.

\item Subsequently, the dynamics are manifested by a formation of transient
		$K$-patterns with the following characteristics: each one is a distinct
		spatiotemporal region  separated by nonlinear caustics
		which bound the pattern of the transient spatiotemporal oscillations.
		Particularly, a structured ``lattice''  of extreme events occurs, occupying a
		triangular region formed between the caustics. Progressively, their amplitude is
		decreasing, as we may conjecture due to the effect of damping. This is evident
		in the cases $K=2,3,4$. Each of these patterns is reminiscent of the pattern
		forming in the case of the integrable NLS {\em in its semiclassical limit
		\begin{equation}
		\label{eq1sl}
		\mathrm{i}\epsilon{{u}_{t}}+\frac{\epsilon^2}{2} {{u}_{xx}}+|u|^2u =0,
		\end{equation}
		for $\epsilon\rightarrow 0$ \cite{BM1,BM2},  when supplemented with
		vanishing boundary conditions}.   We recall in Figure \ref{figure1_0},
	\begin{figure}[!htb]
	\centering
	\includegraphics[scale=0.25]{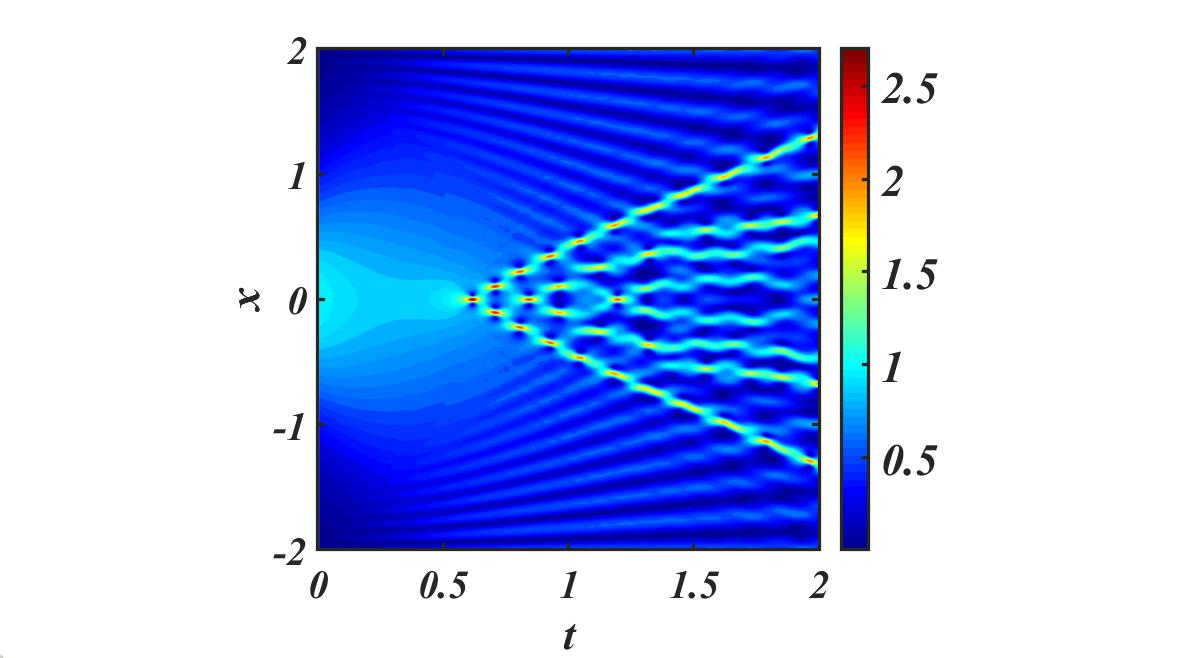}

	\caption{(Color Online) Spatiotemporal evolution of the density $|u|^2$ for the integrable NLS in its semiclassical
limit \eqref{eq1sl} when $\epsilon=0.02$, and the initial condition is \eqref{btin}.   }
	\label{figure1_0}
\end{figure}
the characteristic spatiotemporal pattern of the evolution of the initial condition
\begin{equation}
	\label{btin}
u(x,0)=A(x)\exp\bigg[\frac{\mathrm{i}}{\epsilon}\Phi(x)\bigg], \;\;A(x)=\exp(-x^2),\;\;\Phi'(x)=-\tanh x,
\end{equation}	
for $\epsilon=0.02$ in 	\eqref{eq1sl}. This single lattice, as it was analysed in \cite{BM1,BM2,Rev1_b}, is occupied by
PRW
		structures defined by spots which correspond to the poles of the
		tritronqu{\'e}e solution of the Painlev{\'e}-I equation.  A similar single pattern was also
observed in the case of the damped and forced NLS for algebraically or exponentially vanishing
initial data when the driver is spatiotemporally localized
$\lim_{|x|^2+|t|^2\rightarrow\infty}f(x,t)=0$, \cite{D1}, and explained in terms of decaying
estimates in \cite{D2}, but differs dramatically form the pattern observed in \cite{ZNA2019} in the
dynamics of \eqref{eq1} initiated from the same vanishing initial conditions.  Another interesting
		feature observed in the damped and forced case, is that the points of intersection of the caustics are points of
wave interference, occupied by localized structures of large
		amplitude.

		\item The ``area'' occupied by the above ``Christmas tree'' patterns \cite{BS}, is
		decreasing as the initial wave number is increasing; a compression of the
		triangular regions is observed in the case $K=9$. However, this is only an
		effect of the increased excited modes against the fixed spatial region
		compressing the aforementioned pattern formation. This can be clarified further, when
the numerical experiments for increased length $L$ of the spatial
		interval and some large $K$ are presented. The left panel of Fig. \ref{figure3} shows the
		contour plot of the spatiotemporal evolution of the density when $K=8$ and the
		same set of parameters as in Fig. \ref{figure2}, but for $L=200$.

\begin{figure}[!htb]
\centering
				\includegraphics[scale=0.4]{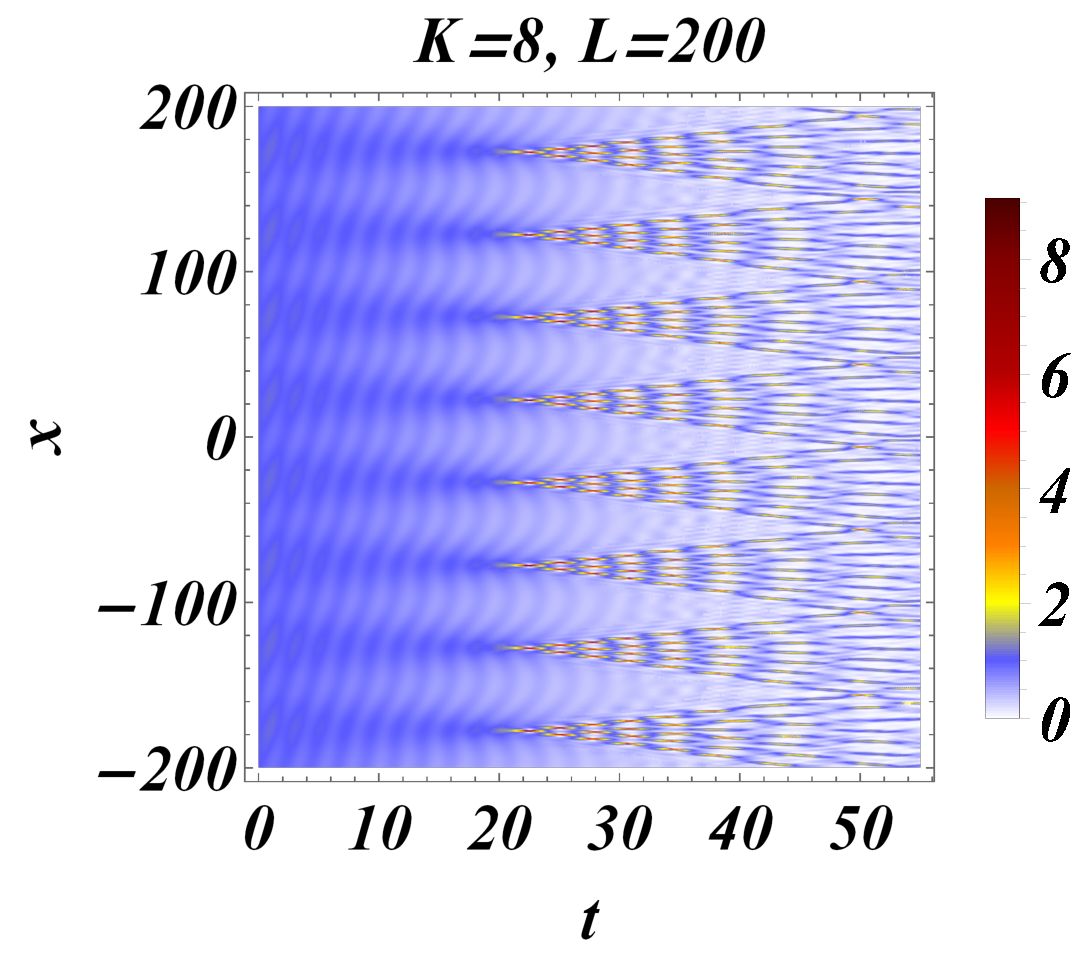}
				\includegraphics[scale=0.385]{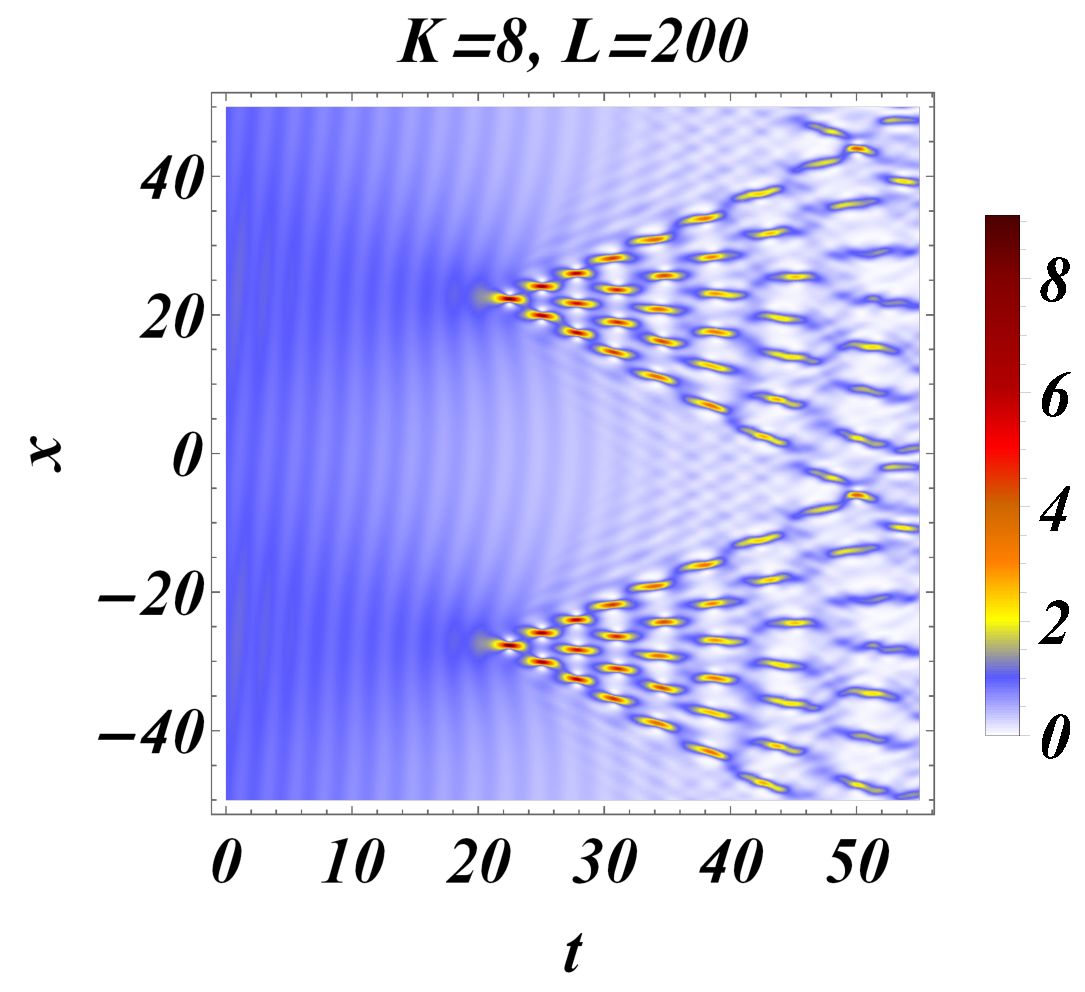}
			\caption{(Color Online) Dynamics of the cw-initial condition \eqref{eq2} for
				$A=1$, $K=8$, $L=200$. The left panel shows the spatiotemporal
				evolution of the density for $x\in [-L,L]$ and $t\in [0,60]$. The right panel
				offers a magnification of the patterns observed in the left panel, for $x\in
				[-50,50]$ and $t\in [0,60]$. Other parameters for the problem
				\eqref{eq1}-\eqref{eq3} are:  $\nu=\sigma=1$, $\gamma=0.01$, $\Gamma=0.1$, $\Omega=-2$.}
			\label{figure3}
\end{figure}	

We clearly
		observe the formation of $K=8$-copies of the triangular pattern with the
		characteristics discussed above, as it is further clarified by the magnification
		offered in the right panel of Fig. \ref{figure3} (showing two of them within the
		region $x\in [-50,50]$, $t\in [0,60]$). This study also reveals that the
		emergence of the above patterns is robust against variations with respect to the
		half-length $L$.
	
		\item Do the first extreme events share characteristics of the PRW? To answer
		this question  we shall implement fitting arguments already used in
		\cite{BS,All2,ZNA2019}, against the analytical PRW of the integrable NLS:
		\begin{eqnarray}
		u_{\mbox{\tiny 				
PS}}(x,t;P_0)=\sqrt{P_0}\left[1-\frac{4\left(1+\frac{2\mathrm{i}t}{\Lambda}\right)}{1+\frac{4x^2}{K_0^2}+\frac{4t^2}{\Lambda^2}}\right]\exp\left(\frac{\mathrm{it}}{\Lambda}\right).
		\label{sprw}
		\end{eqnarray}
		Recall that $P_0$ is the amplitude (power) of the continuous background
		supporting the PRW \eqref{sprw}, and the parameters
		$\Lambda=\frac{1}{\sigma\,P_0}$, $K_0=\sqrt{\nu\,\Lambda}$. We denote the
		space-time translations of \eqref{sprw} by $u_{\mbox{\tiny
				PS}}(x-x^*,t-t^*;P_0)$. The location $x^*$, time $t^*$ and power $P_0$, are
		numerically calculated: The power of its background $P_0$ is locked, so that the amplitude of the
		analytical PRW \eqref{sprw} coincides with the maximum amplitude of the extreme
		localized structures portrayed in \ref{figure4} attained at time $t^*$. The
		position $x^*$ is detected so that the center of the analytical PRW coincides
		with the center of a selected extreme event (from the group of the $K$-identical
		localized structures).

\begin{figure}[!htb]
\centering
				\includegraphics[scale=0.4]{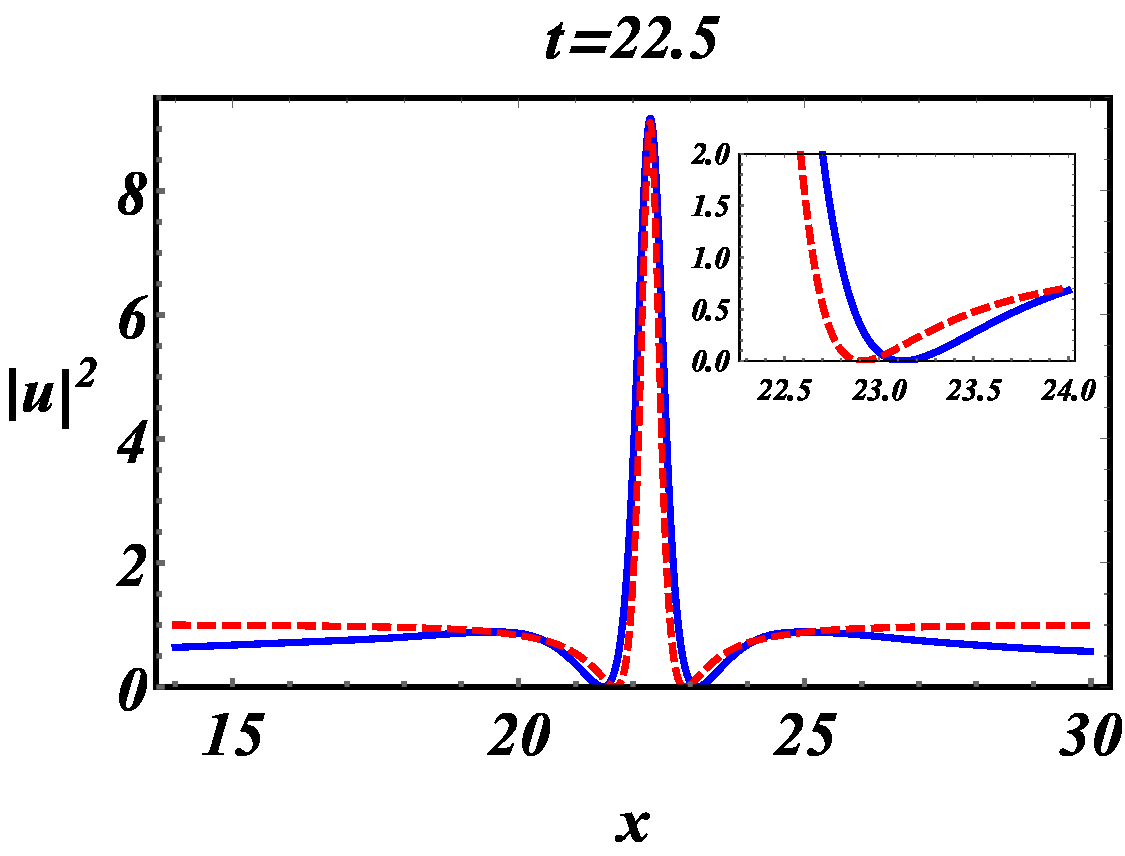}
				\includegraphics[scale=0.4]{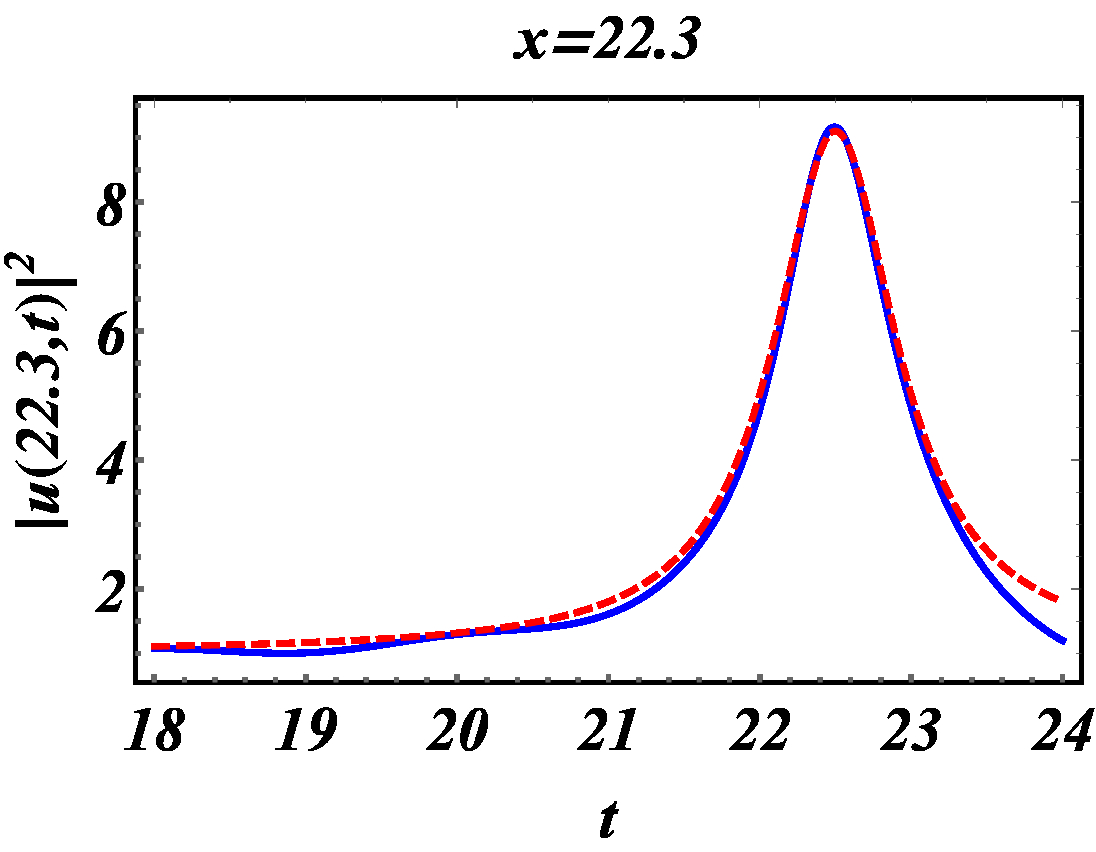}\\
				\includegraphics[scale=0.25]{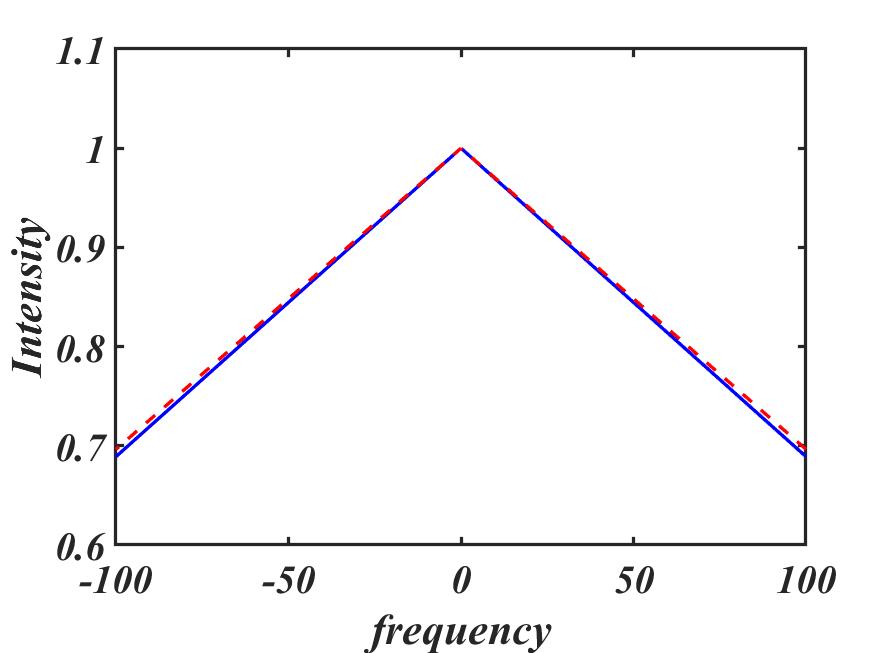}
			\caption{(Color Online) Top panels:  Comparison of the right PRW-type waveform observed in
				upper left panel of \ref{figure1} [continuous (blue) curve], against the
				space-time translation of the analytical PRW of the integrable limit
				$u_{\mbox{\tiny PS}}(x-22.3,t-22.5;1.01)$ where $K_0=0.70$ and $\Lambda=0.99$
				[dashed (red) curve]. The left panel compares the profiles. The right panel
				shows a comparison of the evolution of the density of their peaks
				$|u(22.3,t)|^2$ and $u_{\mbox{\tiny PS}}(0,t-22.5;1.01)$ respectively, for
				$t\in[18,24]$. Bottom panel: comparison of the spectrum of the PRW-type waveform of the top panel, against
the spectrum of the  corresponding analytical PRW.}
			\label{figure4}
\end{figure}
		
		The left panel of Fig. \ref{figure4} justifies the similarity of the profile of
		the right extreme event of maximum amplitude attained at $t^*=22.5$ [continuous
		(blue) curve] to the profile of the PRW \ref{sprw} $u_{\mbox{\tiny
				PS}}(x-22.3,t-22.5;1.01)$, with $K_0=0.70$ and $\Lambda=0.99$. The inset shows a
		detail around the left of the symmetric minima; the slopes of the two curves
		show that the extreme event for the damped and forced NLS should posses, around
		the core, a spatial decay rate close to that of the PRW. The right panel of Fig.
		\ref{figure4} depicts the comparison of the evolution of the density of the
		centers of the two waves; their time growth is very close for $t\in [18,21]$,
		their coincidence becomes almost exact for $t\in [21.5, 23.5]$ and afterwards we
		observe their divergence. Thus, the emerged waves for the damped and forced
		model posses for significant time intervals the algebraic in time, growth and
		decay rate of a PRW.  Additional evidence is provided in the bottom panel of Fig. \ref{figure4} comparing the
spectrum of the emerged event and the analytical PRW, suggesting an almost exact agreement.
		
\item {\it Dynamics of the integrable limit $\gamma=\Gamma=0$.} The above
    ``Christmas tree'' pattern formation is not exhibited in the integrable
    limit but only in the damped and forced case of \eqref{eq1}, for certain
    parametric regimes (see below). This is verified in Fig. \ref{figure7}
    showing the spatiotemporal dynamics in this limit and for the rest of
    parameters fixed as in Fig.  \ref{figure1}.

\begin{figure}[!htb]
\centering
		\includegraphics[scale=0.33]{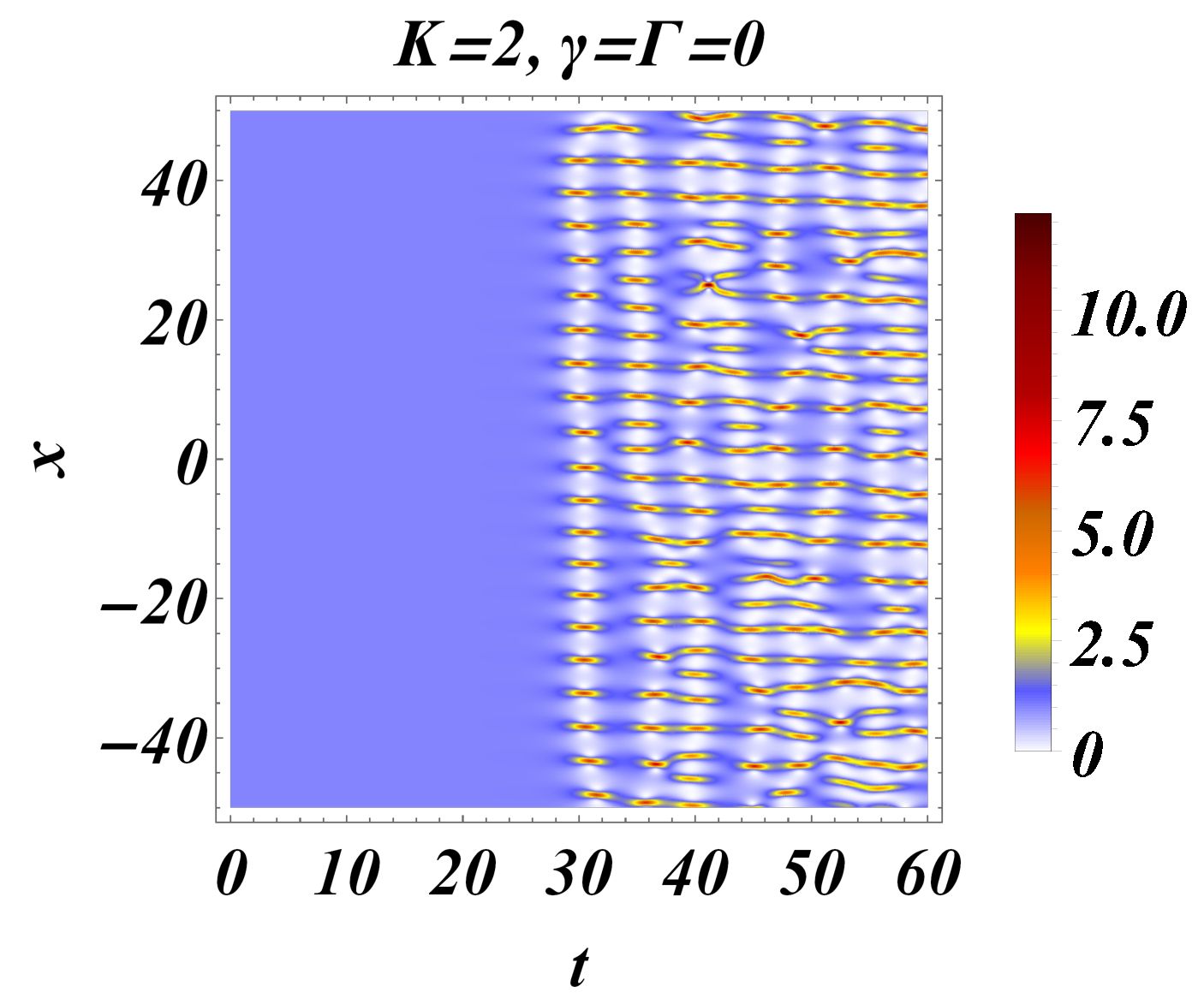}
		\includegraphics[scale=0.33]{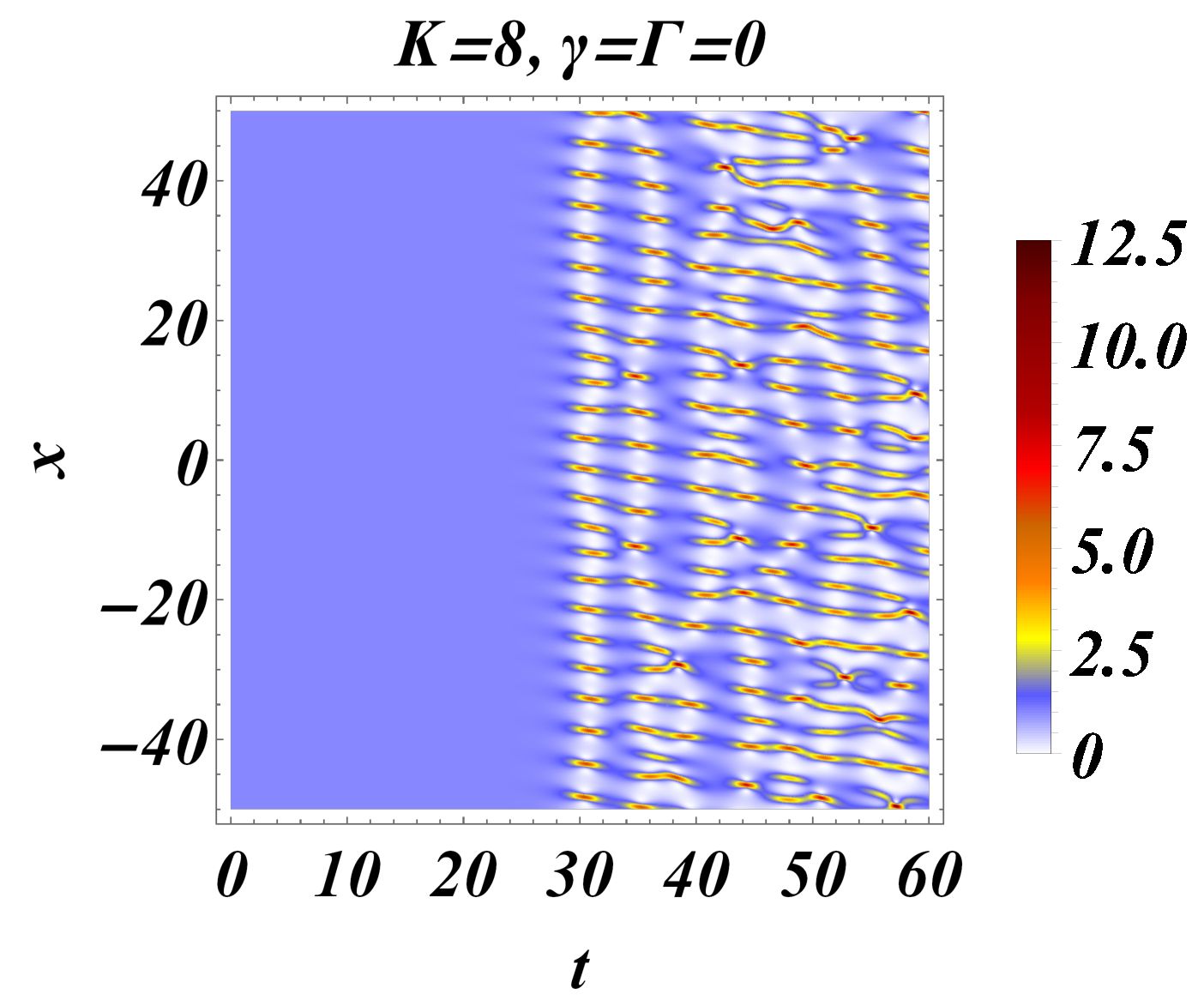}
	\caption{(Color Online) Contour plots of the spatiotemporal evolution of the density in the case
of the integrable limit $\Gamma=\gamma=0$, for two cases of initial wave-numbers. Values of the rest of the
parameters:  $A=1$, $\nu=\sigma=1$, $L=60$.}
	\label{figure7}
\end{figure}

   In both cases, the dynamics are almost identical, characterized by the
    emergence of the  vertical ``first line'' of extreme events, separating
    the spatiotemporal domain in two distinct regions: before this line, and
    for $t\lesssim 30$ the manifestation of MI is weak, while from the
    ``line'' and above $t\gtrsim 30$, the manifestation of MI is
    characterized again by the occurrence of extreme events; the line itself
    resembles {\it a spatially periodic state reminiscent of an Akhmediev
    breather} (see \cite[Fig 3, pg. 103901-3]{Akhb2016} for similar patterns).

While in the integrable limit, we comment on the robustness of
the numerical scheme and its outcomes. As complicated dynamics are discussed,
it is important to make sure that numerically induced complex behavior \cite{HA,AH}, due
to the numerical scheme used is not present,  perturbing artificially its homoclinic structure. This is done through
checking  the behavior of the
conserved quantities in the integrable limit $\gamma=\Gamma=0$. Indeed, in Fig. \ref{figure6a} the first three
integrals of motion $M(t)$ [Eq.~\eqref{mom2}], $P(t)$ [Eq.~\eqref{l2f}] and $H(t)$ [Eq.~\eqref{Hameneg}] of the
unperturbed, integrable NLS equation are shown to
be constant throughout the evolution. As such, our findings are numerically
supported with no artificially induced phenomena entering the dynamics.

\begin{figure}[!htb]
\centering
\includegraphics[scale=0.4]{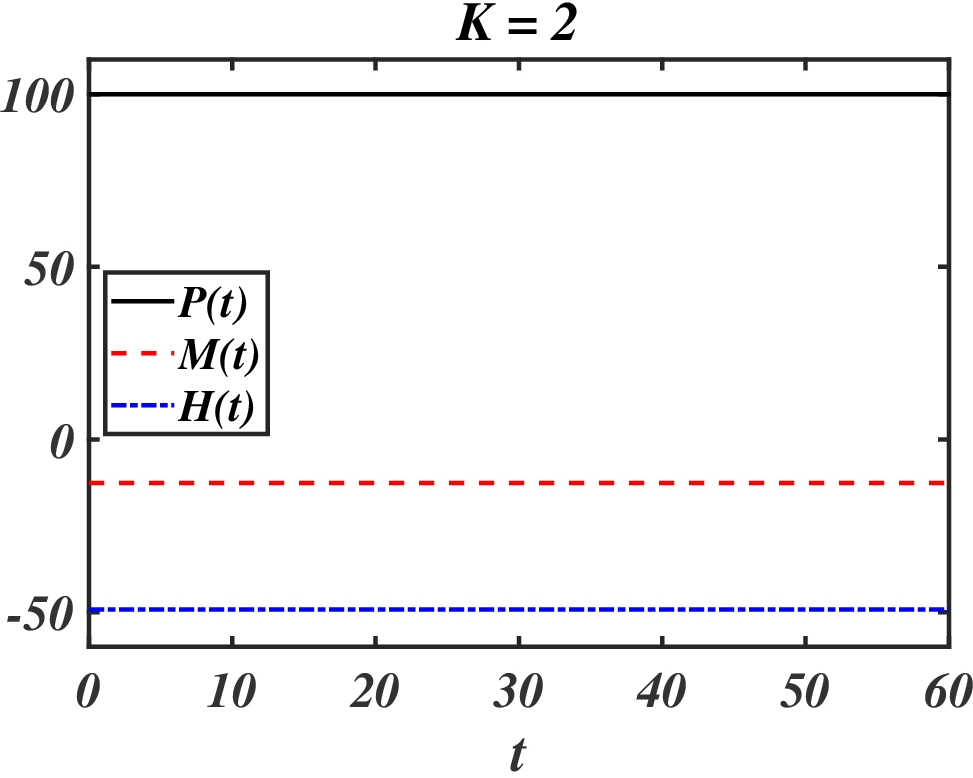}
\includegraphics[scale=0.39]{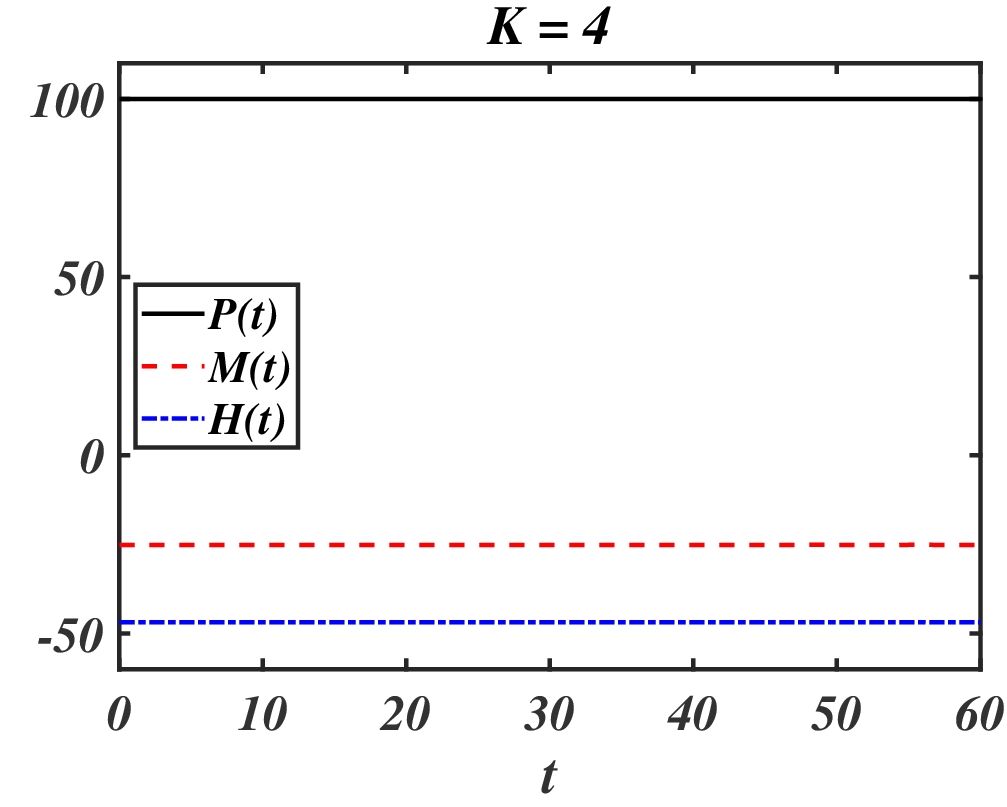}
\caption{(Color Online) The evolution of the first three conserved quantities $P(t)$ [square of the $L^2$-norm or wave
action (black solid curve), $M(t)$ [momentum (dashed red curve)] and $H(t)$ [Hamiltonian (dotted blue curve)] for
the integrable equation Eq. (\ref{eq1}) with $\gamma=\Gamma=0$, for different values of $K$.
The rest of the parameters are as in Fig. \ref{figure7}. No numerical discrepancies are present as all remain constant
for the duration of the evolution.}
	\label{figure6a}
\end{figure}
\item {\it Robustness of the dynamics with respect to parameters.} An
    important question we examine is the potential robustness of the dynamics
    as the various parameters are varied. Generically, it persists up to
    certain thresholds identified for the parameters, as suggested by the
    results presented in Figure \ref{figure6}.
Concerning variations of the damping strength and amplitude of the forcing, we found that the
phenomenon emerges when $\Gamma>\Gamma_{\mathrm{thresh}}$ (for suitable fixed  $\gamma$) and persists
for $\gamma<\gamma_{\mathrm{thresh}}$ (for suitable fixed  $\Gamma$). Particularly, we found that the
dynamics are sensitive for small increments of the damping strength $\gamma\sim
\mathcal{O}(10^{-3})$, as shown in the top left panel for $\gamma=0.02$ where the effects are weaken.
Yet in this regime, the emergence of the triangular patterns persists up to a certain threshold
$\gamma_{\mathrm{thresh}}$.
Similarly, the threshold  $\Gamma_{\mathrm{thresh}}\sim
\mathcal{O}(10^{-3})$ was found relatively small for the given fixed set of parameters; for
$\Gamma<\Gamma_{\mathrm{thresh}}$ the dynamics were close to that of the integrable limit.  For
increased values of $\Gamma$, the amplitude of the first extreme events, as well as, the amplitude
and the spatiotemporal density of the extreme waves within the triangular regions (and on the
caustics) is increased, as shown in the  top middle panel corresponding to the case of $\Gamma=0.8$.
Relevant thresholds $\sim \mathcal{O}(1)$ were identified for the rest of parameters. For
$\Omega<\Omega_{\mathrm{thresh}}$ the dynamics are like the one observed for large $\Gamma$, while
for $\Omega>\Omega_{\mathrm{thresh}}$ approaches that of the integrable limit, as shown in the top right
panel for $\Omega=20$; this is explained as in the limit of large $\Omega$, the time-period of the
oscillations which is dictated by the frequency of the driver should tend to zero. A behavior similar
to the one of large $\Omega$ is observed for $\sigma>\sigma_{\mathrm{thresh}}$ and larger amplitudes
of the initial condition $A>A_{\mathrm{thresh}}$ where the effects are enhanced; see the bottom left
and middle panels for $\sigma=1.5$ and $A=1.5$, respectively. The bottom right panel for $\nu=3$
verifies that for increased $\nu>\nu_{\mathrm{thresh}}$, the triangular patterns are disordered due
to the increased dispersion.
\item {\em Robustness of the spatiotemporal patterns under noisy perturbation of the initial condition}. We also examine
    the stability of the observed spatiotemporal patterns when the plane wave initial condition \eqref{eq2} is perturbed
    by noise. The left panel of Figure \ref{figure6A} depicts the dynamics for the damped and forced NLS \eqref{eq1} for
    the same set of parameters as in the top left panel of Figure \ref{figure1}, and for the initial condition \eqref{eq2}
    with $K=2$ perturbed by random additive noise of $1\%$. The results suggest that the ``Christmas tree'' patterns of Figure \ref{figure1} are
    unstable under noisy perturbations. In the presence of noise the resulting dynamics are now reminiscent to that of the
    noisy induced MI \cite[Fig. 2, pg. 234102-3]{Rad1}.

This instability motivated us to perform a numerical study for the semiclassical limit NLS \eqref{eq1sl} with the same
parameters as in  the study of Figure \ref{figure1_0}, but with the initial condition \eqref{btin} which created the
triangular pattern \cite{BM1,BM2}, now perturbed by the same noise as above.  We observe that even in the semiclassical limit NLS, the
rogue-wave lattice  observed previously is unstable; The new pattern suggests the replacement of the above lattice by the
formation of breather type dynamics, while the whole pattern is more reminiscent of the evolution observed in
\cite{BioMan}.

Both results suggest that the presence of noise enhances the instability effects on the background supporting the rogue
waves, an effect which was analysed in \cite{PNJ}.
\end{enumerate}
\begin{figure}[tbp!]
	\centering
	\includegraphics[scale=0.23]{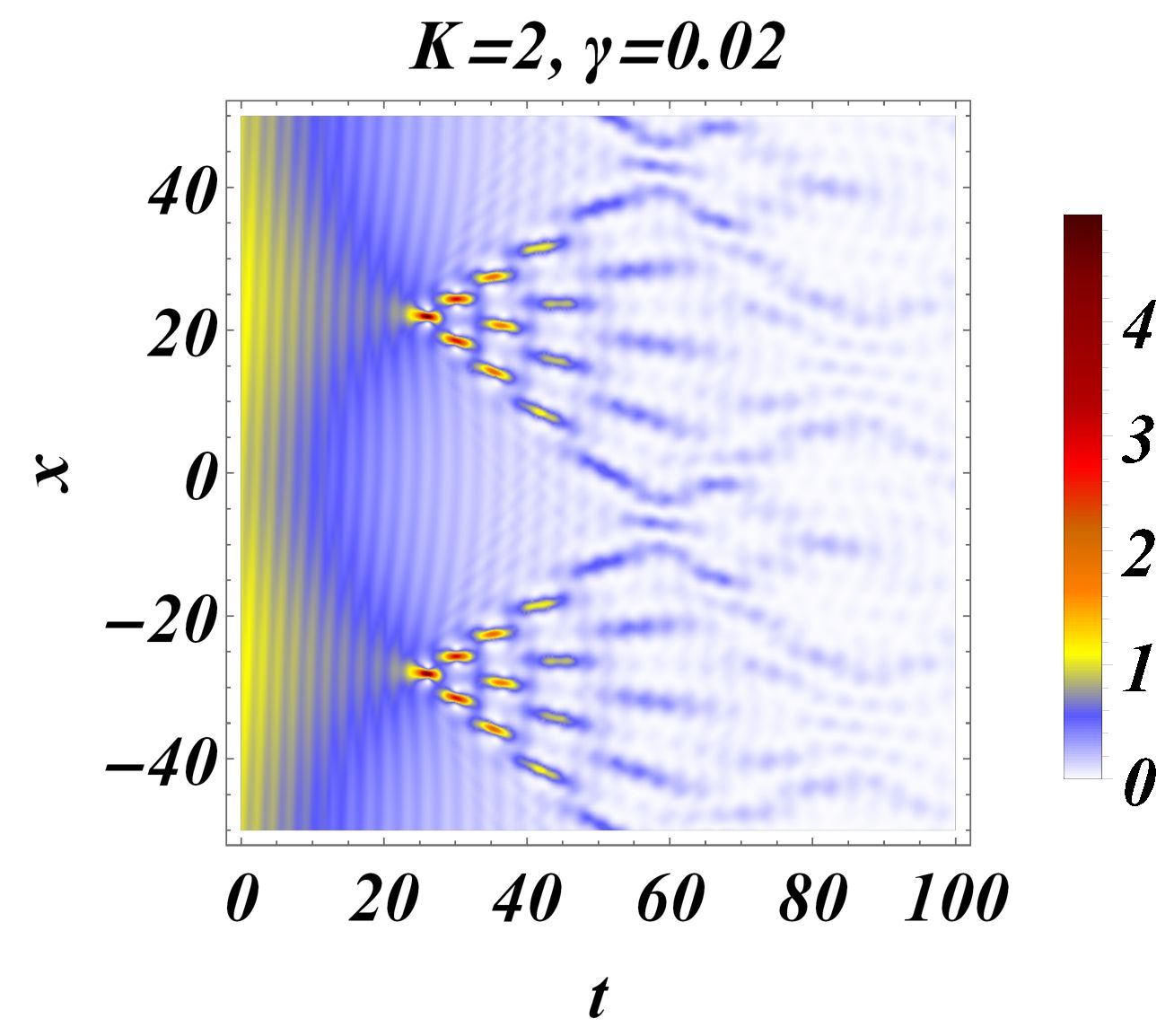}
	\includegraphics[scale=0.23]{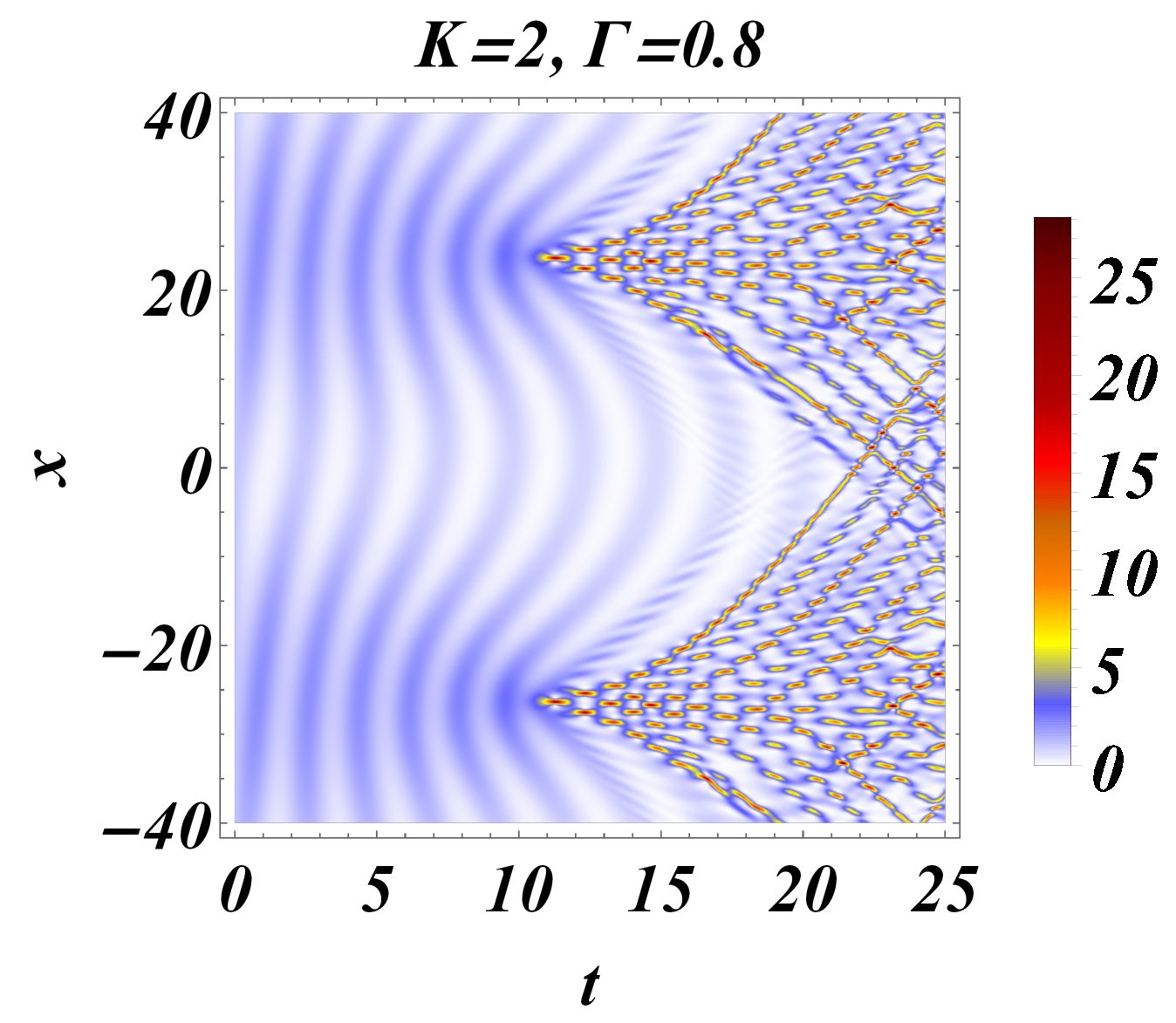}
	\includegraphics[scale=0.23]{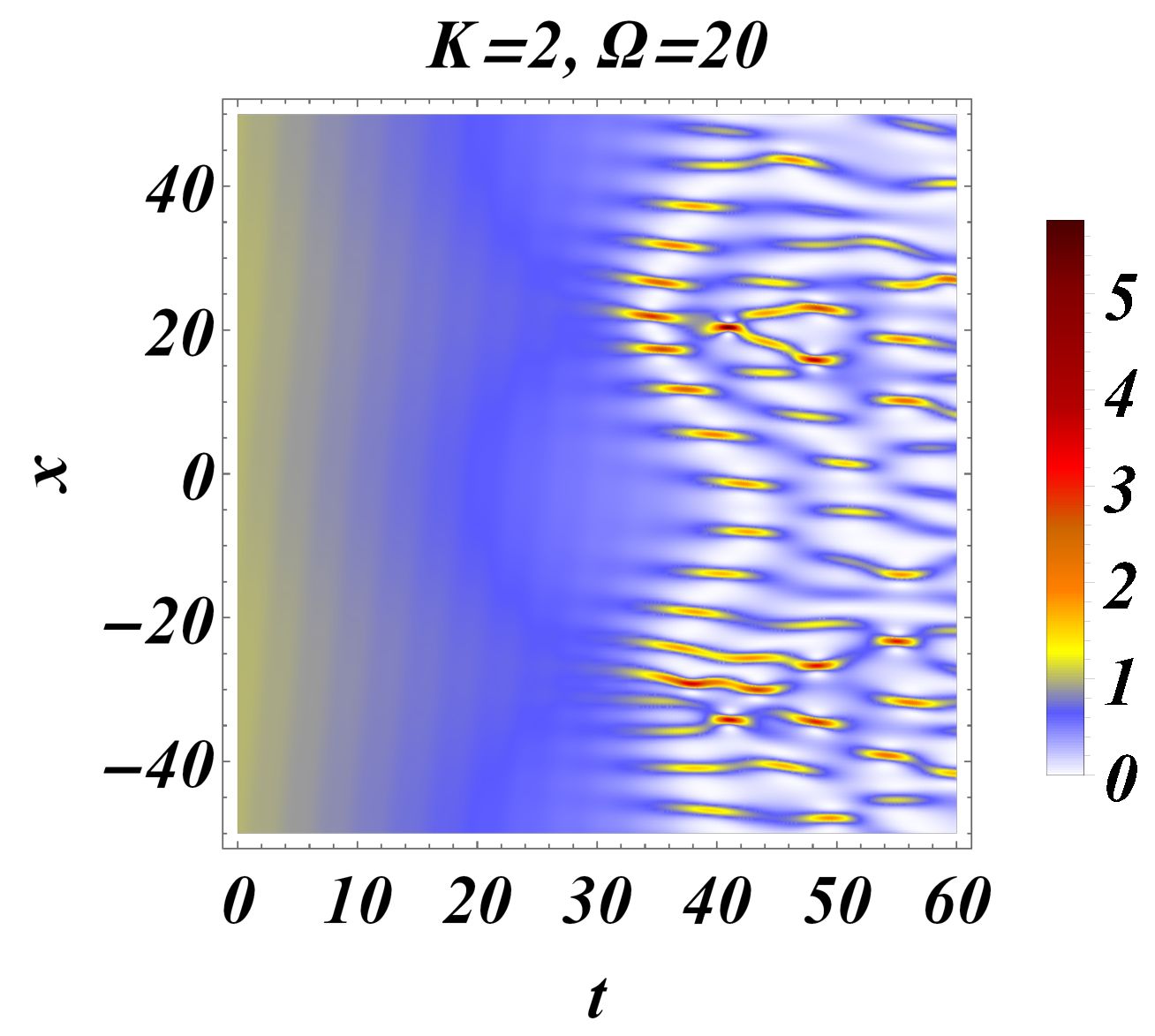}\\
	\includegraphics[scale=0.235]{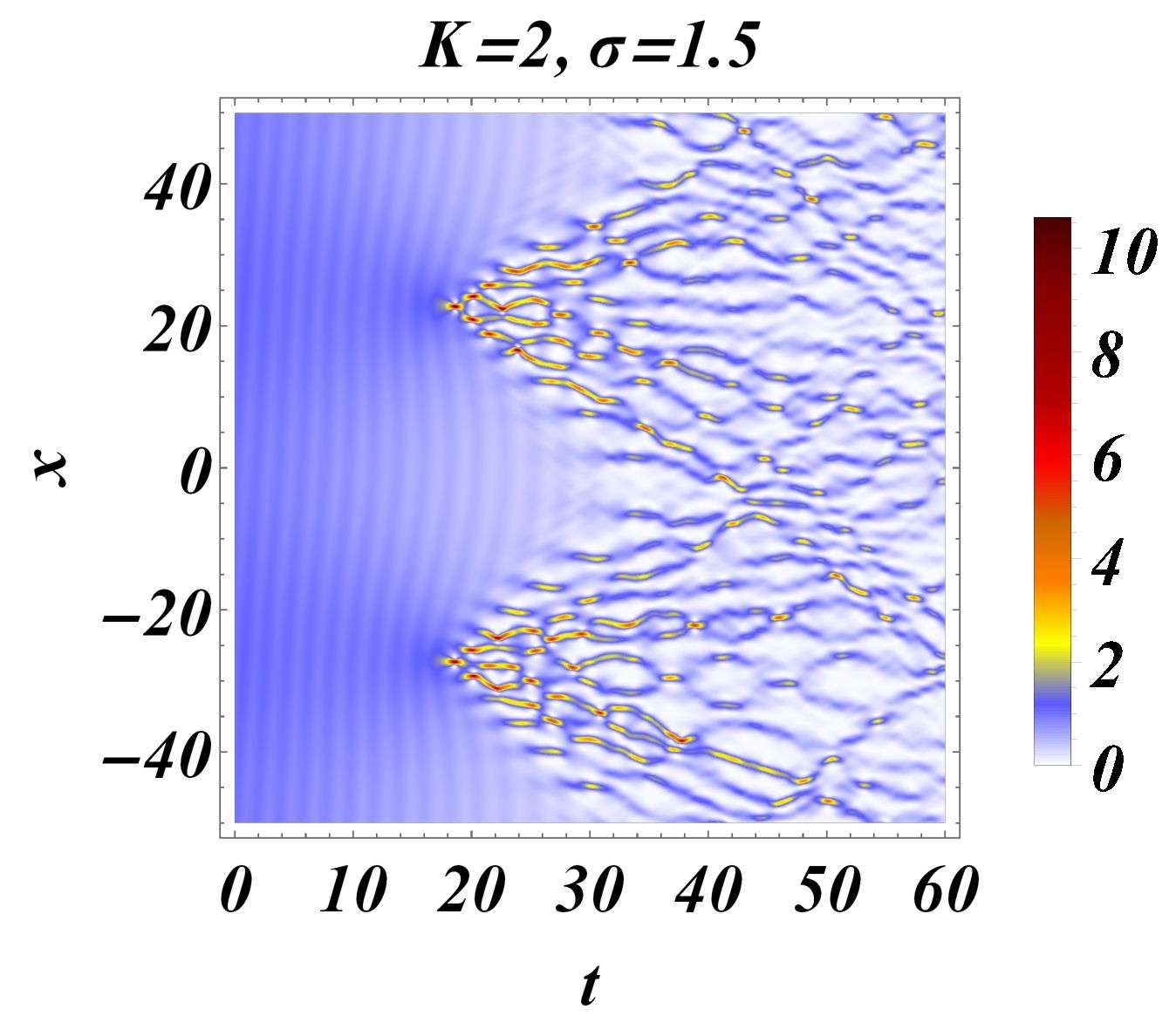}
	\includegraphics[scale=0.23]{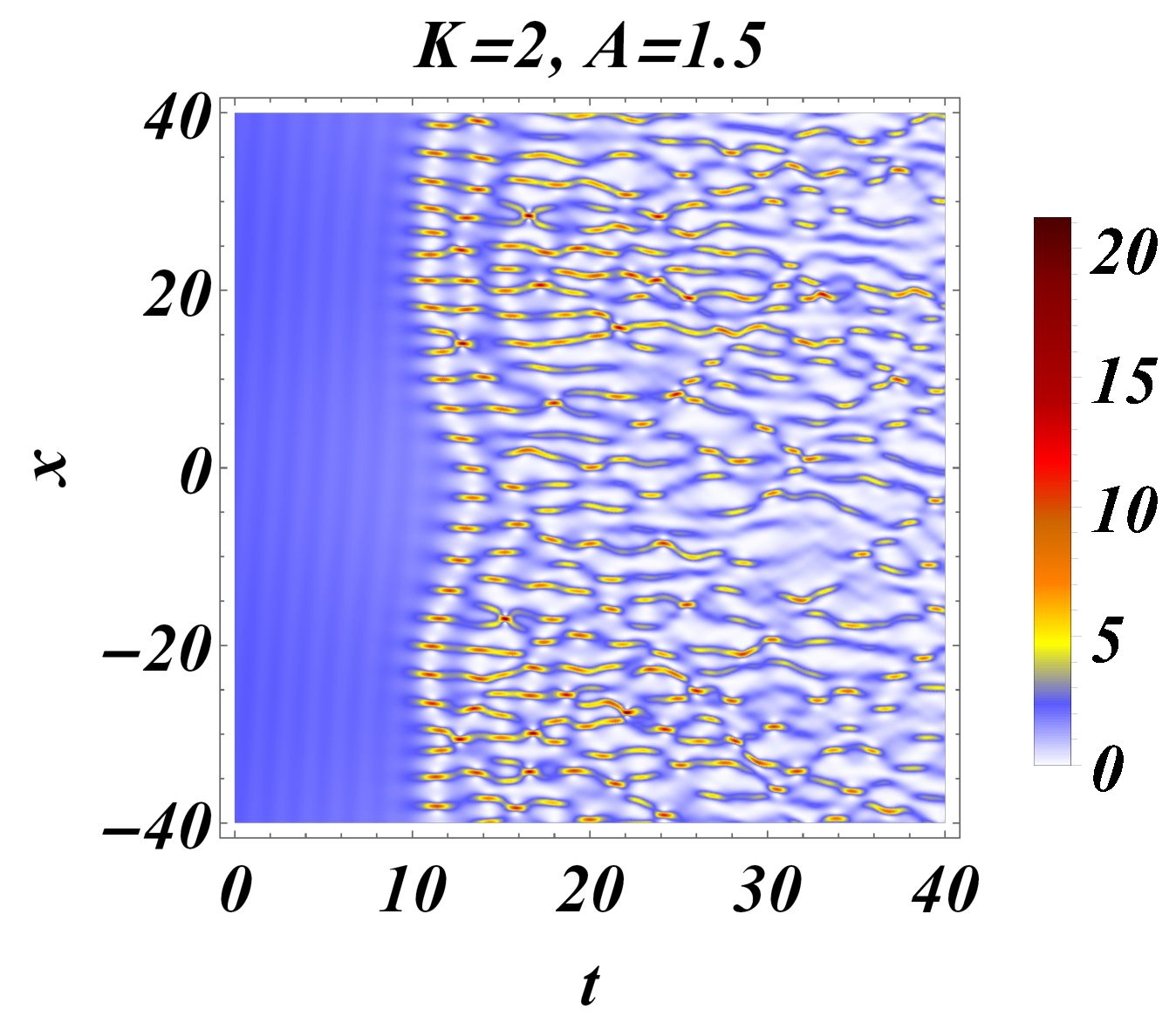}
	\includegraphics[scale=0.23]{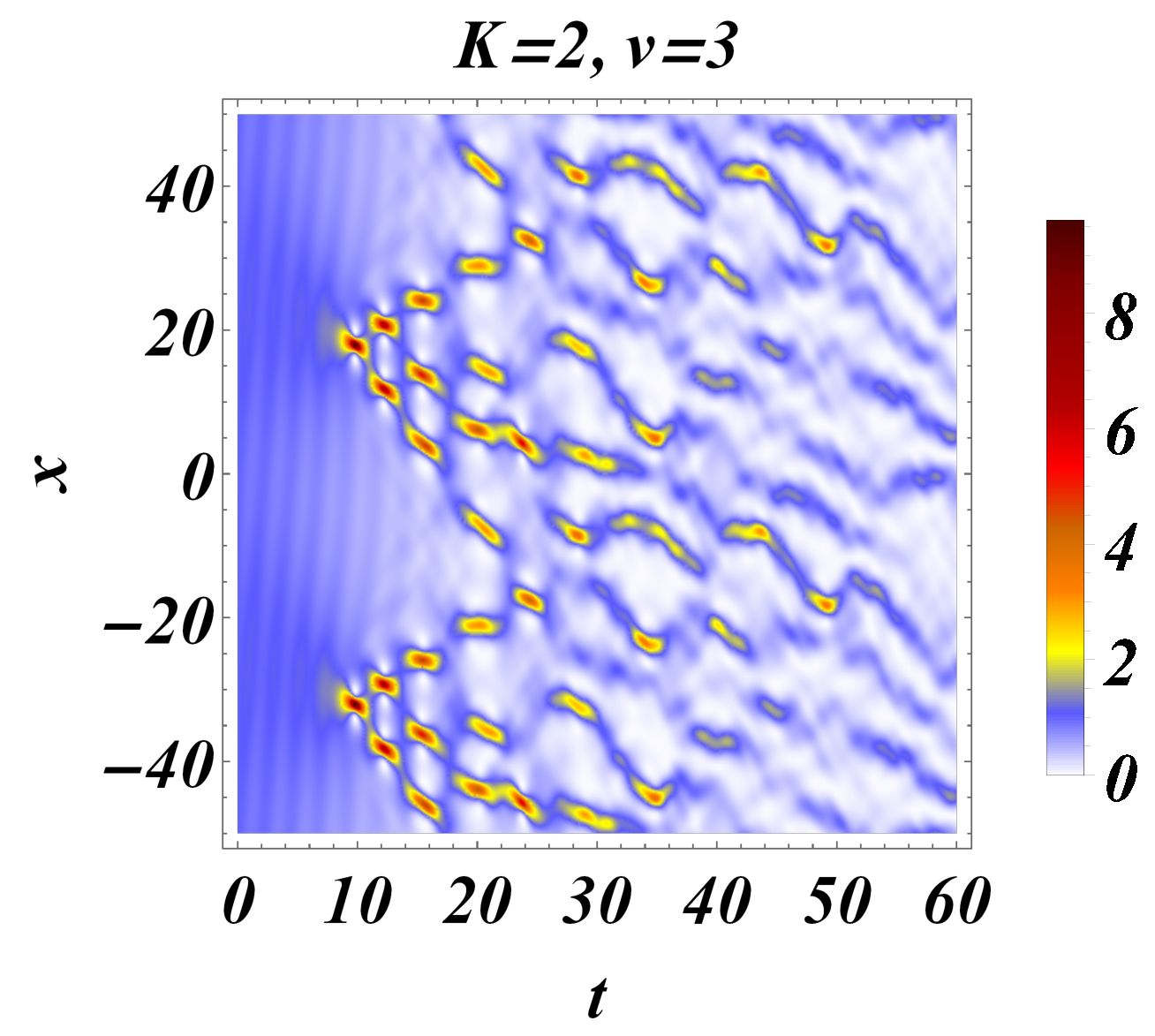}	
	\caption{(Color Online) Contour plots of the spatiotemporal evolution of the density, when a
		parameter is varied while the rest of them are fixed as in Fig. \ref{figure1}.}
	\label{figure6}
\end{figure}
\begin{figure}[tbp!]
	\begin{center}
\hspace{1cm}	\includegraphics[scale=0.31]{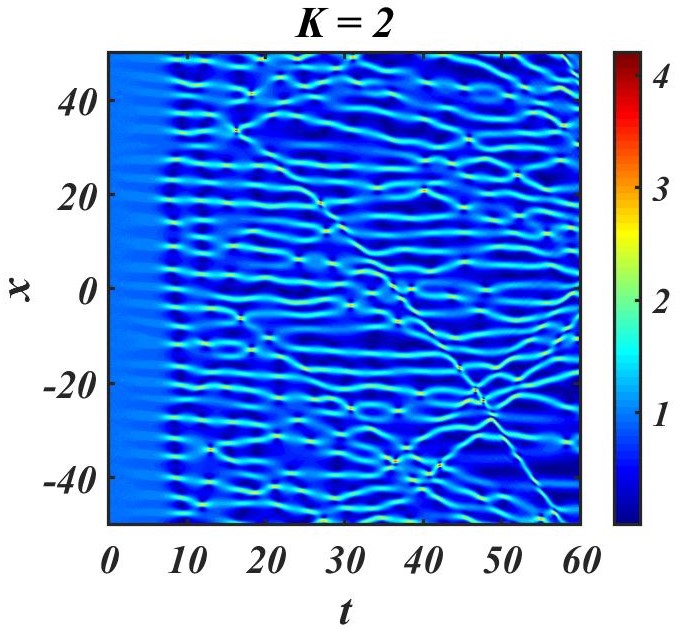}
	\includegraphics[scale=0.215]{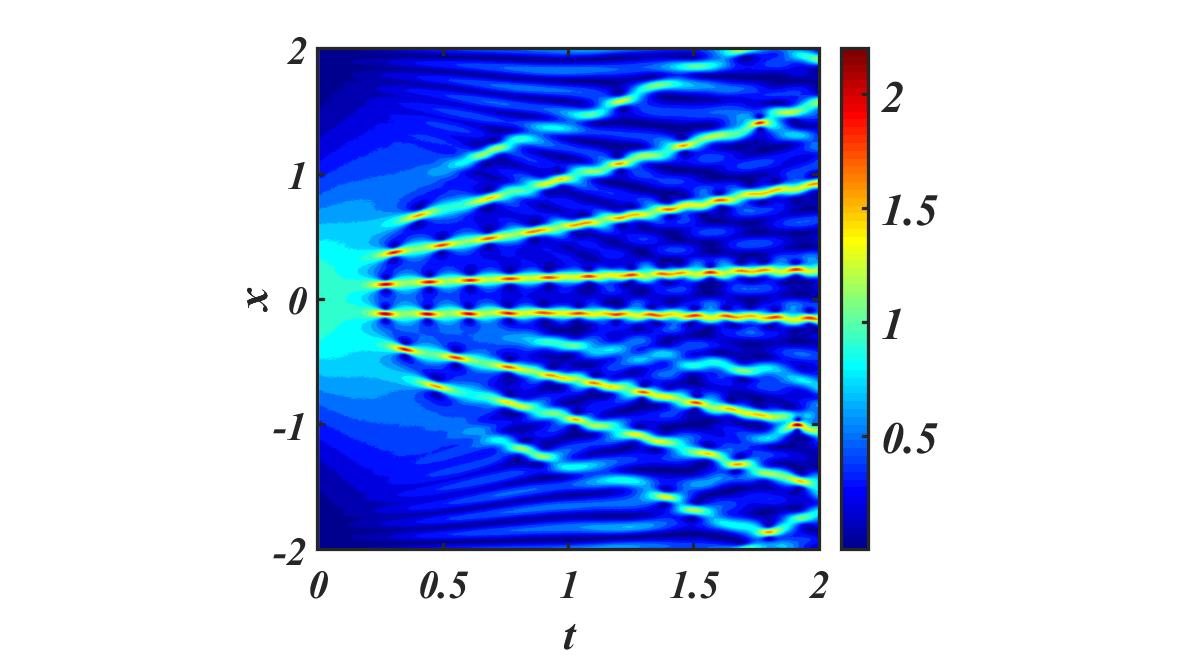}
	\end{center}
	\caption{(Color Online) Left panel: Dynamics of the cw-initial condition \eqref{eq2} for
		$A=1$, and wavenumber $K=2$ when perturbed by random additive noise of $1\%$ for the damped and forced NLS \eqref{eq1}, with parameters
parameters $\nu=\sigma=1$, $\gamma=0.01$,
		$\Gamma=0.1$,  $\Omega=-2$  and $L=50$ (same as in Figure \ref{figure1}). Right panel: Dynamics of the integrable
NLS in the semiclassical limit \eqref{eq1sl} for $\epsilon=0.02$, when the initial condition \eqref{btin} is perturbed by
random additive noise of $1\%$.}
	\label{figure6A}
\end{figure}
\section{Behavior of functionals in the regime of rogue waves and asymptotic behavior}
\subsection{Conserved quantities, balance laws and attractors}

We start the analytical considerations by discussing the conserved quantities
and energy balance laws for \eqref{eq1}. These laws will enable us to justify
the behavior of the functionals which detect the emergence of rogue waves as extrema of functionals. We also recall that
from the energy balance laws,
suitable parametric estimates can be derived  for the solutions, leading to the
well known result of the existence of a global attractor for the corresponding
infinite dimensional dynamical system  \cite{Ghid88, Goubet1, Goubet2}.
This it will be particularly useful in analysing the long-time asymptotics for
certain parametric regimes of interest. 	

The problem is formulated in the Sobolev spaces of $2L$- periodic functions on
the fundamental interval $\mathcal{Q}=[-L,L]$. For the shake of completeness,
we recall their definition:
	\begin{eqnarray}
	\label{defSob}
	H^k_{\mathrm{per}}(\mathcal{Q})&=&\{u:\mathcal{Q}\rightarrow \mathbb{C},\;\;
	u\;\mbox{and}\; \frac{\partial^ju}{\partial x^j}\in L^2(\mathcal{Q}),\;\;
	j=1,2,...,k;\nonumber\\
	&&u(x),\;\;\mbox{and}\;\;\frac{\partial^ju}{\partial x^j}(x)\;\mbox{for
		$j=1,2,...,k-1$, are $2L$-periodic}\}.
	\end{eqnarray}
	
	The change of variables $u\rightarrow
	u\exp(-\mathrm{i}\Omega t)$, transforms the non-autonomous \eqref{eq1}, to its
	autonomous form:
	\begin{eqnarray}
	\label{eq1aut}
	\mathrm{i}{{u}_{t}}+\frac{\nu}{2}{{u}_{xx}}+\sigma|u|^2u =\Omega
	u-\mathrm{i}\gamma u+\Gamma.
	\end{eqnarray}
	
We proceed to the study of the behavior of various functionals for the solutions.	First, it is important to recall that
the functional
	\begin{eqnarray}
	\label{mom1}
	I(t)=\frac{\exp(2\gamma t)}{2L }\operatorname{Im}\int_{\mathcal{Q}}{u\left( x,t
		\right)\overline{{{u}_{x}}}\left( x,t \right)dx},
	\end{eqnarray}
is a conserved quantity for \eqref{eq1}, implying that
 the momentum functional $M(t)$ defined in Eq.~\eqref{mom2}
decays exponentially:
		\begin{eqnarray}
		\label{mom3}
		M(t)=M(0)\exp{(-2\gamma t)},\;\;\;\lim_{t\rightarrow\infty}M(t)=0,
		\end{eqnarray}
as it was proved in \cite{Bar8}.
	\begin{figure}[!htb]
		\centering
		\includegraphics[scale=0.4]{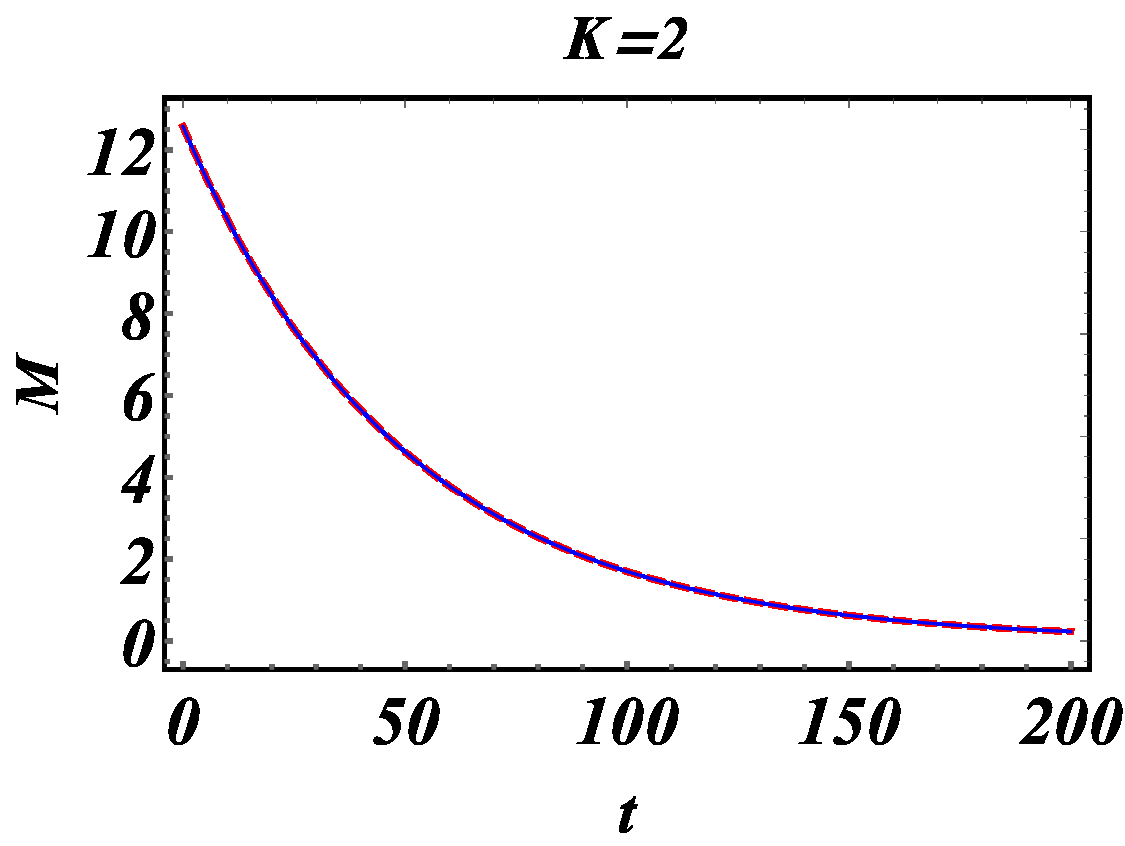}
		\includegraphics[scale=0.40]{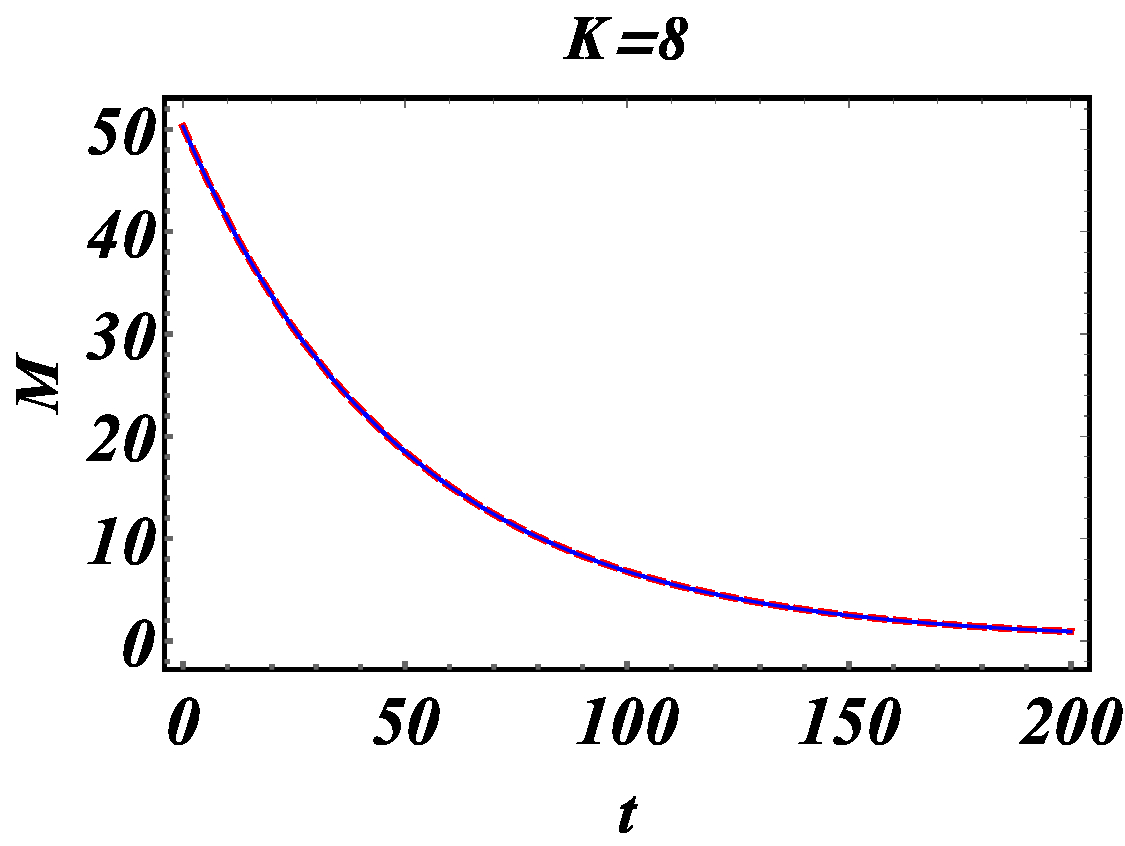}\\
		\caption{(Color Online) Top row: Evolution of the momentum functional $M(t)$
			for two cases of the wave number $K$ of the initial condition \eqref{eq2}. Left
			panel for $K=2$ and right panel for $K=8$. The continuous (blue curve) traces
			the analytical exponential decaying law \eqref{mom3}, and the dashed (red) curve
			the numerical evolution of $M(t)$.  }
		\label{figure10}
	\end{figure}
The consderation of the momentum $M(t)$  is important for the justification of the
	numerical results on the complex dynamics discussed above: Recalling  again \cite{HA,AH},
	numerical complex (even chaotic) behavior may be produced by numerical schemes
	which do not preserve the conservation laws of the system. This is not the case in our numerical
	results for the damped and forced case. Fig. \ref{figure10} depicts the comparison of the evolution of the
	momentum $M(t)$ as given analytically by \eqref{mom3} [continuous (blue) curve],
	against its numerical values [dashed (red) curve] for two cases of wave numbers
	$K=2$ and $K=8$; the rest of parameters are fixed as in the study of Fig.
	\ref{figure1}. The system is integrated for $t\in [0,200]$, and in both cases of
	wave numbers  we observe the excellent agreement of the numerical evolution of
	the momentum with the analytical prediction.
	%
	In the same point of view we may examine the  behavior of
	some other conserved quantities of the integrable limit in the damped and forced
	NLS \eqref{eq1}. The power $P(t)$  satisfies when  $\gamma, \Gamma>0$ the  estimate
	\begin{eqnarray}
	\label{eqlem2}
	P(t)\leq P(0)\exp(-\gamma t)+\frac{2\Gamma^2L}{\gamma^2}[1-\exp(-\gamma
	t)],\;\;\mbox{for all}\;\;t>0.
	\end{eqnarray}
	From the estimate \eqref{eqlem2}, one may extract uniform boundedness of the solutions of \eqref{eq1}
	in the $L^2$-norm.  For the other fundamental conserved quantity of the integrable
	limit, the Hamiltonian energy, it can be shown that in the case of
	the damped and forced NLS \eqref{eq1}, it
	satisfies the balance law
	\begin{eqnarray}
	\label{eqlem5}
	\frac{1}{4}\frac{d}{dt}H(t)=\mathrm{Re}\int_{\mathcal{Q}}\mathrm{i}\gamma
	u\bar{u}_t\,dx+\Omega\mathrm{Re}\int_{\mathcal{Q}}u\bar{u}_tdx
	-\Gamma\mathrm{Re}\int_{\mathcal{Q}}\bar{u}_tdx.
	\end{eqnarray}
Next, we may consider the functional
	\begin{eqnarray}
	\label{eqlem8}
F(t)=\frac{\nu}{4}\int_{\mathcal{Q}}|u_x|^2dx-\frac{\sigma}{4}\int_{\mathcal{Q}}|u|^4dx-\frac{\Omega}{2}\int_{\mathcal{Q}}|u|^2dx+\Gamma\mathrm{Re}\int_{\mathcal{Q}}udx,
	\end{eqnarray}
	for which, the following differential inequality holds:
	\begin{eqnarray}
	\label{eqlem11}
	\frac{d}{dt}F(t)+\gamma F(t)\leq
\frac{\gamma\Omega}{2}||u||_{L^2}^2+\frac{3\gamma\sigma}{8L}||u||_{L^2}^4+\frac{9\gamma\sigma^2}{4\nu}||u||_{L^2}^6.
	\end{eqnarray}
	The differential inequality \eqref{eqlem11} if combined with the uniform in time
	estimates for $P$, it also provides uniform in time estimates in the
	$H^{1}_{\mathrm{per}}(\mathcal{Q})$-norm. Then, one may use the energy balance
	equation \eqref{eqlem5} and the energy method of \cite{JB04}, \cite{XW95}
	(together with the compactness of of the embedding
	$H^{1}_{\mathrm{per}}(\mathcal{Q})\subset L^2(\mathcal{Q})$), to prove the
	following result.
	\begin{theorem}
		\label{ga}
		Let $\nu,\sigma,\gamma>0$  and $\Gamma, \Omega\in\mathbb{R}$. Then, for the
		dynamical system $\varphi_t: H^{1}_{\mathrm{per}}(\mathcal{Q})\rightarrow
		H^{1}_{\mathrm{per}}(\mathcal{Q})$ defined from equation \eqref{eq1} when
		supplemented with the periodic boundary conditions \eqref{eq3}, there exists a
		global attractor $\mathcal{A}$ in $H^{1}_{\mathrm{per}}(\mathcal{Q})$.
	\end{theorem}
The global attractor is invariant under the flow, i.e., $\varphi_t(\mathcal{A})=\mathcal{A}$, for all
$t\geq 0$, and
\begin{eqnarray}
\label{basin1}
\mathrm{dist}(\varphi_t(u_0),\mathcal{A})\rightarrow 0,\;\;\mbox{as $t\rightarrow\infty$, for all
$u_0\in H^{1}_{\mathrm{per}}(\mathcal{Q})$}.
\end{eqnarray}
The distance in \eqref{basin1} is defined as a distance between a point $u$ and a set $\mathcal{S}$ of
$H^{1}_{\mathrm{per}}(\mathcal{Q})$:
\begin{eqnarray}
\label{basin2}
\mathrm{dist}(u,\mathcal{S})=\mathrm{inf}_{w\in\mathcal{S}}||u-w||_{
H^{1}_{\mathrm{per}}(\mathcal{Q})}.
\end{eqnarray}
For the parametric regimes we are interested in our simulations, the structure of the global attractor
will be revealed in the next section.

\subsection{Rogue waves detected by extrema of functionals.}	

Although the momentum
functional $M(t)$ involves the gradient of the solution $u_x$, it does not
capture the manifestation of the PRW-extreme events, which are characterized by
the emergence of steep gradients. Hence, we examine numerically the evolution
of the functionals $P(t)$ and
	$H(t)$, as well as, the components of the Hamiltonian $D(t)$ and $||u(t)||_{L^4}^4$ given in Eq.~\eqref{derN} and
Eq.~\eqref{L4N}, respectively.

\begin{figure}[!htb]
\centering
			\includegraphics[scale=0.4]{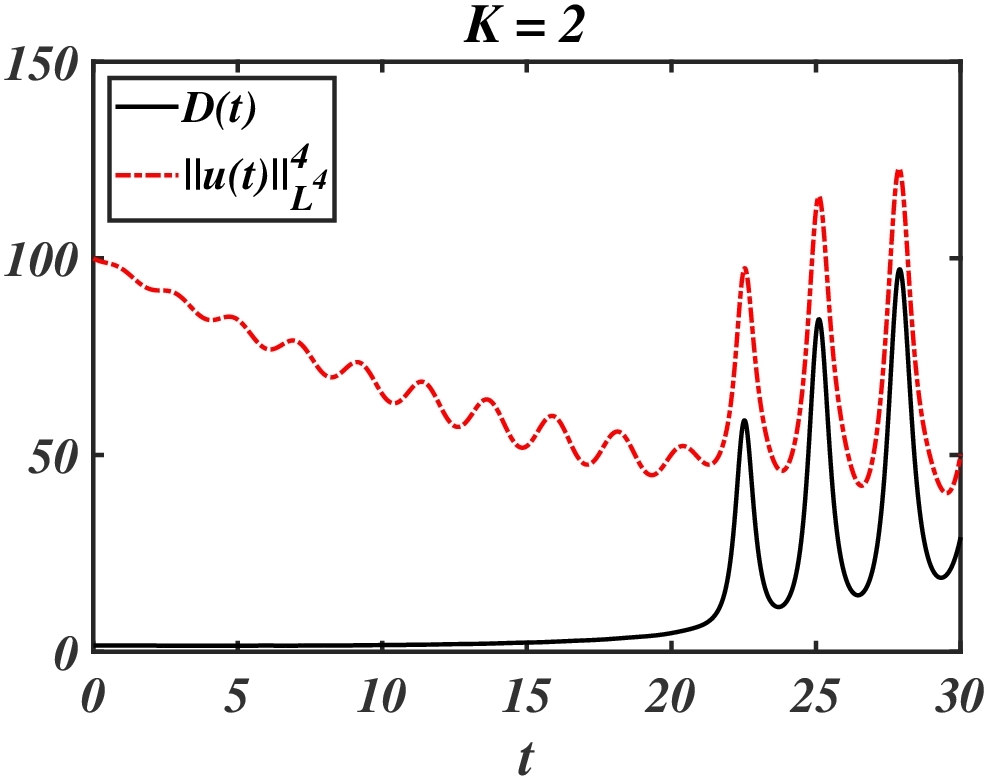}
			\includegraphics[scale=0.4]{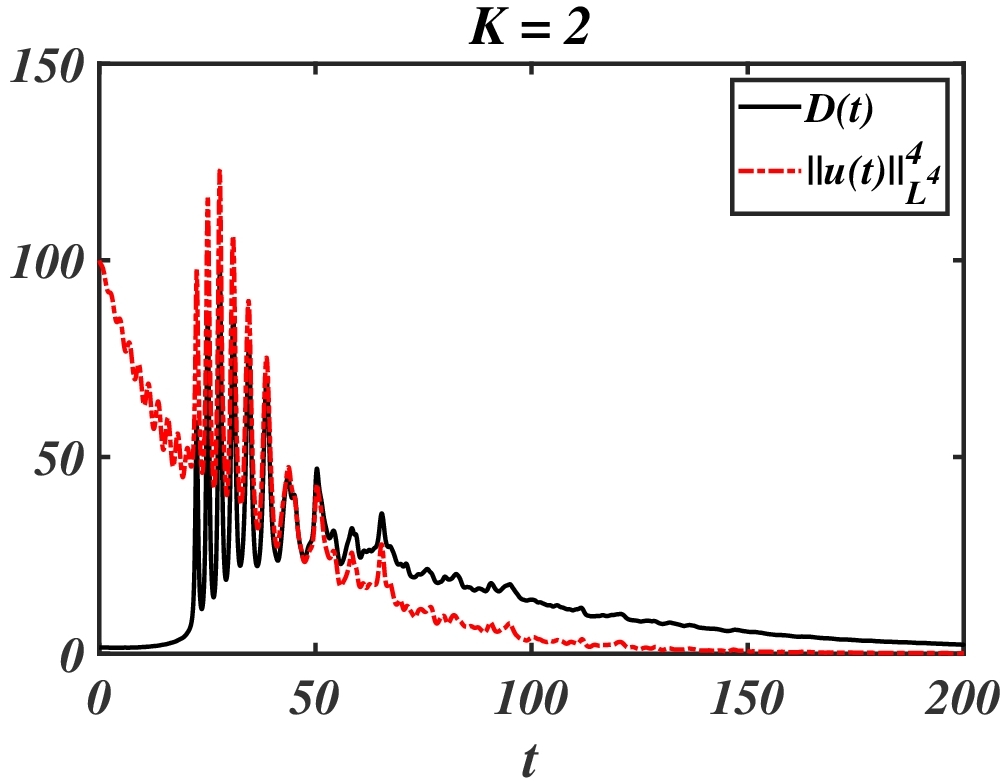}\\
			\hspace{0.19cm}\includegraphics[scale=0.4]{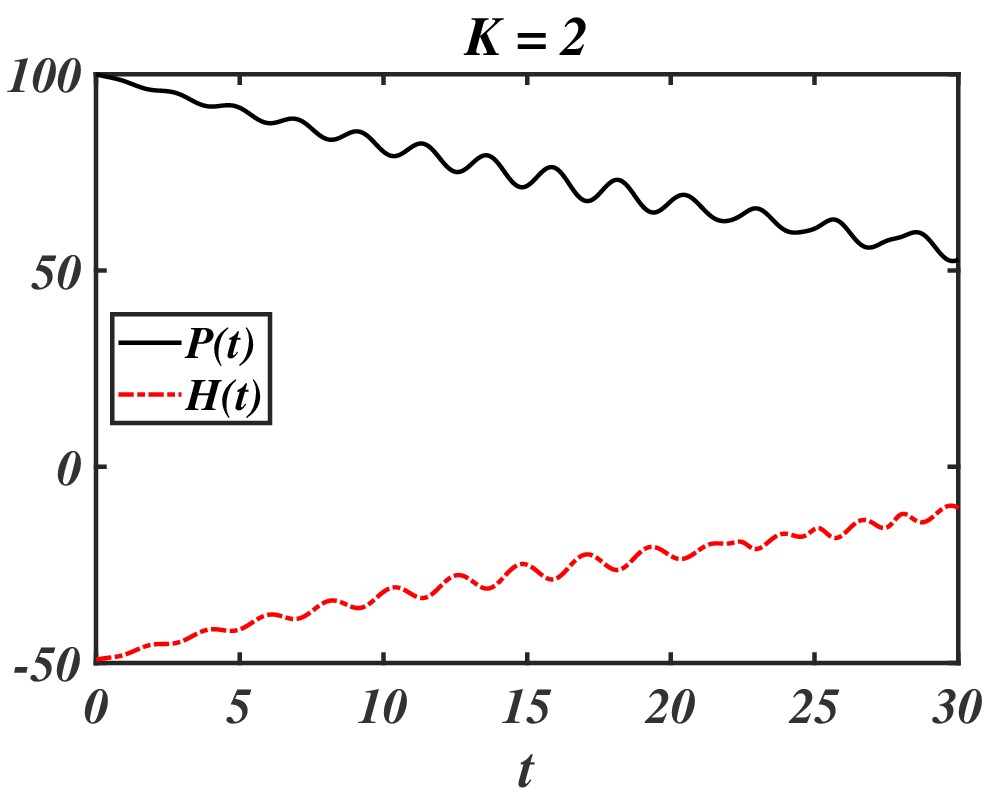}
			\includegraphics[scale=0.4]{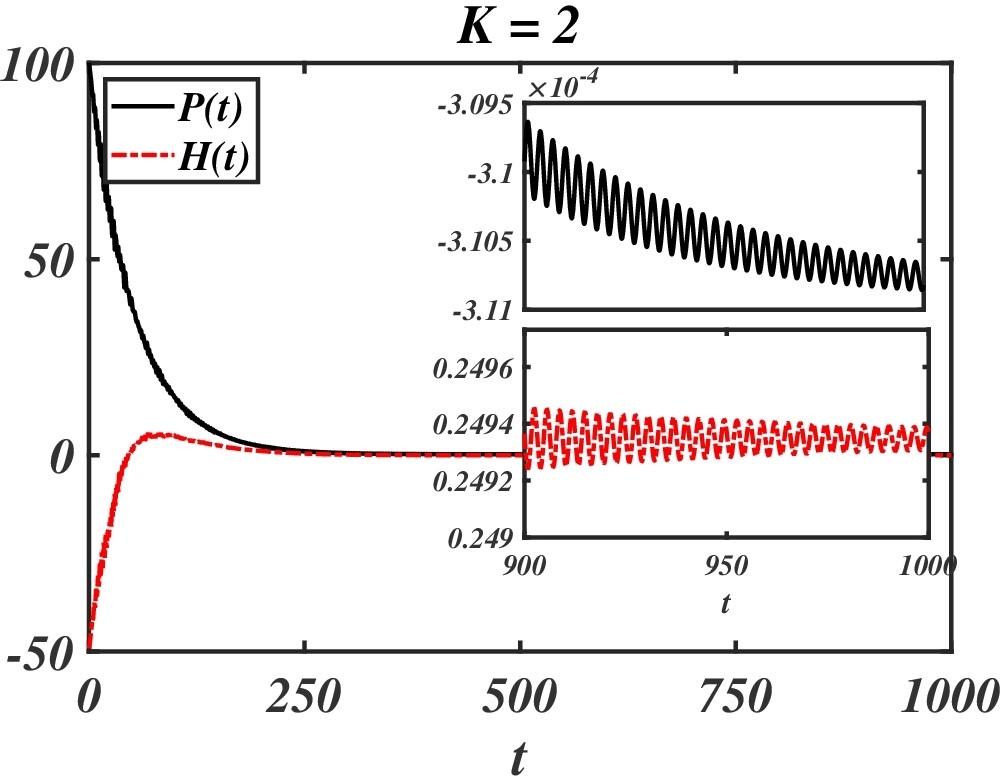}
		\caption{(Color Online) Top left panel: evolution of $D(t)$ (see \eqref{derN}) [continuous (black) curve] and of
			 $||u(t)||^4_{L^4}$ [dashed (red) curve],  for wave number $K=2$ of the initial
			condition \eqref{eq2},  when $t\in [0,30]$. Top right panel: Same as above but
			for $t\in [0,200]$. Bottom left panel: Evolution  of the power $P(t)$ [continuous (black) curve] and of the
			Hamiltonian $H(t)$ [dashed (red) curve], for $K=2$, when $t\in [0,30]$. Bottom right panel: Same as above for
			$t\in [0,1000]$. The insets portray a zoom-in of the evolution when $t\in
			[900,1000]$. }
		\label{figure11}
\end{figure}

\begin{figure}[!htb]
\centering

			\includegraphics[scale=0.4]{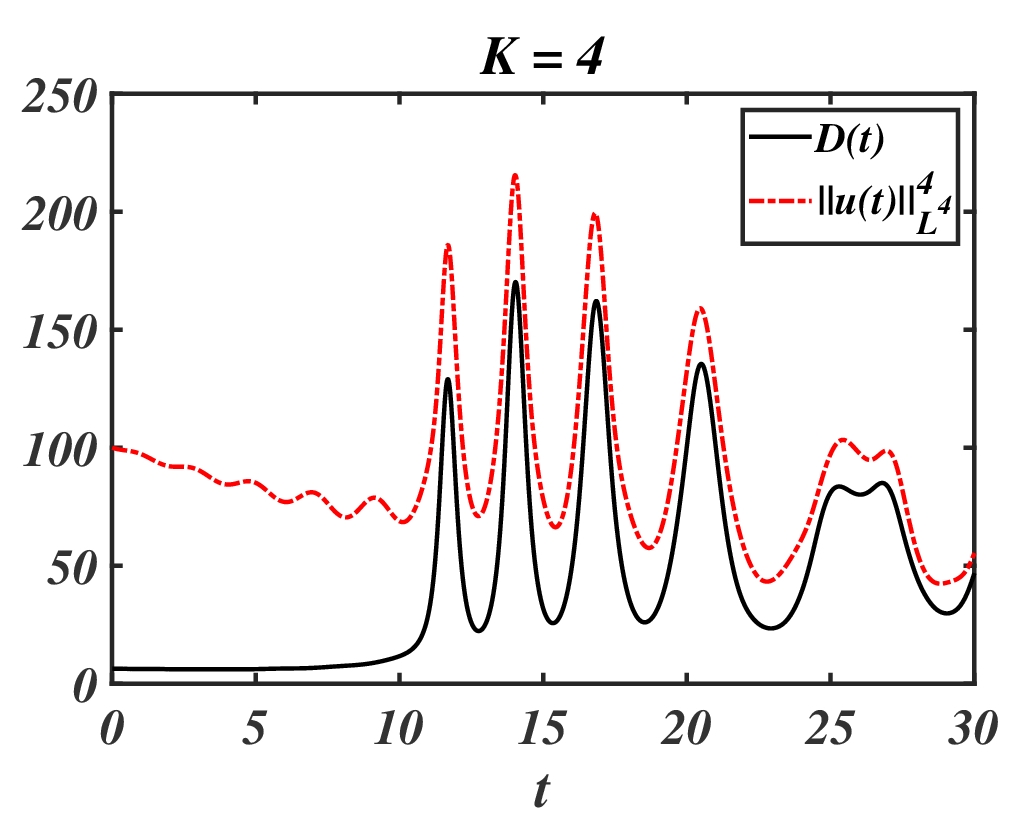}
			\includegraphics[scale=0.4]{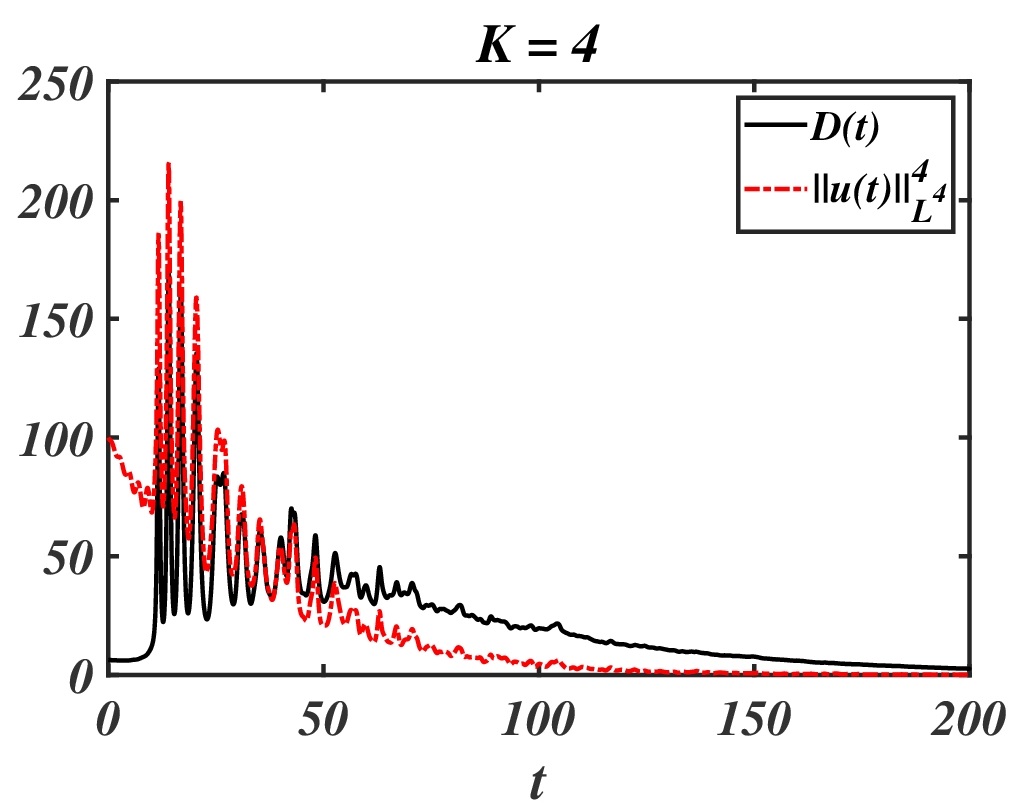}\\
			\hspace{0.12cm}\includegraphics[scale=0.4]{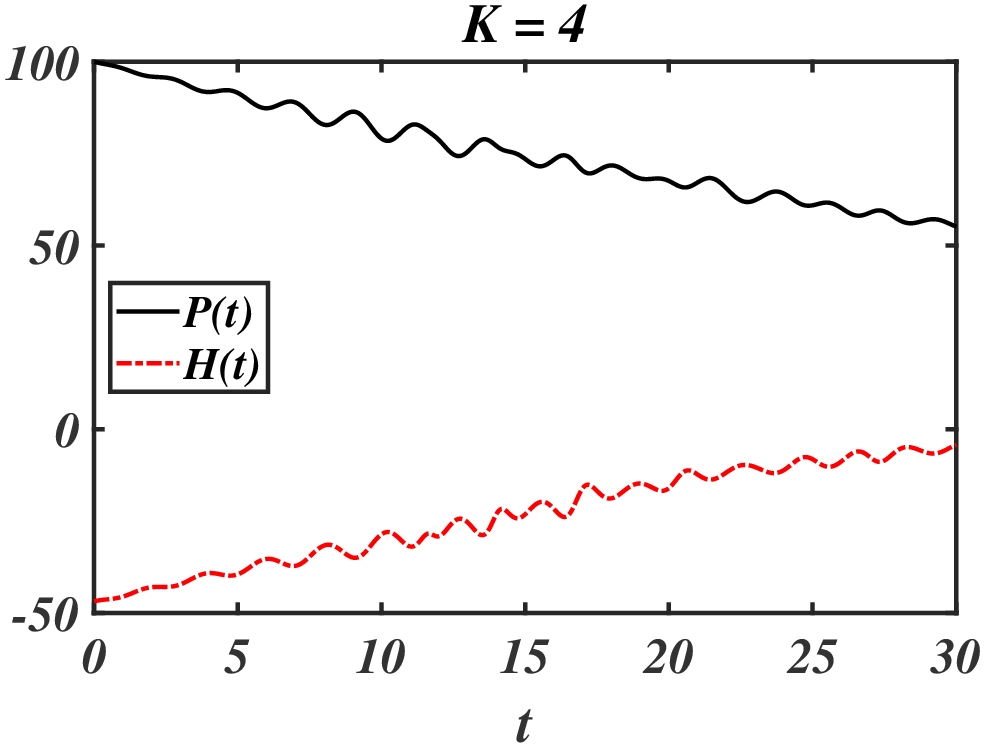}\hspace{0.19cm}
			\includegraphics[scale=0.4]{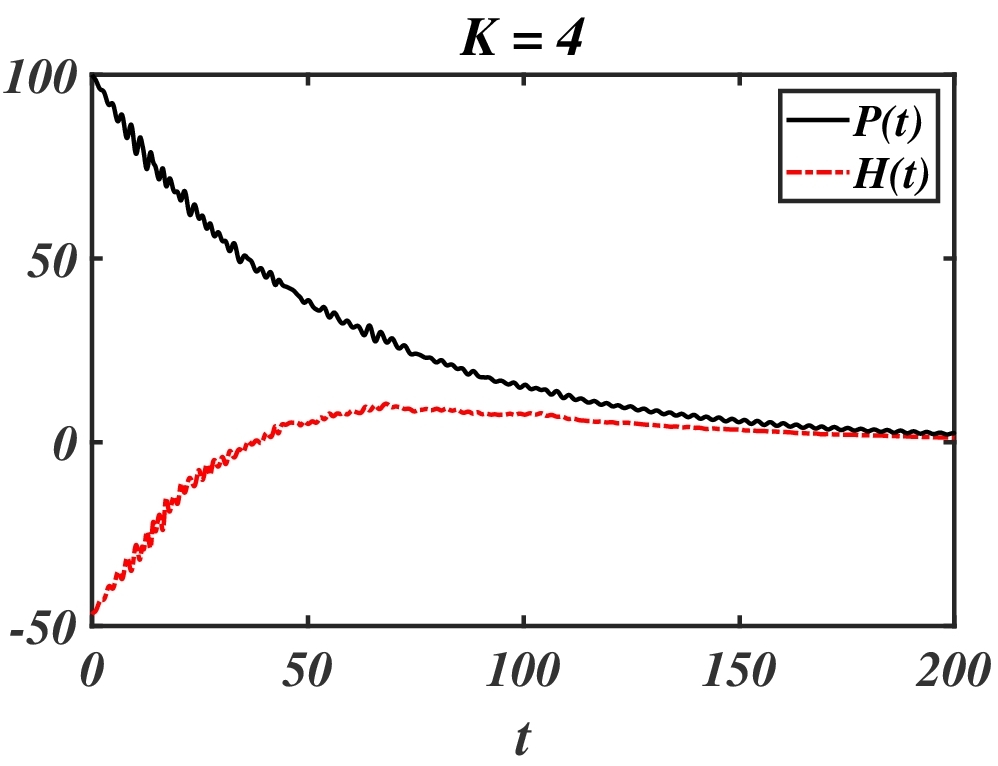}
		\caption{(Color Online) Top left panel: evolution of  $D(t)$ (see \eqref{derN}) [continuous (black) curve] and of
			$||u(t)||^4_{L^4}$ [dashed (red) curve], for wave number $K=4$ of the initial
			condition \eqref{eq2},  when $t\in [0,30]$. Parameters: $\gamma=0.01$,
			$\Gamma=0.1$, $\Omega=-2$. Top right panel: Same as above but for $t\in
			[0,200]$. Bottom left panel: Evolution  of the power $P(t)$ [continuous (black) curve] and of the Hamiltonian
$H(t)$ [dashed (red) curve],
			for $K=4$, when $t\in [0,30]$. Bottom right panel: Same as above for $t\in
			[0,200]$. }
		\label{figure12}
\end{figure}

The top panels of Fig. \ref{figure11} and Figure
\ref{figure12} depict the dynamics of the functionals $D(t)$ [continuous
(black) curve] and of $||u(t)||^4_{L^4}$ [dashed (red) curve], for two
cases of the wave number of the initial condition, $K=2$ and $K=4$
respectively. The rest of parameters are still fixed as in the study of
Fig. \ref{figure1}. The  bottom panels of  Figure \ref{figure11} and Figure
\ref{figure12} depict the dynamics of the power $P(t)$ [continuous (black)
curve] and of the Hamiltonian $H(t)$  [dashed (red) curve].


We observe important differences but also similarities between the
behavior of  $D(t)$ and $||u(t)||^4_{L^4}$, against $P(t)$ and $H(t)$. The differences are the following:
\begin{itemize}
	\item The functionals $D(t)$ and $||u(t)||^4_{L^4}$ detect the emergence of extreme events at the
same time $t^*$ of their occurrence, as it can be easily inferred by a comparison of the top panels of
Figures \ref{figure11}-\ref{figure12} and Figures \ref{figure1}-\ref{figure2} (top and bottom left
panels). This detection is manifested by the \emph{local maxima possessed by $D(t)$ and
$||u(t)||^4_{L^4}$, at the time of occurrence of the rogue waves.} Furthermore, both functionals
detect the time interval where a transient enhanced instability occurs, accompanied by the emergence
of extreme events, while afterwards, both functionals demonstrate stabilization manifested by a
converging behavior.
	\item The power $P(t)$ and the Hamiltonian $H(t)$ do not exhibit extrema at the time
$t^*$ of occurrence of the extreme events. Both functionals exhibit (modulo oscillations) a uniform
with respect to time converging behavior, without detecting the interval of the above enhanced
transient instability.
\end{itemize}
The simulations suggest some other features:
\begin{itemize}
\item Prior to the detection of the emergence of the first rogue wave
    accompanied by the manifestation of the first local maximum, the
    functional $D(t)$ is almost constant for a finite time interval ($[0,
    T_s]$ ($T_s\approx 5$ in the case $K=2$ and $T_s\approx 10$ in the case
    $K=4$). This is the time interval where the instability is not yet
    strongly manifested; the initial condition has the form of a plane wave
    and within this time interval, the solution resembles closely a plane
    wave.  For instance, we may assume that for short time intervals prior
    the manifestation of the instability, like $[0, T_s]$, the solution
    can be approximated as
\begin{eqnarray}
\label{aws}
u(x,t)\approx Ae^{i\theta(x,t)},\;\;\theta(x,t)\approx k x-\omega t,\;\;k\approx
\frac{K\pi}{L},\;\;\omega\approx \Omega.
\end{eqnarray}
The approximation \eqref{aws} implies that $D(t)\approx \frac{2A^2K^2\pi^2}{L}$, for $t\in [0,T_s]$.
For $t\gtrsim T_s$, the instability growth becomes exponential and reaches a local maximum due to the
presence of the first rogue wave. The subsequent appearance of rogue waves is detected within the
time interval where $D(t)$ exhibits its large amplitude oscillations. The appearance of the local
maxima of $D(t)$ is explained by the steep gradients of the emerged rogue waves.

\item For $t\in [0,T_s]$ and within the time interval of the enhanced instability (where the local
    maxima appear of both functionals),  the graph $||u(t)||^4_{L^4}$ is above the graph of $D(t)$. We
    also note the decreasing oscillatory behavior of $||u(t)||^4_{L^4}$ prior to the appearance of
    its first local maximum.  Both effects can be explained by the balance law for the Hamiltonian
    $H(t)$ given in Eq. \eqref{eqlem5} and the approximation \eqref{aws}: Substitution of \eqref{aws}
    into the right-hand side of \eqref{eqlem5} yields that
\begin{eqnarray}
\label{2ndap}
\frac{1}{4}\frac{d}{dt}H(t)\approx -2\gamma A \omega L+\frac{A\Gamma\omega}{K}\sin(kL)\sin(\omega
t)\approx -2\gamma A\Omega L.
\end{eqnarray}
For the parameter values $\gamma=0.01$, $\Gamma=0.1$, $\Omega=-2$ considered
in the numerical studies of Figures \ref{figure11} and  \ref{figure12}, the
right-hand side of \eqref{2ndap} is positive for $t\in (0, T_s]$; Thus for
$t\in (0, T_s]$, the Hamiltonian $H(t)$ is increasing as found in the
numerical simulations. Returning to \eqref{2ndap},  by integration with
respect to time for arbitrary $t\in (0, T_s]$, we get
\begin{eqnarray}
\label{3rdap}
H(t)&\approx& H(0)-8\gamma A\omega L t +\frac{16A\Gamma\sin(kL)}{k^2}\cos{\omega
t}-\frac{8A\Gamma\sin(kL)}{k^2}\nonumber\\
&\approx& H(0) -8\gamma A\Omega L t,\;\; t\in (0, T_s).
\end{eqnarray}
For the given set of parameters for Figs. \ref{figure11} and \ref{figure12},
the right-hand side of \eqref{3rdap} captures qualitatively the
superposition of linear and oscillatory growth of $H(t)$ observed in the
numerics for the short time intervals $(0, T_s]$. Furthermore, from
\eqref{3rdap} and the definition of the Hamiltonian \eqref{Hameneg}, we see
that $||u(t)||^4_{L^4}$ is given by
\begin{eqnarray}
\label{4thap}
||u(t)||_{L^4}^4\approx D(t)-H(0)+8\gamma A\Omega L t.
\end{eqnarray}
Comparing again the graphs of $D(t)$ and $||u(t)||_{L^4}^4$ in Figs.
\ref{figure11} and \ref{figure12}, we see that for the given set of
parameters the approximative formula \eqref{4thap} captures the translation
of  $||u(t)||_{L^4}^4$ over $D(t)$ and its decrease prior to
the time interval of enhanced instability, since $H(0)<0$ and $\Omega<0$. Importantly, this translation
explains that the extrema of $||u(t)||_{L^4}^4$ occur at the same times as of
$D(t)$.

\item We also examined the behavior of another important quantity, the variance
\begin{eqnarray}
\label{vir1}
V(t)=\int_{\mathcal{Q}}|x|^2|u(x,t)|^2dx,
\end{eqnarray}
which  may exhibit a similar behavior as $P(t)$ due to the bound
\begin{eqnarray}
\label{vir2}
V(t)\leq L^2P(t),
\end{eqnarray}
and the estimate \eqref{eqlem2}. Indeed, we observe in Figure \ref{figure13} a converging behavior of
$V(t)$ for the same set of parameters and wavenumbers of the initial condition \eqref{eq2} as in the
study of Figures figure \ref{figure11} and \ref{figure12}.
\begin{figure}[!htb]
\centering
		\includegraphics[scale=0.4]{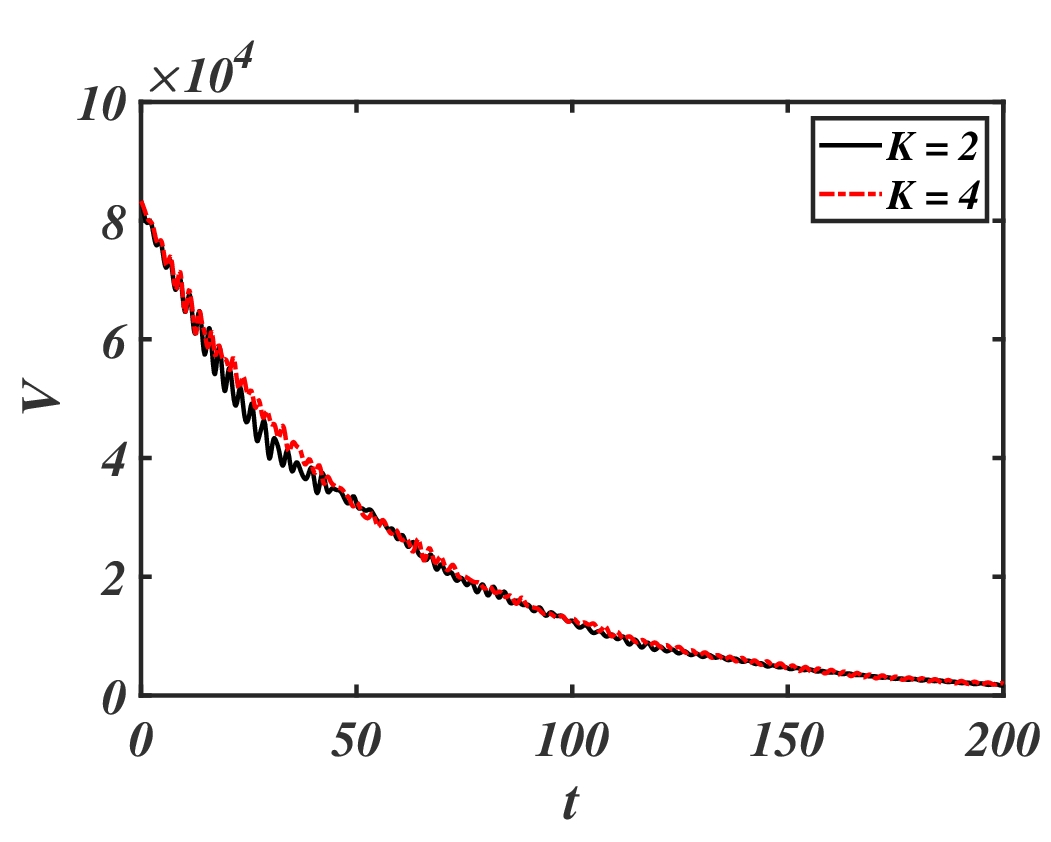}
		\includegraphics[scale=0.4]{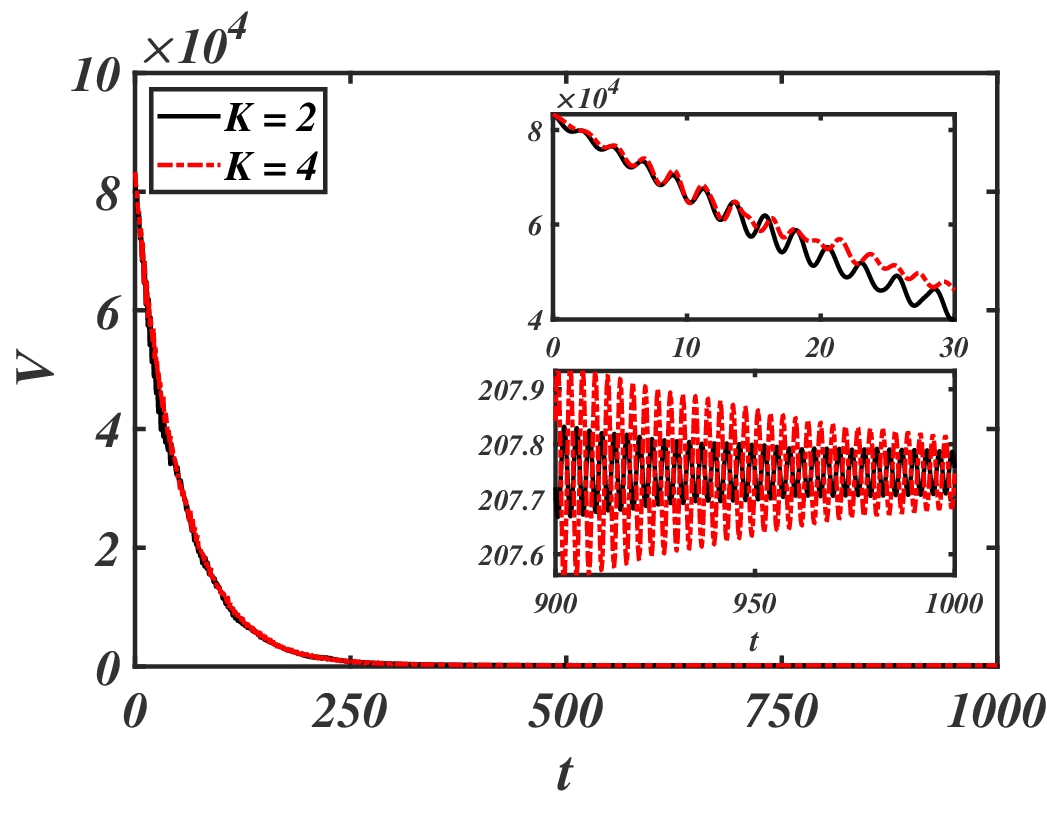}
	\caption{(Color Online) Parameters: $\gamma=0.0$, $\Gamma=0.1$, $\Omega=-2$. Time evolution of
the variance $V(t)$ for the initial condition \eqref{eq2}, when $K=2$ [continuous (black)
curve] and $K=4$ [red (dahsed) curve]. Left panel: time evolution for $t\in [0,200]$. Right panel:
time evolution for $t\in [0,1000]$]. The insets portray a zoom of the evolution for $t\in [0,30]$
(upper inset) and $t\in [900, 100]$ (bottom inset). }
	\label{figure13}
\end{figure}
\end{itemize}	
Summarizing all the above, we found the following: Neither of the main conserved quantities of the
integrable limit $\gamma=\Gamma=0$,  may detect the emergence of rogue waves as extreme events in the
transient dynamics of the initial condition \eqref{eq2} for the damped and forced NLS \eqref{eq1}
$\gamma,\Gamma>0$. Instead, \emph {each part of the Hamiltonian energy $D(t)$ and  $||u(t)||_{L^4}^4$
detects the emergence of rogue waves as local maxima at the times of their appearance.}

\begin{figure}[!htb]
\centering
			\includegraphics[scale=0.4]{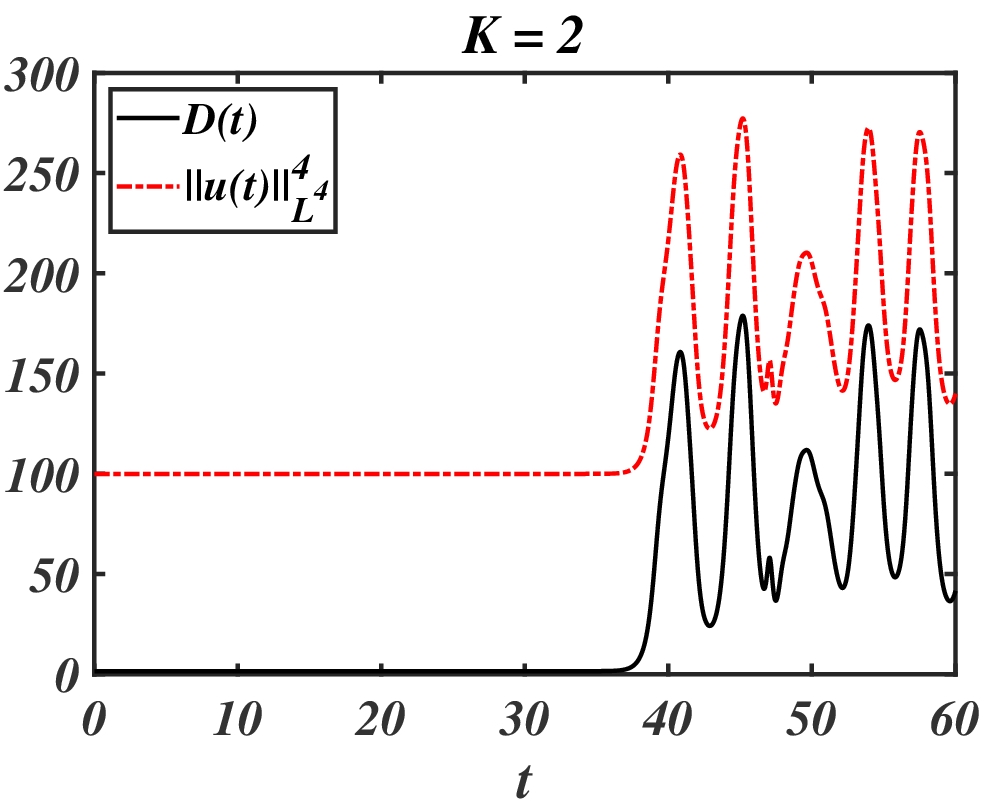}
			\includegraphics[scale=0.4]{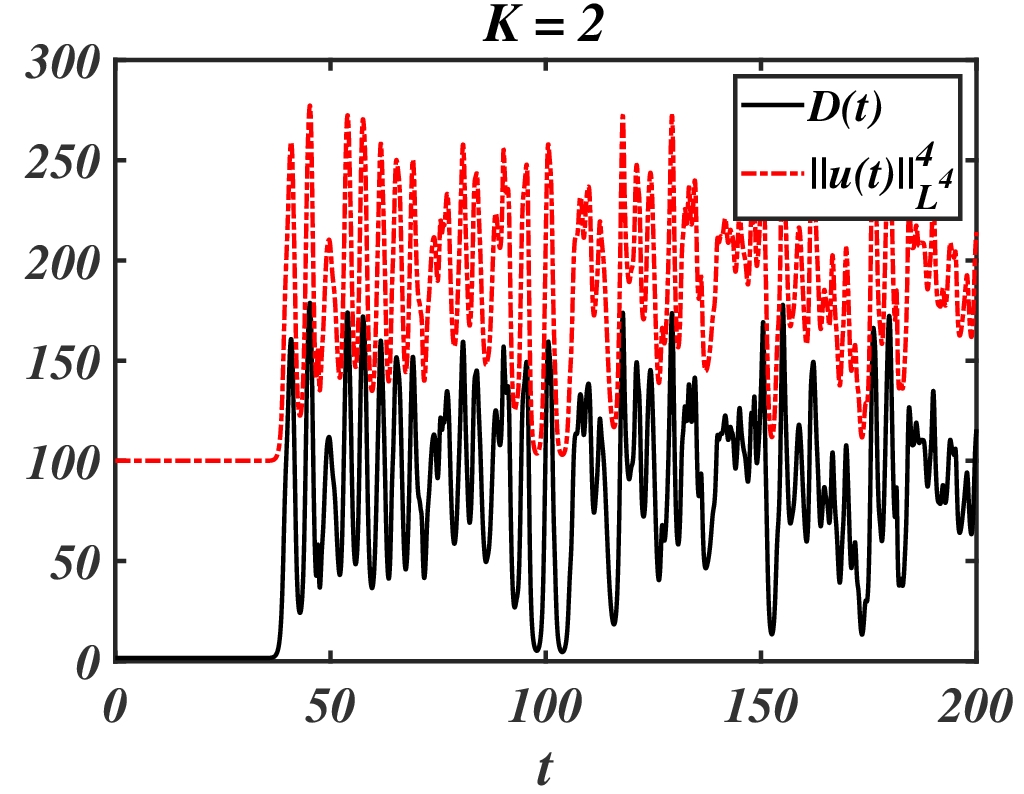}
		\caption{(Color Online) The relative study of Fig. \ref{figure11} in the case of the integrable limit
$\gamma=\Gamma=0$ (see the dynamics depicted in the left panel of Fig. \ref{figure7}).}
\label{figure11a}
\end{figure}

We conclude with a general comment. The diagnostic tools $D(t)$ and
$||u(t)||_{L^4}^4$ may be used as general tools for the detection of extreme
events in NLS-type equations. While we will explore  further this effect elsewhere, we
depict the evolution of these functionals for the integrable NLS equation in
Fig. \ref{figure11a} with the rest of parameters fixed as in the study of Fig. \ref{figure11}, corresponding to the
spatiotemporal dynamics depicted in the left panel Fig. \ref{figure7}.
What is important to note here, is that the two curves are self-similar and only
differ in height consistently. This is due to the conservation of energy which can be rewritten as
\begin{eqnarray*}
||u(t)||_{L^4}^4=D(t)-H(0).
\end{eqnarray*}
Thus, while the Hamiltonian is constant throughout
the evolution and negates any extreme events prediction, this is not the case for its components as we also analysed for the
damped and forced case.
\subsection{Long-time asymptotics in the regime of rogue waves}
\label{lta}
In order to understand the large-time behavior of the solutions as suggested
by above functionals, it is important to discuss the stability of stationary
states of \eqref{eq1}. We will carry out a stability analysis following essentially the
lines of \cite{barashenkov}. We have set for simplicity $\mu=\nu=1$ in
\eqref{eq1}, and summarize the scenarios for modulational stability/instability in the following proposition.
\setcounter{proposition}{1}
	\begin{proposition}
		\label{stab}
		Spatially uniform continuous wave solutions of the form 	
		\begin{equation}\label{cw}
		u_s(t) = \phi_0e^{\rmi\Omega t}, \quad
		\phi_0\in\mathbb{C},\;\;\Omega\in\mathbb{R},
		\end{equation}
		when they exist for \eqref{eq1}, posses the following modulational, linear stability
		properties:
		\begin{enumerate}
			\item
			\begin{enumerate}
				\item When $\Omega>0$, they are stable with respect to small perturbations of
				the form $y(x,t)=z(x)e^{\lambda t}$, $z(x)=e^{-\mathrm{i}k x}$, i.e.,
				modulationally stable, for all $k\in\mathbb{R}$, if $|\phi_0|^2\le \Omega/3$.
				\item When $\Omega<0$ and $|\phi_0|^2<\gamma$, there exists a stability band
				$I_{k^2}=[0,2(-\Omega+|\phi_0|^2)]\cup [2(-\Omega+3|\phi_0|^2),\infty)$, such that
				$u_s$ is modulationally stable if $k^2\in I_{k^2}$. Otherwise is unstable.
			\end{enumerate}	
			\item  When $\Omega>0$ and  $|\phi_0|^2>\Omega/3$, we distinct between the
			following cases:
			\begin{enumerate}
				\item They are unstable if $|\phi_0|^2>\max\left\{\gamma,\Omega/2\right\}$.
				\item When  $\Omega/3 < |\phi_0|^2 < \Omega/2$, they are unstable if
				$\gamma<\Omega/2$.
			\end{enumerate}
		\end{enumerate}
	\end{proposition}
	\emph{Proof:} \emph{A. Existence}.  Substituting \eqref{cw} into \eqref{eq1}, we
	see that $\phi_0$ satisfies the equation,
	$$
	-\phi_0\Omega + |\phi_0|^2\phi_0 = \Gamma - \rmi\gamma \phi_0.
	$$
	Now, writing $\phi_0 = \alpha e^{\rmi\theta}, \alpha\in\mathbb{R},\theta\in
	(-\pi,\pi]$,  the above equation leads to
	$$
	\alpha^3 - \alpha\Omega = \Gamma e^{-\rmi\theta} - \rmi\gamma \alpha.
	$$
	Separating the real and imaginary parts, gives us the system
	\begin{align*}
	\alpha^3 - \alpha\Omega  & = \Gamma\cos\theta,\\
	\gamma\alpha & = -\Gamma\sin\theta.
	\end{align*}
	By eliminating $\theta$, we get
	$$
	\alpha^6 - 2\alpha^4\Omega + (\gamma^2 + \Omega^2)\alpha^2 = \Gamma^2,
	$$
	and setting $\rho = \alpha^2$, we end up with a cubic algebraic equation
	\begin{equation}\label{cubic}
	F(\rho;\gamma,\Gamma,\Omega)\equiv \rho^3 - 2\Omega\rho^2 + (\gamma^2 +
	\Omega^2)\rho - \Gamma^2 = 0.
	\end{equation}
	Substituting $\rho = - q, q>0$, will give us the sum of four negative terms.
	Hence, equation \eqref{cubic} can have either three positive real solutions or
	two complex-conjugate solutions and one real, positive. Any positive root of equation
	\eqref{cubic} defines a solution
	\begin{eqnarray}
	\label{ampsol}
	\phi_0 = \sqrt{\rho}e^{\rmi\theta},\;\;\tan\theta = \frac{\gamma}{\Omega -\rho}.
	\end{eqnarray}
	Through the substitution $\rho = y + 2\Omega/3$ (see \cite[Sec. 2.3.2]{Zwin}),
	equation \eqref{cubic} can be transformed to the ``reduced'' form
	\begin{equation}\label{cubic_reduced}
	y^3 + py + q = 0,
	\end{equation}
	where
	$p = \gamma^2 - \Omega^2/3$ and $q=2/3\Omega(\gamma^2 + \Omega^2/9) - \Gamma^2$.
	
	Analysis of a such cubic equation suggests the distinction between two cases,
	$p\ge 0$ and $p< 0$.
	\begin{itemize}
		\item If $p\ge 0$, then $\gamma^2\ge \Omega^2/3$ and the derivative of the
		equation \eqref{cubic_reduced} will be positive. Hence, \eqref{cubic_reduced} is
		monotone increasing and therefore, has one real solution.

		\item If $p< 0$, then $\gamma^2< \Omega^2/3$ and the number of the real roots
		can be determined by the sign of the following quantity
		\begin{eqnarray}
		\label{Qquan}
		Q = \left(\frac{ \gamma^2 - \Omega^2/3}{3}\right)^3 + \left(\frac{\Omega}{3}
		\big( \gamma^2 + \frac{\Omega^2}{9}\big) - \frac{\Gamma^2}{2}\right)^2.
		\end{eqnarray}
		If $Q<0,$ there are three real roots and if $Q>0,$ there is one. Further
		analysis shows that there exist three roots if $\Gamma_{-}(\gamma) \leq \Gamma
		\leq \Gamma_{+} (\gamma)$, where
		$$
		\Gamma_{\pm}(\gamma,\Omega) = \left\{\frac{2\Omega}{3}\left(\gamma^2 +
		\frac{\Omega^2}{9}\right) \pm\frac{2}{3}
		\sqrt{\frac{1}{3}\left(\frac{\Omega^2}{3} - \gamma^2\right)^3}\right\}^{1/2},
		$$
		and one root otherwise.
	\end{itemize}
	Summarizing all the above, we conclude for the existence of stationary solutions,
	with the following cases:
	\begin{itemize}
		\item  For $\gamma^2 \ge \frac{\Omega^2}{3}$, one real root.
		\item For $\gamma ^2< \frac{\Omega^2}{3}$,
		$\begin{cases}
		\text{three real roots if} \quad  \Gamma_{-}(\gamma) \leq \Gamma \leq
		\Gamma_{+} (\gamma),\\
		\text{one real root, otherwise}.
		\end{cases}$
	\end{itemize}
	In the case where 3 positive roots exist,  there are three branches of solutions
	defined by the critical points of $F$ \eqref{cubic},  namely,
	\begin{equation}\label{critical_points}
	\rho_{\pm}(\gamma,\Omega) = \frac{2\Omega}{3} \pm \frac{1}{3}\sqrt{\Omega^2 -
		3\gamma^2}.
	\end{equation}
	The branches are:  $0 < |\phi_0|^2 \le \rho_{-}(\gamma), \rho_{-}(\gamma) <
	|\phi_0|^2 < \rho_{+}(\gamma)$ and $|\phi_0|^2 \ge \rho_+(\gamma)$.

	\emph{B. Stability}.
	Next, we examine the stability of these solutions.
	Substituting, $\tilde{u}(x,t) = e^{i\Omega t}u(x,t)$ into equation \eqref{eq1}
	we obtain
	\begin{equation}\label{nls_df2}
	\rmi u_t - u\Omega +\frac{1}{2}u_{xx} + |u|^2u = \Gamma - \rmi\gamma u.
	\end{equation}
	
	Let $\xi(x,t)$ be a small perturbation. We consider
	\begin{equation}\label{pert_sol}
	u(x,t) = u_s(x) + \xi(x,t),
	\end{equation}
	where $u_s(x)$ is a stationary solution of equation \eqref{nls_df2}.
	Substituting \eqref{pert_sol} into \eqref{nls_df2}, linearizing and separating
	the real and imaginary parts, yields the following system
	\begin{equation}\label{system}
	J(y_t + \gamma y) = Hy,
	\end{equation}
	where
	$$
	J=
	\begin{pmatrix}
	0 & -1\\
	1 & 0\\
	\end{pmatrix}, \quad\\
	y(x,t)=
	\begin{pmatrix}
	\re~\xi	\\
	\im~\xi
	\end{pmatrix}, \quad\\
	H =
	\begin{pmatrix}
	-\frac{1}{2}\partial_x^2 +\Omega - 3u_R^2 - u_I^2  & - 2u_Ru_I\\
	- 2u_Ru_I & -\frac{1}{2}\partial_x^2 + \Omega - u_R^2 - 3u_I^2
	\end{pmatrix}.
	$$
	Note that, $u_s(x)  = u_R + \rmi u_I$.
	Substituting  $y(x,t) = z(x)e^{\lambda t}$ into \eqref{system}, yields the
	eigenvalue problem
	\begin{equation*}
	Hz(x) = \mu Jz(x),
	\end{equation*}
	where $\mu = \lambda + \gamma $.
	The solution $u_s(x)$ will be stable if the above system does not have
	eigenvalues $\mu$ with real part greater that $\gamma$. In the case of the
	homogeneous solution $u_s(x) = \phi_0$, the eigenvalue $\mu$ and the eigenvector
	$z(x)$ can be found explicitly. Now, by writing $z(x) = z_0e^{-\rmi kx}$ and
	$\phi_0  = u_R + \rmi u_I$, we get
	$$
	(H_k - \mu J)z_0 = 0,
	$$
	where
	$$
	H_k =
	\begin{pmatrix}
	\frac{k^2}{2} + \Omega - 3u_R^2 - u_I^2  & -2u_Ru_I\\
	-2u_Ru_I & \frac{k^2}{2} + \Omega - u_R^2 - 3u_I^2
	\end{pmatrix}.
	$$
	For non-trivial solutions, we require the  determinant of $(H_k - \mu J)$ to be
	zero, implying the equation
	\begin{eqnarray}
	\label{ins}
	-\mu^2 = \left(\frac{k^2}{2}  + \Omega - |\phi_0|^2\right)\left(\frac{k^2}{2}  +
	\Omega - 3|\phi_0|^2\right)\equiv G(k^2).
	\end{eqnarray}
	\begin{itemize}
		\item
		When $\Omega>0$, for $|\phi_0|^2\le \Omega/3, G(k^2)\geq 0$. Hence, $\re\mu =
		0$ and the solution is  modulationally stable.
		\item When $\Omega<0$, we consider
		\begin{eqnarray}
		\label{derk2}
		G'(k^2)= \frac{k^2}{2} + \Omega - 2|\phi_0|^2,	
		\end{eqnarray}
		where the prime denotes differentiation with respect to the variable $k^2$. The
		critical point is
		\begin{eqnarray}
		\label{kkrit}
		k^2_{crit} = 2(2|\phi_0|^2 - \Omega)>0,
		\end{eqnarray}
		since $\Omega<0$, and is a global minimum since
		\begin{eqnarray}
		\label{derk22}
		G''(k^2) = \frac{1}{2}>0.	
		\end{eqnarray}
		Next, we find that $G_{min}=G(k^2_{crit})=-|\phi_0|^4$. Hence, $\re\mu
		=|\phi_0|^2$ and the solution will be unstable for
		$|\phi_0|^2>\gamma$.
		
		In the case where $|\phi_0|^2<\gamma$, we have that $G(k^2)\geq 0$ for $k^2\in
		I_{k^2}$ and the solution is modulationaly stable.
		\item When $\Omega>0$ and $|\phi_0|^2>\Omega/3$, we distinguish between the
		following cases:
		\begin{enumerate}
			\item If $|\phi_0|^2 > \Omega/2$, we work again with
			\eqref{derk2}-\eqref{kkrit}-\eqref{derk22} as in the case $\Omega<0$, to show
			that for
			$|\phi_0|^2>\gamma$ the solution is unstable.
			\item For $\Omega/3 < |\phi_0|^2 < \Omega/2,$ the minimum occurs at $k^2 = 0$
			and is equal to $G_{min} = (|\phi_0|^2 - \Omega)(3|\phi_0|^2 - \Omega)$.
			Therefore, the solution is unstable for
			$$
			-(|\phi_0|^2 - \Omega)(3|\phi_0|^2 - \Omega) > \gamma^2.
			$$
			Further analysis will show that the inequality above is valid when
			$$
			|\phi_0|^2_- < |\phi_0|^2 < |\phi_0|^2_+,
			$$
			where
			$$
			|\phi_0|^2_\pm = \frac{2\Omega}{3} \pm \frac{1}{3}\sqrt{\Omega^2 - 3\gamma^2}.
			$$
			Note that the amplitudes $|\phi_0|^2_\pm$ are equivalent to the critical points of $F$,
			\eqref{critical_points}. Moreover, since $|\phi_0|^2 < \Omega/2$ then
			$|\phi_0|^2_- < \Omega/2$ and after some calculations, we may conclude that this
			instability may occur in the region $\gamma < \Omega/2$.\ \ $\Box$
		\end{enumerate}
	\end{itemize}
\paragraph{Rationalization of the asymptotic behavior for $\gamma=0.01$,$\Gamma=0.1$, $\Omega=-2$.} 	Proposition \ref{stab}
is particularly
useful in explaining the long-time asymptotics observed in Fig. \ref{figure11}.
For this purpose, we recall the notion of orbital stability and orbital asymptotic
stability. The positive orbit  of the flow $\varphi_t$ starting at $u_0$ at
$t_0$ is the set $\mathcal{O}^{+}(u_0)=\cup_{t\geq t_0}\phi_t(u_0)$.
\setcounter{definition}{2}
\begin{definition}
	\label{dos}
The solution $u_*(t)\in H^1_\mathrm{per}(\mathcal{Q})$  is said to be orbitally stable,
if given $\epsilon> 0$, there exists some $\delta= \delta(\epsilon)>0$, such that for any other
solution
$u(t)$  satisfying $||u^*(t_0)- u(t_0)||_{H^1_\mathrm{per}(\mathcal{Q})} < \delta$, then
$\mathrm{dist} (u(t),\mathcal{O}^+(u^*(t_0)) < \epsilon$
for $t >t_0$. It is said to be asymptotically orbitally stable if it is orbitally stable and in
addition, there exists a constant $\tilde{\delta}>0$ such that if $||u^*(t_0)-
u(t_0)||_{H^1_\mathrm{per}(\mathcal{Q})} < \tilde{\delta}$, then
$\lim_{t\rightarrow\infty}\mathrm{dist} (u(t),\mathcal{O}^+(u^*(t_0))=0$.
\end{definition}
Combining Theorem \ref{ga} and Proposition \ref{stab} with Definition \ref{dos}, we have
\setcounter{theorem}{3}
\begin{theorem}
		\label{fc}
		Let $\nu=\sigma=1$ and assume that the parameters $\gamma>0$,
		$\Gamma,\Omega\in\mathbb{R}$ are chosen so that there exists a unique modulationally stable cw
		solution $u^s$ of the form \eqref{cw}. Then it should
		be globally, orbitally asymptotically stable.
	\end{theorem}
	\emph{Proof:} Let $\hat{u}(t)$ denote the solution starting at $t_0=0$ from the initial condition
$\hat{u}_0=u^s(0)+\xi(x,0)$. When the unique cw solution $u_s$ is modulationally stable, we have
	$\lambda<0$, and as its small perturbations decay, the  solution $\hat{u}(t)=\varphi_t(\hat{u}_0)$
comes
	arbitrarily close to $u_s$ in the sense
\begin{eqnarray}
\label{os1}
\mathrm{dist} \left(\hat{u}(t),\mathcal{O}^+(u^s(0))\right)=\mathrm{dist}
\left(\varphi_t(\hat{u}_0),\mathcal{O}^+(u^s(0))\right) < \epsilon.
\end{eqnarray}	
However, according to Theorem \ref{ga}, there exists
a unique global attractor $\mathcal{A}$ attracting all initial conditions $u_0$,
	independently of the size of their $H^1_\mathrm{per}(\mathcal{Q})$- norm and thus the limiting relation
\eqref{basin1} holds. Hence, for all initial conditions $u_0\in H^1_\mathrm{per}(\mathcal{Q})$, there
exists $T_0(u_0)>0$ such that the solutions $u(t)=\varphi_t(u_0)$,with $u(0)=u_0$ satisfy
\begin{eqnarray}
\label{os2}
\mathrm{dist} (\varphi_t(u_0),\mathcal{A}) < \epsilon,\;\;\mbox{for all $t>T_0$}.
\end{eqnarray}		
Since \eqref{os2} holds also for the initial condition $\hat{u}_0$ and
equation \eqref{eq1} is actually autonomous due to \eqref{eq1aut}, the
uniqueness of the attractor and \eqref{os1} imply that
$\mathcal{A}=\mathcal{O}^+(u^s(0))$. Yet, \eqref{basin1} shows that
\begin{eqnarray*}
\label{os3}
\lim_{t\rightarrow\infty}\mathrm{dist} \left(\varphi_t(u_0),\mathcal{O}^+(u^s(0))\right)=0,
\end{eqnarray*}
and that $u^s$ is globally orbitally asymptotically stable.
This asymptotic orbital stability should hold even in
the case where a specific stability band $I_{k^2}$ exists as in Proposition
	\eqref{stab} $1 (b)$, where $G(k^2)\geq 0$ ($\mathrm{Re}\mu=0$) when $k^2\in
	I_{k^2}$ and $G(k^2)<0$ when $k^2\in\mathbb{R}^+\setminus I_{k^2}$
	($\mathrm{Re}\mu\neq 0$): even in the latter case, where $\mathrm{Re}\mu\neq 0$,
	yet $\mathcal{A}=\mathcal{O}^+(u^s(0))$ should attract all initial conditions and
	the divergence from $u_s$ should be only transient; after finite time of
	divergence from $u_s$, the initial condition should approach
	$\mathcal{A}=\mathcal{O}^+(u^s(0))$. \ $\Box$
	
	For the parameters $\gamma=0.01$, $\Gamma=0.1$ and $\Omega=-2$ used in the
	simulations, the algebraic equation \eqref{cubic} has only one real root
	\begin{eqnarray}
	\label{root1}
	\rho_s\approx 0.0025=|\phi_0^s|^2.
	\end{eqnarray}
	This is in accordance with the analysis of Proposition \ref{stab}, since for the
	above choice of parameters the quantity $Q$ defined in \eqref{Qquan} is found to equal
	$Q\approx 0.003>0$. Thus, there exists a unique stationary solution defined by
	\eqref{cw}, \eqref{ampsol} and \eqref{root1},
	\begin{eqnarray*}
		u_s(t) = \phi_0^s e^{\rmi\Omega t},\;\;	
		\phi^s_0 = \sqrt{\rho_s}e^{\rmi\theta_s},\;\;\tan\theta_s = \frac{\gamma}{\Omega
			-\rho_s}.
	\end{eqnarray*}
	with  $\sqrt{\rho_s}\approx 0.05,\;\;\tan\theta_s\approx -0.005$. Now, this
	small amplitude cw $u_s$ satisfies the stability criterion of Theorem
	\ref{stab} 1(b), since $\Omega<0$ and $|\phi_0^s|^2<\gamma$. Then, applying
	Theorem \ref{fc}, we extract its global asymptotic stability. Thus, all
	initial conditions should converge to the small amplitude cw $u_s$, and this
	also holds for the initial condition \eqref{eq2} with $A=1$ and $K=2$,  as shown
	by the convergence of the energy and norm quantities discussed in Figures
	\ref{figure11} and \ref{figure12}.
	
\begin{figure}[!htbp]
\centering
			\includegraphics[scale=0.33]{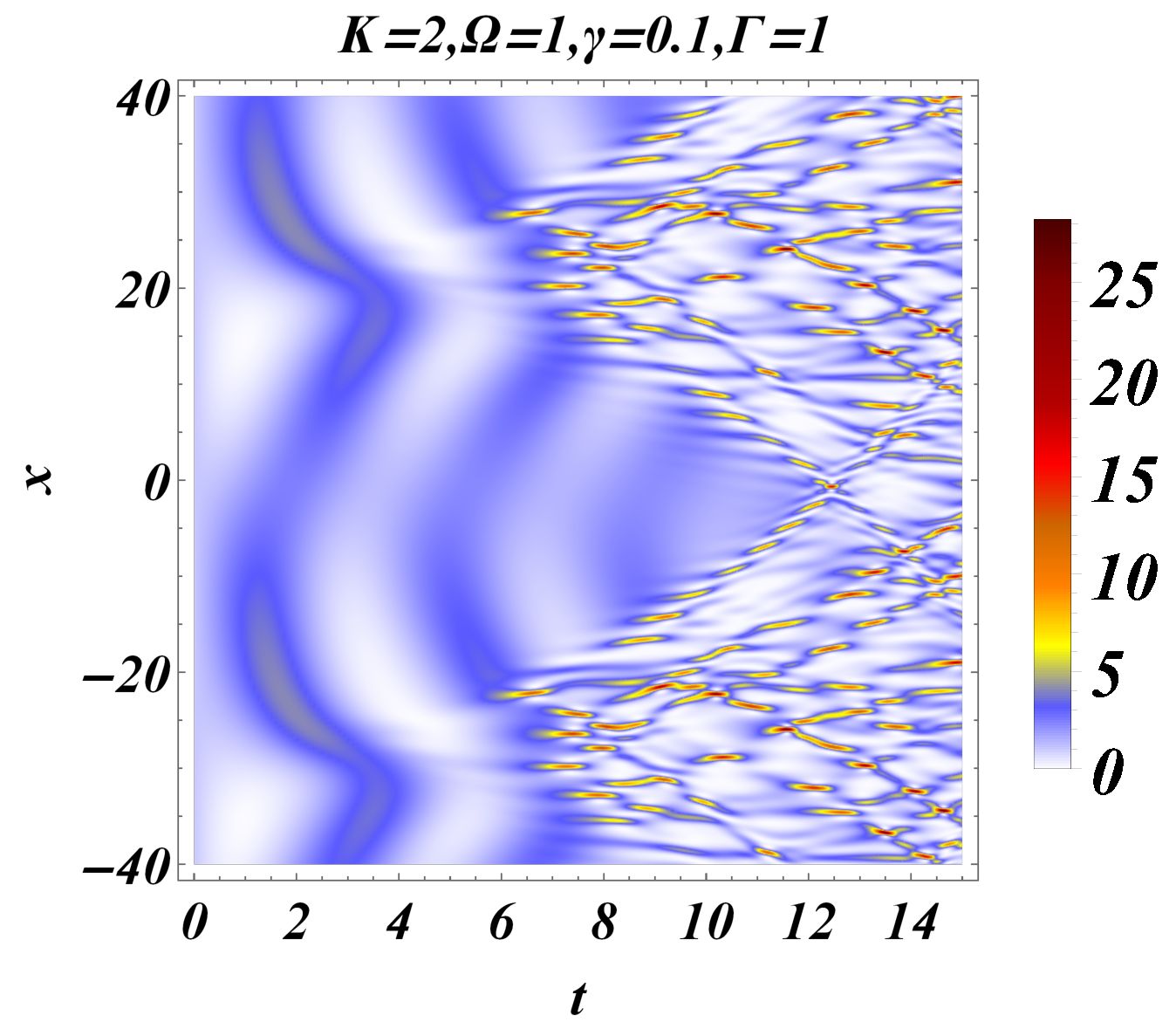}
			\includegraphics[scale=0.336]{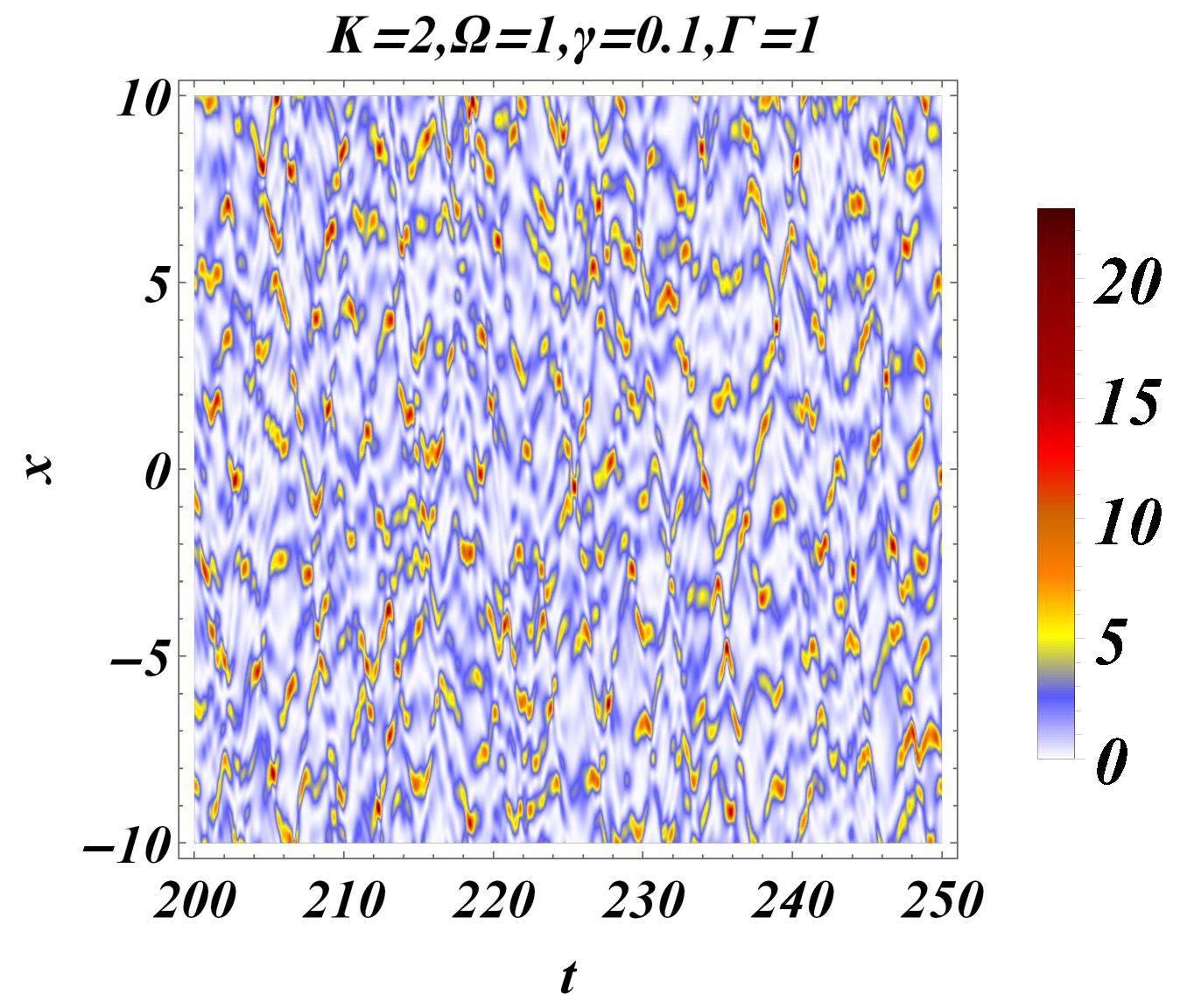}
		\caption{(Color Online) Parameters \eqref{eq1}-\eqref{eq3}:  $\nu=\sigma=1$,
			$\gamma=0.1$, $\Gamma=1$, $\Omega=1$. Dynamics of the cw-initial condition
			\eqref{eq2} for $A=1$, $K=2$, and $L=50$. The left panel shows the
			spatiotemporal evolution of the density for $x\in [-40,40]$ and $t\in [0,15]$. The
			right panel offers a magnification of the patterns observed in the left panel,
			for $x\in [-10,10]$ and $t\in [200,250]$.  }
		\label{figure8}
	\end{figure}

	We present a numerical illustration of the results stated in Proposition
	\ref{stab} and Theorem \ref{fc}.  Figure \ref{figure8} shows the
	spatiotemporal behavior of the solutions when $\nu=\sigma=1$, $\gamma=0.1$,
	$\Gamma=1$, $\Omega=1$. The initial condition is \eqref{eq2} with $A=1$, $K=2$
	and $L=50$.  For this set of parameters there exist again one spatially uniform
	cw solution defined by the real root of the equation \eqref{cubic}
	$$\rho_u=1.69=|\phi_0^u|^2.$$

Since $|\phi_0^u|^2>\max\left\{\gamma,\Omega/2\right\}=0.5$, the solution is
unstable according to the instability criterion of Proposition \ref{stab} $2(b)$.
In the contour plot shown in the left panel for $t\in [0,16]$, we still observe
the emergence of extreme events on the top of two distinct ``tree''-patterns.
However, they do not posses the structure  observed in Figure
	\ref{figure1} and \ref{figure3}. The system exhibits complex spatiotemporal
	behavior as clearly shown in the right panel, for $t\in [200,250]$.

\begin{figure}[!htbp]
	\centering
	\includegraphics[scale=0.33]{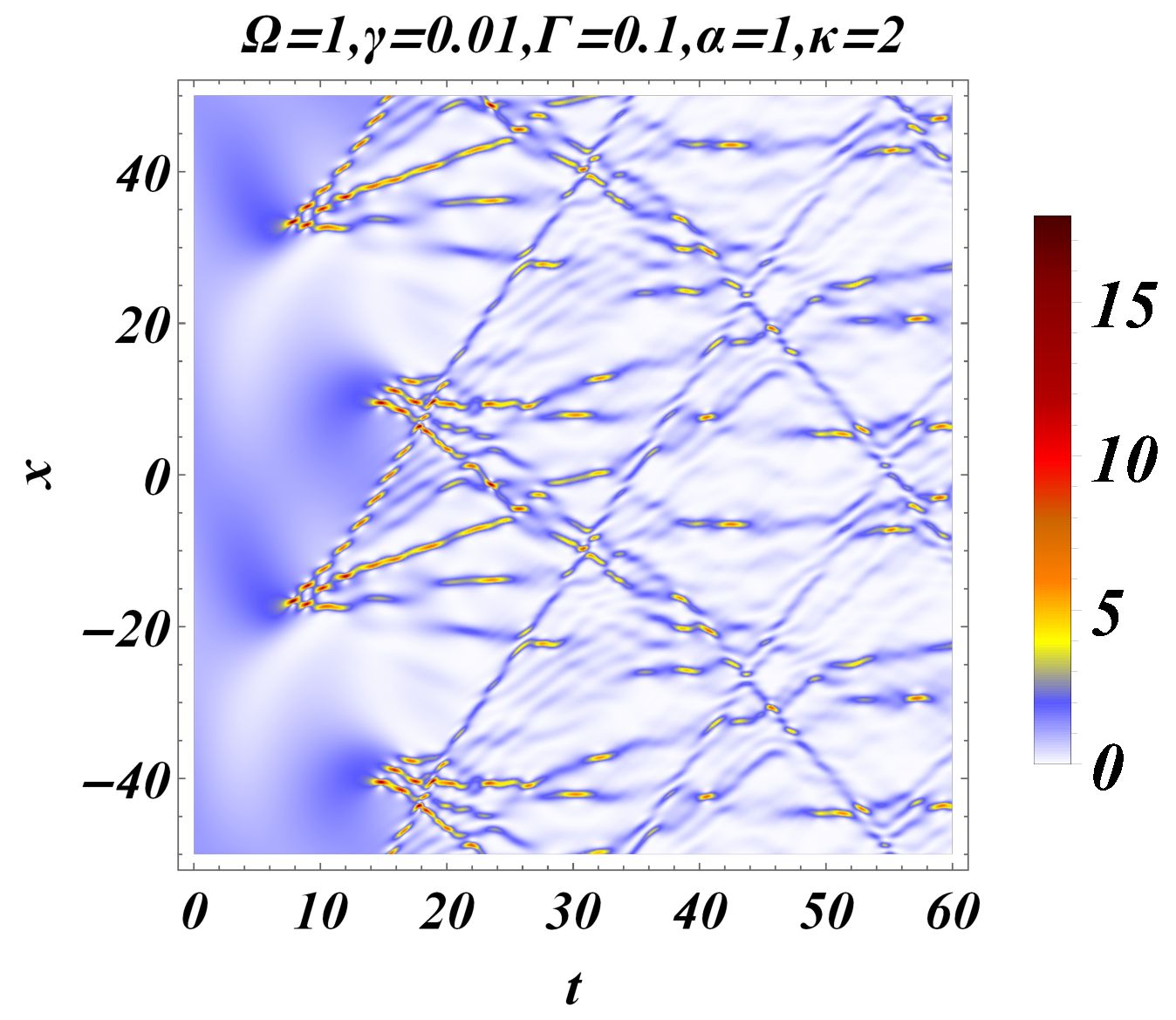}
	\includegraphics[scale=0.33]{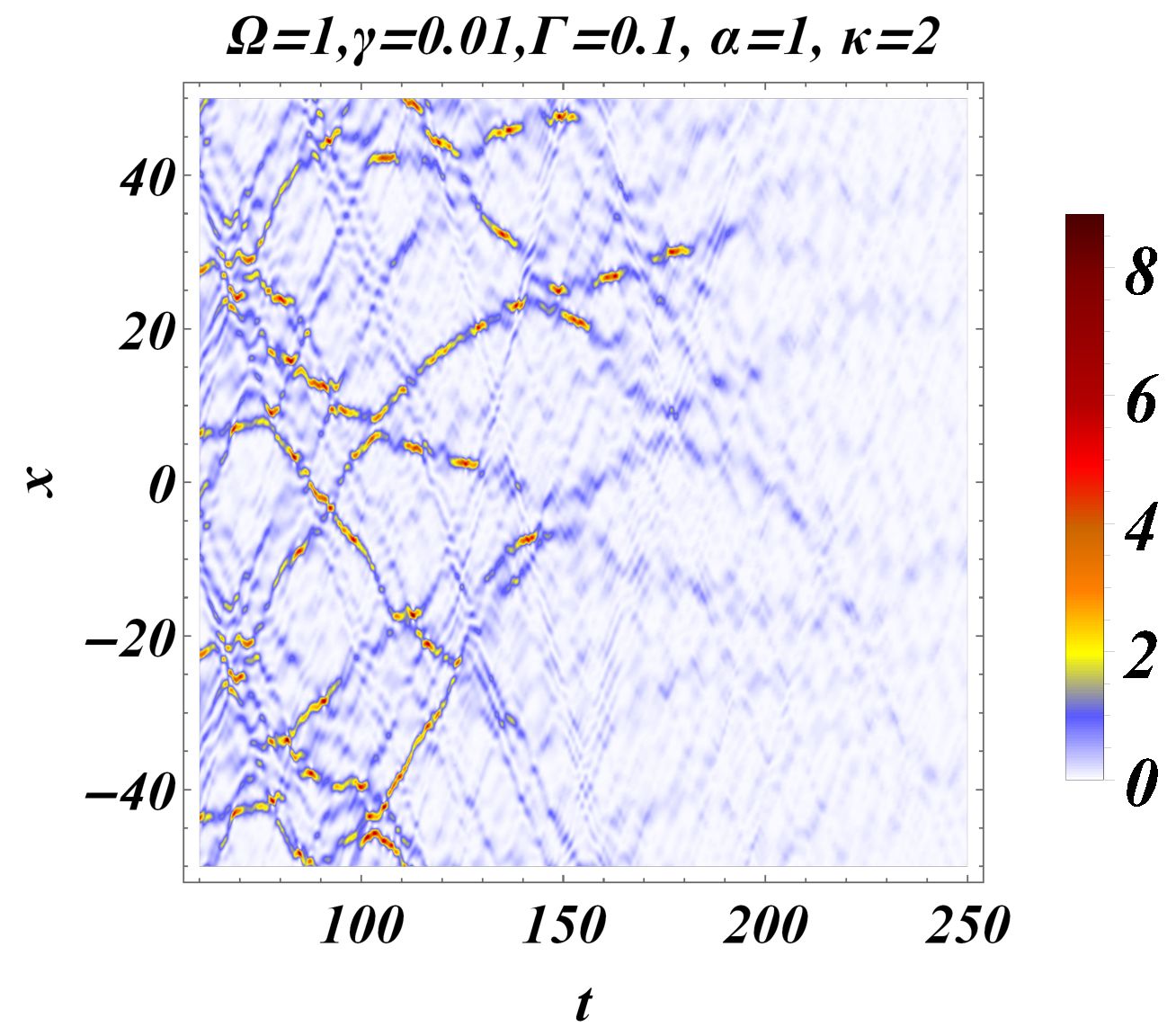}
	\caption{(Color Online) Parameters \eqref{eq1}-\eqref{eq3}:  $\nu=\sigma=1$,
		$\gamma=0.01$, $\Gamma=0.1$, $\Omega=1$. Dynamics of the initial condition
		\eqref{pertin} for $a=1$, $\kappa=2$, and $L=50$. The left panel shows the
		spatiotemporal evolution of the density for $x\in [-L,L]$ and $t\in [0,15]$. The
		right panel depicts the spatiotemporal evolution,
		for $x\in [-L,L]$ and $t\in [60,250]$.  }
	\label{figure9}
\end{figure}

\begin{figure}[!htbp]
\centering
			\includegraphics[scale=0.5]{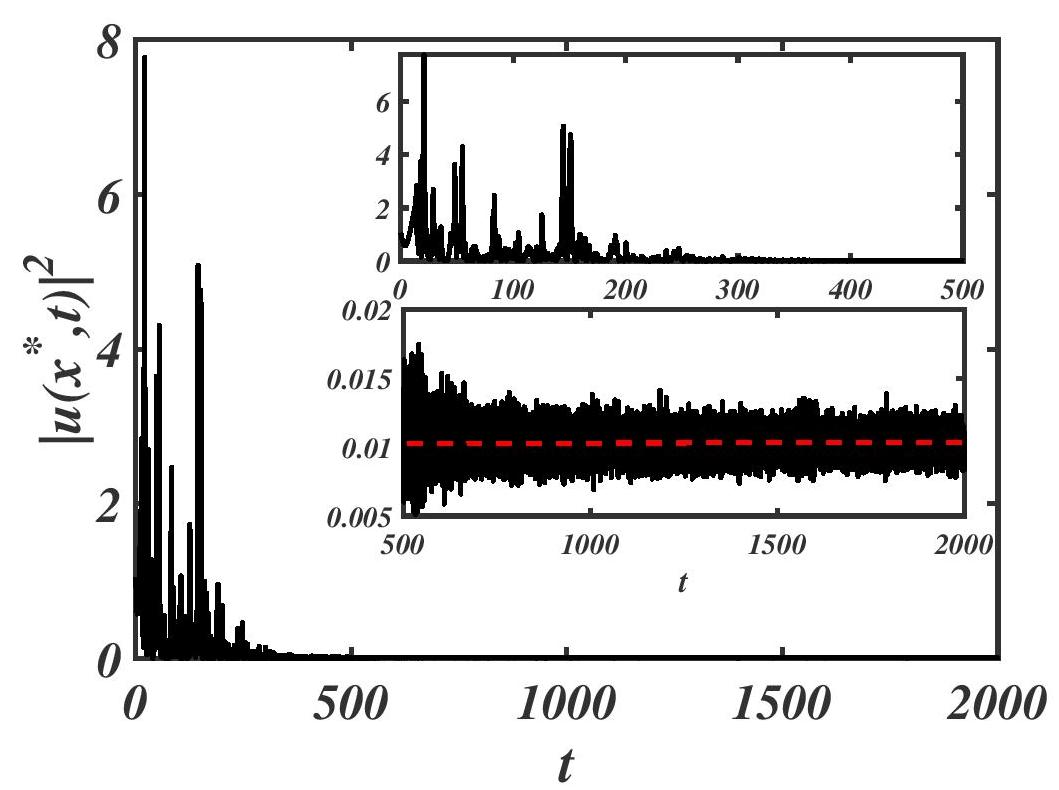}
		\caption{(Color Online) Dynamics of the density $|u(x^*,t)|^2$ [continuous (black) curve], for fixed $x^*=10$ and
the initial condition and parameters of Fig. \ref{figure9}. Asymptotic orbital stability of the cw $u_s^{s,1}$ with
amplitude $\rho^s_1= 0.01=|\phi_0^{s,1}|^2$ marked by the red (dashed) line (see the discussion for the amplitudes given in
\eqref{exors}). }
		\label{figure12a}
	\end{figure}

\begin{figure}[!htbp]
	\centering
	\includegraphics[scale=0.33]{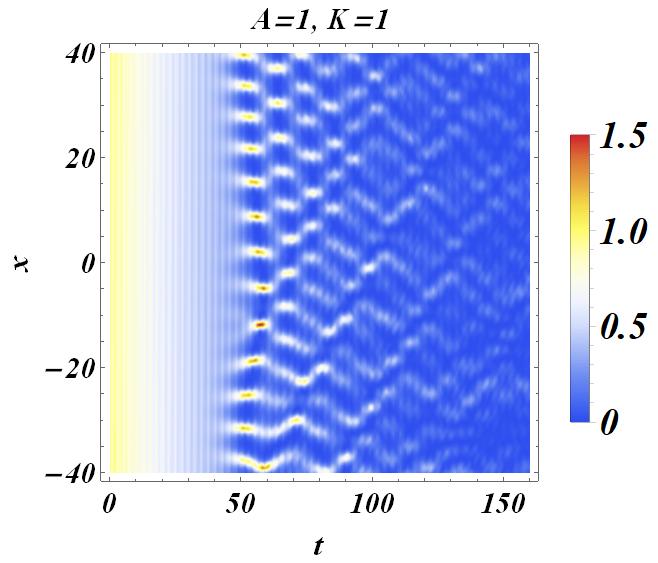}
	\includegraphics[scale=0.33]{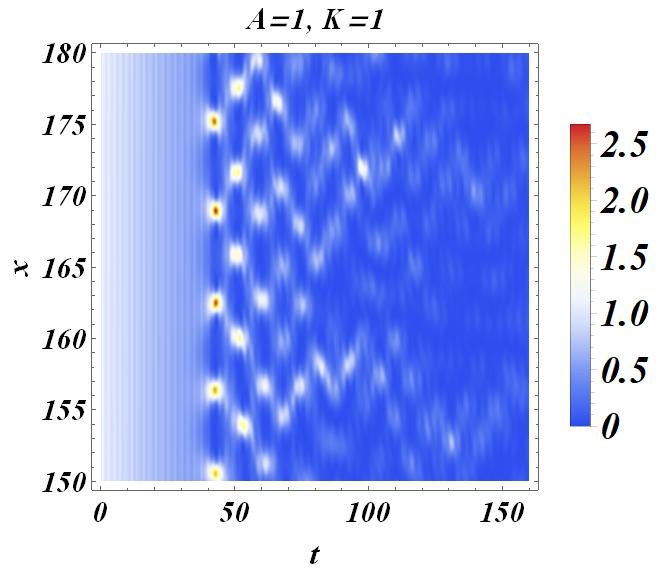}
	\caption{(Color Online) Parameters \eqref{eq1}-\eqref{eq3}:  $\nu=\sigma=1$,
		$\gamma=0.01$, $\Gamma=0.1$, $\Omega=-2$. Dynamics of the initial condition
		\eqref{eq2} for $A=1$, $K=1$, and $L=200$. The left panel shows the
		spatiotemporal evolution of the density for $x\in [-40,40]$ and $t\in [0,160]$. The
		right panel depicts the spatiotemporal evolution,
		for $x\in [150,180]$ and $t\in [0,160]$.  }
	\label{figure16}
\end{figure}

A second numerical example considers the case of parameters $\nu=\sigma=1$,
	$\gamma=0.01$, $\Gamma=0.1$, $\Omega=1$. For these parameters, the quantity $Q$
	defined in \eqref{Qquan} is found $Q\approx -0.0003<0$ and the equation
	\eqref{cubic} has three real roots
\begin{eqnarray}
\label{exors}
\rho^s_1= 0.01=|\phi_0^{s,1}|^2,\;\;\rho^u_2=0.98=|\phi_0^{u,2}|^2,\;\;
\rho^u_3=1=|\phi_0^{u,3}|^2.
\end{eqnarray}
	The above roots define three cw solutions of the form \eqref{cw}, namely,
	$u^{s,1}_s$, $u^{u,2}_s$ and $u^{u,3}_s$ respectively. The solution $u^{s,1}_s$
	is asymptotically stable due to Theorem \ref{fc} as it satisfies
	$|\phi_0^{s,1}|^2<\Omega/3=1/3$, i.e., the stability criterion of Proposition \ref{stab} $1 (a)$. The
other two solutions  $u^{u,2}_s$ and $u^{u,3}_s$ are unstable since they satisfy the instability
criterion $|\phi_0^{u,2}|^2,|\phi_0^{u,3}|^2>\Omega/2=0.5$ of Proposition \ref{stab} $2(a)$. Figure
\ref{figure9} depicts the spatiotemporal evolution of the initial condition
	\begin{eqnarray}
		\label{pertin}
		u_0(x)=u_s^{1,s}(0)+a\exp\left(\frac{\mathrm{i}\kappa\pi x}{L}\right),
\end{eqnarray}	
for $a=1$, $\kappa=2$, $L=50$. The initial condition \eqref{pertin} with $a=1$ is a large amplitude
perturbation of the cw solution $u_s(t)$. The instability illustrated in the left panel, manifested
again by the emergence of extreme events, is transient, as predicted by Theorem \ref{fc}. In the right
panel it is shown that solution approaches the asymptotically stable cw $u_s^{s,1}$. The convergence
becomes even more evident as verified in Fig. \ref{figure12a}, where the evolution of the density of the solution
$|u(x^*,t)|^2$ at a specific point $x=x^*$ is plotted. This $x^*$ is chosen to be the point where the solution exhibits
an extreme spot, here $x^*=10$, as seen in the left panel of Fig. \ref{figure9}. The plot depicts the convergence of the
density to the amplitude of the asymptotically orbitally stable cw, as predicted.
\paragraph{The case $K=1$ of the wave number of the initial condition \eqref{eq2}.} We conclude with a comment for the case
$K=1$ of the wave number of the plane wave initial condition which may also explain the formation of the dendritic patterns
for higher wavenumbers. To assist us on this, we plot in Fig. \ref{figure16} the spatiotemporal evolution of the density
$|u(x,t)|^2$ of the initial condition \eqref{eq2} for $K=1$, while all other parameters are $\Omega= -2$, $\gamma=0.01$,
$\Gamma=0.1$, as in the case of the study of Fig. \ref{figure1} and $L=200$.

The two panels present two different regions. As one can readily see, the patterns observed in Fig. \ref{figure1} are
missing and the dynamics are found to be essentially different compared to the case $K\geq2$. However, these figures may be
used to provide a useful understanding of the potential mechanism that may lead to the
formation of the dendritic patterns observed in the case $K\geq 2$.

Indeed, noting that in the early stages of the evolution
linear effects are more dominant, a plane wave initial condition
will behave as a solution to the linear problem, and as such, will be
subject to linear superposition. The single maximum/minimum initial
state evolves as above. If more maxima/minima are added, the
initial states around the minima/maxima will also undergo such
evolution, but now, constructive and destructive interference may
allow for more intricate patterns to form. The constructive
interference will lead to the peaks of the dendritic structure,
while destructive interference will leave some small amplitude
waves on the constant background (seen in all numerical simulations), which have
no effect on the overall phenomenon.

\begin{figure}[!htbp]
	\centering
	\includegraphics[scale=0.33]{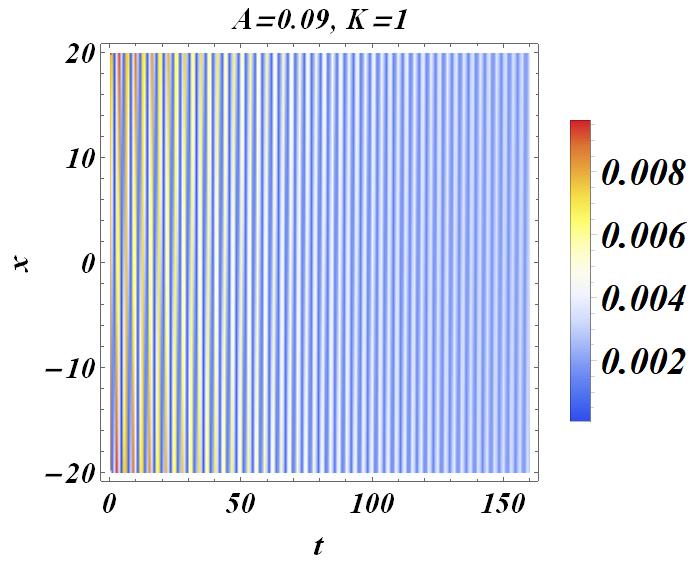}
	
	\caption{(Color Online) Parameters \eqref{eq1}-\eqref{eq3}:  $\nu=\sigma=1$,
		$\gamma=0.01$, $\Gamma=0.1$, $\Omega=-2$. Dynamics of the initial condition
		\eqref{eq2} for $A=0.09$, $K=1$, and $L=200$. The  panel shows the
		spatiotemporal evolution of the density for $x\in [-20,20]$ and $t\in [0,160]$. }
\label{figure17}	
\end{figure}

We also note that the dynamics we observed when $K=1$ is no exception to the findings of
the article suggested by Proposition \ref{stab} and Theorem \ref{fc}, which seem to be rather sharp: Since the amplitude of the initial
condition satisfies $A^2=1>\gamma$, we expect, according to the instability case 1(b) of Proposition \ref{stab}, transient
modulational instabilities and then convergence to the orbitally stable cw state of amplitude \eqref{root1}, as in the case
of $K=2$.
On the other hand, when switching the initial amplitude in order to satisfy the stability criterion  1(b) of Proposition
\ref{stab} $A^2<\gamma$ with $k^2\in I_{k^2}$, we don't expect transient instabilities prior the convergence to the
asymptotically stable cw state. This is indeed the case when $A=0.09$  ($A^2=0.0081<\gamma=0.01$) and $K=1$ which
corresponds to $k^2=0.0002\in I_{k^2}$, as shown in the evolution of the initial condition in Fig. \ref{figure17}.

\section{Conclusions}
In this paper, we examined the dynamics of the linearly damped and
time-periodically driven Schr\"odinger equation, initiated by the simplest
initial condition, namely, a single mode plane wave.  This  gives rise to
highly non-trivial transient dynamics manifested by the birth of rogue waves
reminiscent of the Peregine soliton in terms of its temporal decay and spatial
profile.  We also identified the formation of spatiotemporal patterns
resembling the one observed in the semiclassical Hamiltonian limit. Our
explorations justified that this behavior is far from the one of the integrable
limit and discussed its robustness as the parameters of the model are varied.
We also found that suitable functionals may detect the emerged rogue waves as their extrema. This observation can be
potentially useful in designing analytical pursuits establishing the existence of extreme wave events by variational
methods.

The asymptotic behavior
of solutions for the parametric regimes examined in the simulations was
justified by the orbital stability of stationary states of the model, aided by the existence of the global attractor of the
system.

We also examined the stability of the patterns when the plane wave initial condition is perturbed by noise. We found that
the noisy perturbation of the initial condition destructs the dynamics observed previously. Such an instability was also
found when repeating the experiment \cite{basis} for the semiclassical limit NLS perturbing the relevant vanishing initial
condition, also by noise. These findings suggest that the  semiclassical type dynamics are unstable in the
presence of noisy perturbations, and may inspire further investigations concerning the physical observability of such structured spatiotemporal patterns \cite{Rev1_b}, when random effects are present.

We believe that our findings for the important example of the AC- damped and driven NLS support further the universality of
extreme wave dynamics in systems where the NLS equation is the underlying model. Further studies may include more general
initial conditions in the form of a superposition of such plane waves and more general norm-based diagnostic tools. Relevant
investigations are in progress, and results will be reported in upcoming works.
\section*{Acknowledgment}
\vspace{-0.5cm}
We would like to thank the referees for their constructive comments and suggestions (including the one for the study of
noisy perturbations), which improved considerably the original version of the paper.

\end{document}